\documentclass[submission, Phys]{SciPost}

\usepackage{amsmath}
\usepackage{epic,eepic,color,graphicx}
\usepackage{centernot}
\usepackage{diagbox}
\usepackage[titletoc,title]{appendix}
\usepackage{lscape}
\usepackage{caption, subcaption, float}
\usepackage{cite}

\hypersetup{
    colorlinks,
    linkcolor={red!50!black},
    citecolor={blue!50!black},
    urlcolor={blue!80!black}
}

\usepackage[bitstream-charter]{mathdesign}
\DeclareSymbolFont{usualmathcal}{OMS}{cmsy}{m}{n}
\DeclareSymbolFontAlphabet{\mathcal}{usualmathcal}

\newcommand{\bra}[1]{\ensuremath{\left\langle#1\right|}}
\newcommand{\ket}[1]{\ensuremath{\left|#1\right\rangle}}

\newcommand{\expval}[1]{\ensuremath{\left\langle #1 \right\rangle}}

\begin{document}
\sloppy

\newpage
\setcounter{page}{1}

\begin{center}{\Large \textbf{
Entanglement entropy in quantum spin chains with broken parity number symmetry\\
}}\end{center}

% TODO: write the author list here. Use first name (+ other initials) + surname format.
% Separate subsequent authors by a comma, omit comma and use "and" for the last author.
% Mark the corresponding author with a superscript star.
\begin{center}
A. Jafarizadeh\textsuperscript{1$\star$} and
M. A. Rajabpour\textsuperscript{1}
\end{center}

\begin{center}
{\bf 1} Universidade Federal Fluminense, Niter\'oi, Brazil
\\
%{\bf 2} Universidade Federal Fluminense, Niter\'oi, Brazil
%\\
${}^\star$ {\small \sf arashjafarizadeh@id.uff.br}
\end{center}

\begin{center}
\today
\end{center}

% For convenience during refereeing (optional),
% you can turn on line numbers by uncommenting the next line:
%\linenumbers

\section*{Abstract}
{\bf
Consider a generic quantum spin chain  that can be mapped to free quadratic fermions via Jordan-Wigner (JW) transformation. In the presence of arbitrary boundary magnetic fields, this Hamiltonian is no longer a quadratic Hamiltonian after JW transformation. Using ancillary sites and enlarging the Hamiltonian we first introduce a bigger quadratic Hamiltonian. Then we diagonalize this enlarged Hamiltonian in its most generic form and show that all the states are degenerate because of the presence of a zero mode. The eigenstates of the original spin chain with boundary magnetic fields can be derived after appropriate projection. We study in depth the properties of the eigenstates of the enlarged Hamiltonian. In particular we find: 1) the eigenstates in configuration bases, 2) calculate all the correlation functions, 3) find the reduced density matrices, 4) calculate the entanglement entropy.  We show that the generic eigenstate of the enlarged Hamiltonian (including the eigenstates of the original spin chain) breaks the parity number symmetry and consequently one needs to take care of some technicalities regarding the calculation of the reduced density matrix and entanglement entropy. Interestingly we show that the entanglement structure of these eigenstates is quite universal and independent of the Hamiltonian. We support our results by applying them to a couple of examples.
}

\vspace{10pt}
\noindent\rule{\textwidth}{1pt}
\tableofcontents\thispagestyle{fancy}
\noindent\rule{\textwidth}{1pt}
\vspace{10pt}

\section{Introduction}\label{sec:Introduction}
%Quantum entanglement has a vast number of applications in the emerging technologies of quantum computing \cite{HOLZHEY1994443}, and has been used to realize some quantum effects -like quantum teleportation- experimentally. 
%Quantum entanglement in many body systems has been studied in a great detail in the early 21$^{st}$ century. There are comprehensive reviews on the applications of the entanglement entropy in condensed matter physics \cite{LAFLORENCIE20161}, quantum field theories \cite{Casini_2009}, integrable models \cite{Castro_Alvaredo_2009}, conformal field theory \cite{Calabrese_2009}, and topological systems \cite{Kitaev_2006, Levin_2006}. To review the applications of the bipartite entanglement entropy of the ground state in many-body quantum systems see \cite{Amico_Fazio_2008, Affleck_2009, Refael_2009, Peschel_Eisler_2009, Eisert_2010, Modi_2012, LAFLORENCIE20161, De_Chiara_2018}, and for the ground state of 1D critical and non-critical quantum spin chains see \cite{latorre_2003,Vidal_Kitaev_2003, Hastings_2007}.

There exist many quantum spin chains which can be transformed into quadratic fermion models using Jordan Wigner (JW) transformation, the spin-$1/2$ $XY$ chain is just one example of such spin chains. These fermionic models have been studied thoroughly in the past, and it was shown that they are exactly solvable \cite{LIEB1961407}. 
%For spins (or fermions) with translational symmetry (for instance on a ring), the model is analytically solvable \cite{Pasquale_Antonella_2009}. 
In the context of the non-interacting fermions, the calculation of some quantities such as reduced density matrix (RDM) \cite{Peschel_2001,Peschel_2004}, entanglement \cite{Vidal2001,jin2004quantum,Keating2004,Peschel_Eisler_2009} and formation probabilities \cite{Franchini_2005,NajafiRajab2015, Najaf_Rajab_2020} can be written in terms of correlation functions, which reduces the adversity of such calculations. The connection to the free fermion models also makes it possible to study the R\'enyi entanglement entropy for excited eigenstates of free fermions and related spin chains \cite{Alba_2009,Alcaraz_Sierra_2011,Berganza_2012,Ares_2014,Rigol_2017,PhysRevB.99.075123, Rajabpour_2019,2020arXiv201013973Z,2020arXiv201016348Z}. 

%Free fermion models play an important role in the study of entanglement in many body physics. These models have been studied thoroughly in the past \cite{LIEB1961407, McCoy_1970}. It was first shown in \cite{LIEB1961407} that they are exactly solvable and they can be related to the quantum spin chains in the family of $XY$-model. Integrablity of such models, even in long range interaction, makes them a viable model to study the entanglement. The bipartite entanglement entropy has been well studied for the ground state of 1D critical and non-critical quantum chains \cite{Vidal_Kitaev_2003, Hastings_2007}. An outstanding outcome of these studies was the role of the entanglement entropy to distinguish different phases and classify the critical points of the system in consistent with the traditional classifications. Re\'neyi Entanglement entropy has also been studied for excited eigenstates of free fermions and related spin chains \cite{Rigol_2017, Rajabpour_2019, Alba_2009}.

Most of the above mentioned studies were based on the bulk properties, however, there are also studies regarding the entanglement entropy in systems with boundaries and impurities. In the presence of boundaries analytical and numerical calculations of the entanglement entropy can be more challenging due to the lack of the translational symmetry. The entanglement entropy of a few quantum chains in the presence of the boundaries has been studied with analytical and numerical techniques, see for instance \cite{PhysRevLett.96.100603, PhysRevA.74.050305, PhysRevLett.99.087203, PhysRevB.77.045106, Castro_Doyon_2009, Affleck_2009, PhysRevB.88.075112, Fagotti_2011}. An interesting consequence of presence of the boundary is the breach of connection between spin chains and quadratic fermion models, specially in subsystem entanglement \cite{Igloi2010,Fagotti_2011}. An spin chain (containing $L$ spins) with arbitrary boundary magnetic fields can be modeled as: bulk Hamiltonian plus boundary terms,
\begin{equation}\label{eq:Ham_intro}
    H_{_{\mathrm{Spin Chain}}}=H_{_{\mathrm{bulk}}}+H_{_{\mathrm{boundary}}}.
\end{equation}
The $H_{_{\mathrm{boundary}}}$ resembles the effect of boundary on far end sites or spins. For instance, a general boundary condition produced by external magnetic fields reads as $H_{_{\mathrm{boundary}}}\!\!=\Vec{B}_{_1}.\Vec{S}_{_1}+\Vec{B}_{_L}.\Vec{S}_{_L}$, where $\Vec{S}_1$, $\Vec{S}_L$ are spin operators at the beginning and end of the chain, and $\Vec{B}_L$, $\Vec{B}_1$ denote the preferred direction of alignment of boundary magnetic fields.  While such a non uniform boundary condition can be physically valid for a spin chain, the fermionization of such a spin chain would end up in a non-physical fermion model. A fermion model which violates the parity symmetry will break the locality. Locality forbids a Hamiltonian that does not commute with fermionic parity symmetry \cite{BRAVYI_2002,Banuls_Wolf_2007, moriya_2006}. However, as far as one is concerned with spin model this violation is not a problem.

Some spin models, such as the XXZ chain, with arbitrary direction of boundary magnetic fields (ADBMF) have been already exactly solved by the thermodynamic Bethe ansatz method \cite{Vega_1993,NEPOMECHIE2002,Nepomechie_2003,CAO_2003,CAO_2013,CAO_2013_2,NICCOLI_2013,belliard_2013,Nepomechie_2013_3,Faldella_2014, Kitanine_2014,LI_2014,Pozsgay_2018}. However, calculation of some quantities such as the entanglement entropy seems out of reach at this moment. In this work we take advantage of a method proposed in \cite{Colpa_1979,BARIEV_1991}, see also \cite{Bilstein_1999,Campostrini_2015,XavierRajabpour2020} to transform a generic quantum spin chain; with $H_{_{\mathrm{bulk}}}$ that can be mapped to the Hamiltonian of free fermions; with non uniform magnetic field at the boundaries into a quadratic fermion Hamiltonian. It is done by adding auxiliary spins to the system with coupling to the boundary spins and enlarging the Hilbert space of the original model. Afterward, we would be able to fermionize the spin system via JW transformation and make use of postulation of quadratic fermion Hamiltonians to study the quantities of interest. The quadratic (or bi-linear) form of Hamiltonian in fermionic operators is crucial since the Hamiltonian can be diagonalized exactly and the correlation functions can be reduced to the expectation values of pairs of fermionic operators (Wick theorem \cite{Wick_1950}). To retrieve the eigenstates of original model, we can use particular projection of new model's eigenstates. In \cite{Colpa_1979}, the same method was used to study a fermionic model with linear operator terms, which breaks the parity. In this case, the author couples the auxiliary fermions to every other in the system, and gets a quadratic Hamiltonian.

Starting from the enlarged bi-linear fermionic representation of the Hamiltonian, similar to \cite{Colpa_1979,Bilstein_1999,XavierRajabpour2020}, the degeneracy of (at least) two is expected. This degeneracy results in degenerate ground states with opposing parities, which allows us to find a superposition for the ground state with broken parity number. However, for such an eigenstate, we would not be able to use conventional methods to find the RDM, entanglement and so forth. We study the aspects of this state and how it is related to the ground states of the original boundary magnetic field (BMF) Hamiltonian. Also, the comparison of the correlation of this state with those of the parity symmetry states is investigated. Such an investigation enables us to make an ansatz for RDM based on the correlation matrices. This ansatz was first proposed in \cite{XavierRajabpour2020} for a special type of parity broken state and special type of subsystem. Here, we generalize the previous results and provide a proof for consistency of this ansatz. In addition, the RDM has been calculated in different methods.
We also show that all of these results can be extended to arbitrary enigenstates of the Hamiltonian (\ref{eq:Ham_intro}). 

Having the RDM in terms of correlation matrices facilitates the calculation of entanglement. For the parity broken state the behavior of entanglement entropy with respect to parity number is intriguing. In this paper, we mostly deal with a connected subsystem starting from one end of the chain to some point in the middle of the system. An interesting observation is that based on the way one breaks the parity of the ground state entanglement can be minimized or unaffected. Besides, we show that with an adjustment to the Peschel method \cite{Peschel_2001, Vidal_Kitaev_2003,latorre_2003,Peschel_2004,jin2004quantum}, it is possible to get the entanglement of parity broken state in terms of entanglement of parity protected ground states. With these results, we will be able to study the effect of boundary conditions on the entanglement of subsystem, in particular, change in the direction of magnetic field at boundaries on the value of entanglement entropy. 

It is worth mentioning that one can treat the enlarged Hamiltonian as a model of interest and study its different properties independent of the quantum spin chain with boundaries. This might be an interesting approach to study a phenomena like spontaneous breaking of parity number symmetry.
Having this in mind  most of our studies and results go beyond just the quantum spin chains with ADBMF. We find the correlations, reduced density matrix and entanglement for generic eigenstates of the enlarged Hamiltonian. Quite surprisingly, the general features of the results are quite independent of the bulk Hamiltonian. 

The remainder of this paper is structured as follows: In section \ref{sec:StatementOfResults} we first summarize the main results of the paper.
In section \ref{sec:Ham_and_Diagonaliz}, we start by general quadratic Hamiltonian that one can derive after the addition of extra sites. We, provide a general discussion on diagonalization of such a system, and the connection to the eigenstates of our original model, the BMF Hamiltonian. In section \ref{sec:Correlation}, we use the result of diagonalization of quadratic Hamiltonian to find the correlation matrices for the generic eigenstates including the most interesting ones, i.e. states with $\pm1$ parity and the one with no parity. Section \ref{sec:RDM} contains study of the reduced density matrix (RDM). We first present a general formulation of RDM using Berezin integration of Grassmann variables, followed by the RDM in terms of correlation matrices. We address the RDM not only for typical eigenstates of Hamiltonian, but also eigenstates which break the parity. With the results of former sections, we study the entanglement in section \ref{sec:Entanglement}. Notably, we dig to the behavior of entanglement for a state which defies the parity symmetry. In section \ref{sec:betastate} we give an interesting physical interpretation of the parity-broken state based on three-part system. This gives a simple way to reproduce some of our results in especial cases.
Finally, in section \ref{sec:Examples}, we look at some interesting examples of systems with no parity symmetry. The first being free fermion model with no bulk term, and the second being the $XY$-spin chain with arbitrary boundary magnetic fields.
%%The change in the bulk entanglement entropy of system is due to boundary conditions is called ``\emph{boundary entropy}" (BE). This BE in the context of the entanglement entropy has been studied in quantum spin chains \cite{PhysRevLett.96.100603, PhysRevA.74.050305, PhysRevLett.99.087203, PhysRevB.77.045106, PhysRevB.88.075112, Fagotti_2011, Najafi_2016, Hong_Hao_2017}. In particular, using DMRG technique the authors of \cite{Barthel_Chung_2006} estimate the universal BE for the transverse field Ising chain with particular boundary conditions, mainly boundary magnetic field in the $x$-direction.

\section{Summary of Main Results}\label{sec:StatementOfResults}

Consider the spin chain Hamiltonian 
\begin{equation}\label{eq:Ham_boundary}
    H_{_{\mathrm{Spin Chain}}}=H_{_{\mathrm{bulk}}}+\Vec{b}_{_1}.\Vec{S}_{_1}+\Vec{b}_{_L}.\Vec{S}_{_L},
\end{equation}
where  $\Vec{S}_1$ and $\Vec{S}_L$, are spin operators at the beginning and end of the chain, and $\Vec{B}_L$, $\Vec{B}_1$ are arbitrary boundary magnetic fields. We consider here bulk Hamiltonians, i.e. $H_{_{\mathrm{bulk}}}$, that can be mapped to quadratic free fermions via Jordan-Wigner (JW) transformation.
The above Hamiltonian does not have quadratic form after JW transformation and a priory is not clear that can be solved exactly. However, using the ancillary spins, $S_0$ and $S_{L+1}$ one can transform the above Hamiltonian to
\begin{equation}\label{eq:Ham_boundary-enlarged}
    H=H_{_{\mathrm{bulk}}}+b_{1}^xS^x_{0}S^x_{1}+b_{1}^yS^x_{0}S^y_{1}+b_{1}^zS^z_{1}
    +b_{L}^xS^x_{L}S^x_{{L+1}}+b_{L}^yS^y_{L}S^x_{{L+1}}+b_{L}^zS^z_{L}
\end{equation}
The above Hamiltonian after JW transformation has a quadratic form and can be solved exactly. The Hamiltonian (\ref{eq:FermionHamil}) is considered with this goal in mind.
The eigenstates and eigenvalues of the original Hamiltonian can be found with proper projections. The section \ref{sec:Ham_and_Diagonaliz} of this paper shows how this procedure can be done in its most general form. The idea of using ancillary sites to solve boundary spin chains is already explored before in \cite{Colpa_1979,BARIEV_1991,Bilstein_1999,Campostrini_2015,XavierRajabpour2020}, however, in this work we solve the problem in its most general form without restricting to a particular Hamiltonian.
The eigenstates, eigenvalues and correlation functions are found in the most general cases in sections \ref{sec:Ham_and_Diagonaliz} and \ref{sec:Correlation}.

Interestingly all the eigenstates of the Hamiltonian (\ref{eq:Ham_boundary-enlarged}) have at least the degeneracy of two due to the presence of a zero mode. In the fermionic language the vacuum, i.e. $\ket{0}_\eta$, and the state with the zero mode excited, i.e. $\eta_{0}^\dagger\ket{0}_\eta$,  are degenerate. This means one can define 

\begin{equation}\label{eq:Betastate}
\ket{\beta}=\frac{1}{\sqrt{1+|\beta|^2}}\big(\ket{0}_\eta+\beta\eta_{0}^\dagger\ket{0}_\eta\big);
\end{equation}
as the most generic ground state of this Hamiltonian\footnote{Note that the same can be done with all the eigenstates. Our results can be extended for the $\ket{\beta}$ states made of each eigenstates.}.  In particular, the ground state of the boundary quantum spin chain can be derived out of the especial cases of $\beta=\pm1$. Due to its generality in sections \ref{sec:Ham_and_Diagonaliz} and \ref{sec:Correlation} we study some basic properties of this state such as parity and correlation functions. This state has also an interesting interpretation as three part system which we will explore in section \ref{sec:betastate}.

In section \ref{sec:RDM} we give two different forms for the reduced density matrix of the generic $\ket{\beta}$ state. We first find the reduced density matrix in fermionic coherent basis. The presented form can be used to calculate the R\'enyi entanglement entropy as it was shown in section \ref{sec:Entanglement}. We also think that this form is useful to calculate the formation probabilities in the configuration basis which are although interesting for their own sake, we are not going to explore them in this paper. The second method is a generalization of the results of \cite{XavierRajabpour2020} for generic eigenstates of the generic Hamiltonians with arbitrary $\beta$. It is an extension of a method which is based on making an ansatz for the reduced density matrix and then fix the exact form by matching the correlation functions \cite{Peschel_2003}. Our result for $\beta=\pm1$ gives also the reduced density matrix of the quantum spin chain with arbitrary boundary magnetic fields. Using the results of this section in section \ref{sec:Entanglement} we write an exact formula for the entanglement entropy of the $\ket{\beta}$ state  which its complexity grows polynomially with the size of the subsystem. As before, this also solves the problem of the calculation of the entanglement entropy of the eigenstates of the Hamiltonian (\ref{eq:Ham_boundary}).

In section \ref{sec:betastate} we show that the $\ket{\beta}$ state has a very interesting property. It can be written like a three-part system which helps to understand the entanglement of the  ancillary sites with the rest of the system using this simple interpretation. This property makes studying the Hamiltonian \eqref{eq:Ham_boundary-enlarged} and its entanglement content interesting for its own sake independent of the original motivation of solving the quantum spin chain with arbitrary boundary magnetic fields. This has been the main motivation to start with the fermionic version of the enlarged Hamiltonian in section \ref{sec:Ham_and_Diagonaliz}.

All of the discussions up to section \ref{sec:Examples} are independent of the Hamiltonian and also valid for arbitrary eigenstates of the Hamiltonians (\ref{eq:Ham_boundary}) and (\ref{eq:Ham_boundary-enlarged}). In section \ref{sec:Examples} we give two explicit examples. The first one is a slight generalization of the  Hamiltonian (\ref{eq:Ham_boundary-enlarged}) without a bulk term. In the first sight this seems oversimplification but interestingly a lot of entanglement properties of the generic Hamiltonian have similar features which was the main motivation for its presentation. To give a non-trivial example we also study the entanglement entropy in the XY chain. The boundary entropy of this model is already studied in \cite{XavierRajabpour2020}, however, here we fill a few holes such as the presentation of the exact extra zero modes present in the Ising chain which their form is essential to calculate the entanglement entropy exactly.

\section{The Hamiltonian}\label{sec:Ham_and_Diagonaliz}

In this section, we introduce a general bi-linear fermionic Hamiltonian which is related to spin chain Hamiltonians with boundary magnetic field \eqref{eq:Ham_intro}. We first study the general properties of this model and point out the evident zero mode of the model. In subsection \ref{subsec:QuadHamDiagonaliz}, we tend to present the diagonalization procedure using the standard methods. Later, in subsection \ref{subsec:QuadHamBetaState}, we look into the particular eigenstates of the Hamiltonian which does not respect the parity number symmetry. Finally, subsection \ref{subsec:SelectionRule} presents a set of selection rules to retrieve the eigenvalues of the boundary magnetic field (BMF) model. Interestingly the structure that unfold is very similar to the ones in \cite{XavierRajabpour2020}, see also \cite{Bilstein_1999} with a slightly different notation.

\subsection{General properties}\label{subsec:QuadHamGen}

We are going to study the quadratic fermionic Hamiltonian of the form:
\begin{equation}\label{eq:FermionHamil}
    H=H_0+H_b
\end{equation}
where
\begin{equation}\label{eq:HamQuad}
H_0=\sum_{i,j=1}^{L}\big[c_{i}^{\dagger}A_{ij}c_{j}+\tfrac{1}{2}c_{i}^{\dagger}B_{ij}c_{j}^{\dagger}+\frac{1}{2}c_{i}B_{ji}^{*}c_{j}\big]-\frac{1}{2}\text{tr}(\mathbf{A}^*),
\end{equation}
and
\begin{equation}\label{eq:HamLine}
\begin{aligned}
H_b=\sum_{j=1}^{L}\big[\alpha_j^0(c_{0}c_{j}-c_{0}^\dagger c_{j})+\alpha_j^{L+1}(c_{j}c_{L+1}^{\dagger}+c_{j}c_{L+1})+ \text{H. C.} %{\alpha_{j}^{0}}^*(c_{j}^{\dagger}c_{0}^\dagger-c_{j}^{\dagger}c_{0})+{\alpha_{j}^{L+1}}^*(c_{L+1}c_{j}^{\dagger}+c_{L+1}^{\dagger}c_{j}^{\dagger})
\big].
\end{aligned}
\end{equation}
%and $L$ is the total number of sites. 
%In the above expression, $H_0$ corresponds to a general Quadratic model which has been studied vastly \cite{GHS-note2, }. 
As we mentioned in the introduction, $H_b$ part can be related to the boundary terms. Note that the above Hamiltonian is a bit more general than just the extension of a quantum chain with boundaries because the added extra sites $0$ and $L+1$ are coupled to all the sites. Note also that we are not bounded here to any particular dimension. That means all of the upcoming results, as far as we do not talk about spin chains, are valid in arbitrary dimensions. Here, $\mathbf{A}$ matrix should be Hermitian and $\mathbf{B}$ matrix should be anti-symmetric. The Hamiltonian can be written as:
\begin{equation}\label{eq:Ham matrix form}
\mathbf{H}=\frac{1}{2}\left(\begin{array}{c c}
\mathbf{c}^\dagger & \mathbf{c}
\end{array}\right) \mathbf{M} \left(\begin{array}{c}
\mathbf{c} \\ \mathbf{c}^\dagger
\end{array}\right),
\end{equation}
where $\left(\begin{array}{c c}
\mathbf{c}^\dagger & \mathbf{c}
\end{array}\right)$ stands for $\left(\begin{array}{c c c c c c}
c_0^\dagger &\cdots& c_{L+1}^\dagger & c_0 &\cdots &c_{L+1}
\end{array}\right)$. 
%\begin{equation}%\label{eq:M matrix}
%\mathbf{M}= \left(\begin{array}{c c}
%\mathbf{A}^\prime & \mathbf{B}^\prime \\
%-\mathbf{B}^{\prime*} & -\mathbf{A}^{\prime*}
%\end{array}\right).
%\end{equation}
%One can write the $\mathbf{A}^\prime$ and $\mathbf{B}^\prime$ matrices in terms of $\mathbf{A}$ and $\mathbf{B}$ matrices respectively. In general, 
The matrix $\mathbf{M}$ in \eqref{eq:Ham matrix form} should look like below.\small
\begin{equation}\label{eq:M matrix}
\mathbf{M}=\left(\begin{array}{ccccc|ccccc}
0&-\alpha_1^{0}&\cdots&-\alpha_L^{0}&0&0&-{\alpha_{1}^{0}}^*&\cdots&-{\alpha_{L}^{0}}^*&0\\
-{\alpha_1^{0}}^*&&&&-{\alpha_{1}^{L+1}}^*&{\alpha_{1}^{0}}^*&&&&-{\alpha_{1}^{L+1}}^*\\
\vdots&&\scalebox{2}{$\mathbf{A}$}&&\vdots&\vdots&&\scalebox{2}{$\mathbf{B}$}&&\vdots\\
-{\alpha_L^{0}}^*&&&&-{\alpha_{L}^{L+1}}^*&{\alpha_L^{0}}^*&&&&-{\alpha_{L}^{L+1}}^*\\
0&-\alpha_{1}^{L+1}&\cdots&-\alpha_{L}^{L+1}&0&0&{\alpha_{1}^{L+1}}^*&\cdots&{\alpha_{L}^{L+1}}^*&0\\\hline
0&{\alpha_1^{0}}&\cdots&{\alpha_L^{0}}&0&0&{\alpha_1^{0}}^*&\cdots&{\alpha_L^{0}}^*&0\\
-{\alpha_1^{0}}&&&&{\alpha_{1}^{L+1}}&{\alpha_1^{0}}&&&&{\alpha_{1}^{L+1}}\\
\vdots&&\scalebox{2}{$-\mathbf{B}^*$}&&\vdots&\vdots&&\scalebox{2}{$-\mathbf{A}^*$}&&\vdots\\
-{\alpha_L^{0}}&&&&{\alpha_{L}^{L+1}}&{\alpha_L^{0}}&&&&{\alpha_{L}^{L+1}}\\
0&-\alpha_{1}^{L+1}&\cdots&-\alpha_{L}^{L+1}&0&0&{\alpha_{1}^{L+1}}^*&\cdots&{\alpha_{L}^{L+1}}^*&0
\end{array}\right).
\end{equation}\normalsize

For later use, we define the following spin operators at positions $0$ and $L+1$,
\begin{equation}
\sigma_0^x=c_0+c_0^\dagger\quad\text{and}\quad\sigma_{L+1}^x=(1-2c_0 c_0^\dagger)\prod_{l=1}^{L}(1-2c_l^\dagger c_l)(c_{L+1}+c_{L+1}^\dagger).
\end{equation}
These operators commute with each other and the Hamiltonian \eqref{eq:FermionHamil}; i.e.
\begin{equation}
\left[H,\sigma_{L+1}^x\right]=\left[H,\sigma_{0}^x\right]=\left[\sigma_{0}^x,\sigma_{L+1}^x\right]=0.
\end{equation}
The above relations are important in upcoming sections.

\subsection{Zero mode eigenstates}\label{subsec:QuadHamZeroMode}

The $\mathbf{M}$ matrix in \eqref{eq:Ham matrix form} has at least two eigenvectors with zero eigenvalues. These eigenvectors correspond to the modes with zero energy. 
Later on in this paper, we present a formulation (section \ref{subsec:QuadHamDiagonaliz}) which simplifies the effort to find the correlations (section \ref{sec:Correlation}), reduced density matrix (section \ref{sec:Entanglement}) and other properties of the system such as entanglement. To use such a formulation, one needs to identify and write the zero eigenvectors in a correct form (see the following subsection).

Due to the form of the $\mathbf{M}$ matrix \eqref{eq:M matrix}, we expect to have two zero modes. These zero modes do not depend on the parameters and interactions of the system. They can be written as:
\begin{equation}\label{eq:zeromode evec 1}
\ket{u_0^1}=\left(\begin{array}{c}
\sqrt{a}e^{i\theta_1}\\ 0\\ \vdots\\ 0\\ \sqrt{\tfrac{1}{2}-a}e^{i\phi_1}\\ \sqrt{a}e^{i\theta_1}\\0 \\ \vdots \\0 \\ -\sqrt{\tfrac{1}{2}-a}e^{i\phi_1} \end{array}\right),\qquad \ket{u_0^2}=\left(\begin{array}{c}
\sqrt{\tfrac{1}{2}-a}e^{i\theta_2}\\ 0\\ \vdots\\ 0\\ -\sqrt{a}e^{i\phi_2}\\ \sqrt{\tfrac{1}{2}-a}e^{i\theta_2}\\0 \\ \vdots \\0 \\ \sqrt{a}e^{i\phi_2} \end{array}\right).
\end{equation}
The orthogonality condition of these states requires the following equality to hold for the parameters
\begin{equation}
\theta_1-\theta_2=\phi_1-\phi_2.
\end{equation}
The above zero modes are independent of the parameters of the Hamiltonian. In the rest, for the sake of simplicity, we take all the angles ($\theta_1,\theta_2,\phi_1,\phi_2$) to be zero and put  $a=\tfrac{1}{4}$. Taking other values does not change the upcoming results. Depending on the values of coupling parameters there could arise more zero modes in spectrum of $\mathbf{M}$ matrix\footnote{See section \ref{subsec:XYExample}.}.

\subsection{Diagonalization}\label{subsec:QuadHamDiagonaliz}

In this subsection, we want to briefly review the diagonalization and find the eigenstates of the Hamiltonian. First, we introduce a new operator $\mathbf{J}$ which acts on the eigenstates of the matrix $\mathbf{M}$ as
\begin{equation}\label{eq:JOperator}
\mathbf{J}\left(\begin{array}{c}
u\\ v\\
\end{array}\right)=\left(\begin{array}{c}
v^*\\ u^*\\
\end{array}\right).
\end{equation}
It is easy to show that the operator $\mathbf{J}$ anticommutes with the matrix $\mathbf{M}$, i.e. $\{\mathbf{M},\mathbf{J}\}=0$. As a consequence, if $\ket{V}$ is an eigenvector of matrix $\mathbf{M}$ with eigenvalue $\lambda$ then the vector $\mathbf{J}\ket{V}$ is also an eigenvector with eigenvalue $-\lambda$. It means that finding the eigenstates corresponding to positive eigenvalues is enough; we can get the eigenstates for negative eigenvalues by acting on the eigenstates of the positive eigenvalues with operator $\mathbf{J}$ \cite{vanHemmen1980}.

As a result of operator $\mathbf{J}$, the Hamiltonian \eqref{eq:Ham matrix form} can be diagonalized in the following form:
\begin{equation}\label{eq:HamDiago}
\mathbf{H}=\frac{1}{2}\left(\begin{array}{c c}
\mathbf{c}^\dagger & \mathbf{c}
\end{array}\right)\mathbf{U}^\dagger\mathbf{U} \mathbf{M}\mathbf{U}^\dagger\mathbf{U}  \left(\begin{array}{c}
\mathbf{c} \\ \mathbf{c}^\dagger
\end{array}\right)=\frac{1}{2}\left(\begin{array}{c c}
\mathbf{\eta}^\dagger & \mathbf{\eta}
\end{array}\right)\left(\begin{array}{c c}
\mathbf{\Lambda}&0\\0&-\mathbf{\Lambda}
\end{array}\right) \left(\begin{array}{c}
\mathbf{\eta} \\ \mathbf{\eta}^\dagger
\end{array}\right);
\end{equation}
The matrix $\mathbf{\Lambda}$ is a diagonal matrix with non-negative entries and matrix $\mathbf{U}$ has the block from:
\begin{equation}\label{eq:Diag_U_Mat}
\mathbf{U}=\left(\begin{array}{c c}
g & h\\ h^* & g^*\\ \end{array}\right). 
\end{equation}
The Hamiltonian can also be written with respect to new fermionic operators (Bogoliubov fermions) as
\begin{equation}
\mathbf{H}=\sum_{k}\lambda_k\eta_k^\dagger\eta_k-\frac{1}{2}\text{tr}(\mathbf{\Lambda}),
\end{equation}
where 
\begin{equation}
\left(\begin{array}{c}
\boldsymbol{\eta} \\ \boldsymbol{\eta}^\dagger
\end{array}\right)=\mathbf{U}  \left(\begin{array}{c}
\mathbf{c} \\ \mathbf{c}^\dagger
\end{array}\right).
\end{equation}
From \eqref{eq:JOperator} and special form of $U$ matrix, we get the following constrains on the elements of $\mathbf{g}$ and $\mathbf{h}$ matrices:
\begin{equation}
g_{k,L+1}=h_{k,L+1}\quad\text{and}\quad g_{k,0}=-h_{k,0}\quad\text{for}\quad k\neq0
\end{equation}

To conclude the diagonalization part, we write the explicit expression of $\eta$ operators:
\begin{eqnarray}
\eta_k\!\! &=&\!\!\sum_{j=0}^{L+1}g_{kj}c_j+h_{kj}c_j^\dagger=g_{k0}(c_{0}-c_{0}^\dagger)+\sum_{j=1}^L(g_{kj}c_j+h_{kj}c_j^\dagger)+g_{kL+1}(c_{L+1}+c_{L+1}^\dagger);\;\; k\neq0,\label{eq:eta from c} \\
\eta_0\!\!&=&\frac{1}{2}\big(c_0 +c_{L+1}+c_0^\dagger-c_{L+1}^\dagger\big).
\end{eqnarray}
From above, we can also write the $c$-fermions in terms of $\eta$-operators as:
\begin{eqnarray}
c_k&=&\sum_{j=1}^{L+1}(g_{j,k}^*\eta_j+h_{j,k}\eta_j^\dagger);\quad k\neq0,L+1,\\
c_0&=&\tfrac{1}{2}\eta_0+\tfrac{1}{2}\eta_0^\dagger+\sum_{j=1}^{L+1}(g_{j,0}^*\eta_j-g_{j,0}\eta_j^\dagger),\\
c_{L+1}&=&\tfrac{1}{2}\eta_{0}-\tfrac{1}{2}\eta_{0}^\dagger+\sum_{j=1}^{L+1}(g_{j,L+1}^*\eta_j+g_{j,L+1}\eta_j^\dagger).
\end{eqnarray}
This final result means that the following commutation (anticommutation) relations hold (for $k\neq 0$)
\begin{eqnarray}
\big[\sigma_{L+1}^x,\eta_k\big]&=&\big[\sigma_{L+1}^x,\eta_k^\dagger\big]=0,\label{eq:eta and sigmaX1}\\
\big\{ \sigma_{0}^x,\eta_k\big\}&=&\big\{ \sigma_{0}^x,\eta_k^\dagger\big\}=0.\label{eq:eta and sigmaX2}
\end{eqnarray}
Anticommutations \eqref{eq:eta and sigmaX2} will be useful in subsection \ref{subsec:SelectionRule} for sectorization of eigenstates of the Hamiltonian \eqref{eq:FermionHamil}.

\subsection{Eigenstates in configuration basis}\label{subsec:Configur_Basis}

The vacuum state $\ket{0}_{\eta}$ is the state which is annihilated by the action of all $\eta_k$ operators,
\begin{equation}\label{eq:etavacuum}
\eta_k\ket{0}_\eta=0,\quad\forall\; k.
\end{equation}
One can write the $\ket{0}_\eta$ as a superposition of configurations of the $c$-fermions \cite{Peschel_2001}. It is called the configurational basis of such a state. For even size, if the parity of $\ket{0}_\eta$ is $+1$ then we have:
\begin{equation}\label{eq:VacEtaVacC}
\ket{0}_\eta=\big(\text{det}[\mathbf{I}+\mathbf{R}^\dagger\mathbf{R}]\big)^{-\tfrac{1}{4}}e^{\tfrac{1}{2}\sum_{i,j}R_{ij}c^\dagger_ic_j^\dagger}\ket{0}_c
\end{equation}
where $c_j\ket{0}_c=0$ for all $j$, and the $\mathbf{R}$ matrix is an antisymmetric matrix defined as $\mathbf{g}.\mathbf{R}+\mathbf{h}=0$. The \eqref{eq:VacEtaVacC} works only for the case where parity of the $\ket{0}_\eta$ is $+1$, and/or the matrix $\mathbf{g}$ is invertible. Otherwise, to use the above equation, one needs to do a canonical transformation to change the parity of the vacuum for $\eta$-fermions, and/or to make the $\mathbf{g}$ invertible \cite{Zanardi_2007}.

Such a canonical transformation could be the following change in the creation and annihilation operators:
\begin{equation}
\left\{\begin{aligned}
&c_j\;\rightarrow\;\tilde{c}_j^\dagger, \\
&c_j^\dagger\;\rightarrow\;\tilde{c}_j.
\end{aligned}\right.
\end{equation}
We call this transformation \textit{Tilda transformation}. If we do such a change for even number of the indices, the parity would not change (in the case where $\mathbf{g}$ matrix is not invertible). For odd number of sites, it will change the parity of the state $\ket{0}_\eta$. As an example, when the vacuum state has parity $-1$, we do the tilda transformation
\begin{equation}
c_j\;\rightarrow\;\tilde{c}_j^\dagger\qquad\text{and}\qquad c_j^\dagger\;\rightarrow\;\tilde{c}_j
\end{equation}
for only one specific index $j$. Regardless of selected $j$, the entanglement properties of the vacuum would not change. If we choose $j=0$ then this transformation can be written in operator form as $\widetilde{\mathcal{P}}=e^{i\tfrac{\pi}{2}\sigma_0^x}$. 

As a matter of fact, any eigenstate of Hamiltonian \eqref{eq:FermionHamil} (with parity $+1$) can be written in configurational basis in an exponential form. Excited states are created by exciting different modes on the vacuum \eqref{eq:etavacuum} as:
\begin{equation}\label{eq:ExcitedState1}
\ket{\psi}=\ket{k_1,k_2,\cdots,k_N}=\prod_{k_j\in \mathbb{E}}\eta_{k_j}^\dagger\ket{0}_{\eta};\qquad E_\psi=\sum_{k_j\in\mathbb{E}}\lambda_{k_j}-\frac{1}{2}\text{tr}(\mathbf{\Lambda}),
\end{equation}   
where the set $\mathbb{E}$ can be any subset of indices from $0$ to $L+1$. We denote the set of indices of excited modes as $\mathbb{E}$ and the set of indices which are not excited as $\bar{\mathbb{E}}$ ($\bar{\mathbb{E}}\cup\mathbb{E}=\{0,1,\cdots,L+1\}$). Assume that we can write the following excited state in the configurational form as:
\begin{equation}\label{eq:ExcStRMat}
\ket{\psi}=C^\psi e^{\tfrac{1}{2}\sum_{i,j}R_{ij}^\psi c^\dagger_ic_j^\dagger}\ket{0}_c,
\end{equation}
where $C^\psi=(\det[\mathbf{I}+\mathbf{R^\psi}^\dagger\mathbf{R^\psi}])^{-\tfrac{1}{4}}$. For this excited state we have:
\begin{equation}\label{eq:RExciCondition1}
\begin{aligned}
&\eta_{_{k_j}}\ket{\psi}=0;\qquad k_j\in\bar{\mathbb{E}}, \\
&\eta_{_{k_n}}^\dagger\ket{\psi}=0;\qquad k_n\in\mathbb{E}.
\end{aligned}
\end{equation}
Therefore, from the above equation, we get:
\begin{eqnarray}
g_{k_j l}R^\psi_{lm}+h_{k_j m}=0;\qquad k_j\in\bar{\mathbb{E}},\label{eq:RmatExci1}\\
h_{k_n l}^*R^\psi_{lm}+g_{k_n m}^*=0;\qquad k_n\in\mathbb{E}.\label{eq:RmatExci2}
\end{eqnarray}
The generalized formula for the $\boldsymbol{R^\psi}$ would be:
\begin{equation}\label{eq:RMatExcSt}
\boldsymbol{\mathfrak{g}}\boldsymbol{R^\psi}+\boldsymbol{\mathfrak{h}}=0,
\end{equation}
where $\boldsymbol{\mathfrak{g}}$ and $\boldsymbol{\mathfrak{h}}$ are generalized versions of $\mathbf{g}$ and $\mathbf{h}$ given by
\begin{equation}
\mathfrak{g}_{nm}=\begin{cases}
g_{nm} & if\; n\in\bar{\mathbb{E}}\\
h_{nm}^* & if\; n\in\mathbb{E}\\
\end{cases}\qquad\qquad
\mathfrak{h}_{nm}=\begin{cases}
h_{nm} & if\; n\in\bar{\mathbb{E}}\\
g_{nm}^* & if\; n\in\mathbb{E}\\
\end{cases}
\end{equation}
Note that same as the case of vacuum state, one should make sure that the $\mathbf{\mathfrak{g}}$ has an inverse, and state $\ket{\psi}$ has the right parity. If not one should use the canonical (Tilda) transformation to be able to write a typical eigenstate in configuration basis. The \eqref{eq:ExcStRMat} gives an excited eigenstate in the configuration basis which is advantageous in the study of entanglement in later sections.

\subsection{Parity broken state}\label{subsec:QuadHamBetaState}

The quadratic Hamiltonian \eqref{eq:FermionHamil} commutes with the parity operator defined as $\mathbf{P}=(-1)^{\hat{N}}$ where $\hat{N}=\sum_{l=0}^{L+1}c_l^\dagger c_l$ is the fermion number operator. This means that the eigenstates of the Hamiltonian have fixed parity $P=\pm1$.
One can have eigenstates that do not respect the parity and these types of states are very interesting to study. For instance, these types of states are related to the ground state of spin systems with boundary magnetic fields which one example is given in section \ref{subsec:XYExample}.

We define a parity broken state ($\beta$-defected parity state) as:
\begin{equation}\label{eq:BetaParityBrokenState}
\ket{\beta}=\frac{1}{\sqrt{1+|\beta|^2}}\big(\ket{0}_\eta+\beta\eta_{0}^\dagger\ket{0}_\eta\big);
\end{equation}
where $\beta$ can be a complex number. The expectation value of parity for such a state is given by
\begin{equation}\label{eq:BetaParityBrokenParity}
P(\beta)=\bra{\beta}\mathbf{P}\ket{\beta}=\begin{cases}
\frac{|\beta|^2-1}{|\beta|^2+1}&\quad P_0=-1,\\
&\\
\frac{1-|\beta|^2}{1+|\beta|^2}&\quad P_0=+1,\\
\end{cases}
\end{equation}
where $P_0$ is the parity of vacuum state. The first excited state after the vacuum is created by $\eta_{0}^\dagger\ket{0}_\eta$ which inevitably has the same energy as the vacuum, while this excited state has parity -$P_0$. In fact $\beta$ can be considered as a parameter which can be tuned to break the parity. 

We define the state $\ket{G_{\pm}}$ by taking $\beta=\pm1$ as:
\begin{equation}\label{eq:GPMstate}
\ket{G_{\pm}}=\frac{1}{\sqrt{2}}\big(\ket{0}_\eta\pm \eta_0^\dagger\ket{0}_\eta\big).
\end{equation}
These states are especial cases of $\beta$-broken parity states and have interesting properties which cares for special attention. As an example, these states are related to the ground state of Hamiltonian with boundary magnetic field (see \ref{sec:Introduction}). 
%In general, for the state $\ket{\beta}$, we are not allowed to use the Peschel method \cite{Peschel_2001,Peschel_2003} to calculate the entanglement for this state. Computationally, Peschel method is far more efficient to calculate the entanglement of state using correlation matrices. However, it was demonstrated in \cite{XavierRajabpour2020} that with subtraction of a $\log2$ from the final result, one would get the correct entanglement.
Other useful properties of these two states are
\begin{equation}
\sigma_{0}^x\ket{G_{\pm}}=\pm\ket{G_{\pm}}\quad\text{and}\quad\sigma_{L+1}^x\ket{G_{\pm}}=\delta_{\pm}\ket{G_{\pm}}.
\end{equation}
The $\delta_{\pm}=\pm$ can be calculated with respect to the expectation values of Majorana fermions \cite{XavierRajabpour2020}. For instance, one can write:
\begin{equation}\label{eq:deltaplus}
\delta_{+}=(i)^{L+1}\text{Pf}[\boldsymbol{\mathcal{D}}],
\end{equation}
where
\begin{equation}\label{eq:delta_p_DMat}
\boldsymbol{\mathcal{D}}=\left(
\begin{array}{c c c c c}
0&\expval{\bar{\gamma}_0\gamma_1}&\expval{\bar{\gamma}_0\bar{\gamma}_1}&\cdots&\expval{\bar{\gamma}_0\gamma_{L+1}}\\
\expval{\gamma_1\bar{\gamma}_0}&0&\expval{\gamma_1\bar{\gamma}_1}&\cdots&\expval{\gamma_1\gamma_{L+1}}\\
\expval{\bar{\gamma}_1\bar{\gamma}_0}&\expval{\bar{\gamma}_1\gamma_1}&0&\cdots&\expval{\bar{\gamma}_1\gamma_{L+1}}\\
\vdots&\vdots&\vdots&{}&\vdots\\
\expval{\gamma_{L+1}\bar{\gamma}_{0}}&\expval{\gamma_{L+1}\gamma_{1}}&\expval{\gamma_{L+1}\bar{\gamma}_{1}}&\cdots&0\\
\end{array}
\right).
\end{equation}
in above, $\text{Pf}[\boldsymbol{\mathcal{D}}]$ is the Pfaffian of the matrix $\mathcal{D}$, and Majorana fermions are defined as $\gamma_j=c^\dagger_j+c_j$ and $\bar{\gamma}_j=i(c^\dagger_j-c_j)$. In \eqref{eq:delta_p_DMat}, the $\langle\cdots\rangle$ stands for the expectation value with respect to the vacuum state of the $\eta$-operators ($\ket{0}$).

\subsection{Eigenstates of boundary magnetic field model}\label{subsec:SelectionRule}

Since the eigenstates of boundary magnetic field model is related to those of Hamiltonian \eqref{eq:FermionHamil}, we are going to present selection rules to get the desired eigenstates. In fact, Hilbert space of Hamiltonian \eqref{eq:FermionHamil} is 4 times bigger than BMF Hamiltonian.

The Hilbert space of Hamiltonian \eqref{eq:FermionHamil} can be divided into 4 sub-spaces. Each sub-space can be identified using eigenvalues of operators $\sigma_{0}^x$ and $\sigma_{L+1}^x$ acting on states $\ket{G_\pm}$. We can make the following argument: consider $\delta_{+}=1$, which means $\ket{G_{+}}$ belongs to the sector marked by the pair $(\bra{G_\pm}\sigma_0^x\ket{G_\pm},\bra{G_\pm}\sigma_{_{L+1}}^x\ket{G_\pm})$. %$(+,+)$ defined as
%\begin{equation}
 %   \begin{cases}
 %   \sigma^x_{0}\ket{G_+}=+\ket{G_+}, \\
 %   \sigma^x_{L+1}\ket{G_+}=+\ket{G_+}. \\
 %   \end{cases}
%\end{equation}
Then due to commutation (anticommutation) relations \eqref{eq:eta and sigmaX1} and \eqref{eq:eta and sigmaX2} all the states
\begin{equation}
\prod_{j=1}^{n}\eta_{k_j}^\dagger\ket{G_{+}},\qquad n\;\text{is even.}
\end{equation}
also belong to $(+,+)$. Note that in the above expression $0<k_j<k_{j+1}< L+1$ which means that the dimension of $(+,+)$ sub-space is $2^L$. Next, in this case ($\delta_+=+1$), the state $\ket{G_{-}}$ and the following states belong to $(-,-)$ sector,
\begin{equation}
\prod_{j=1}^{n}\eta_{k_j}^\dagger\ket{G_{-}},\qquad n\;\text{is even.}
\end{equation}
For the two other sectors we have:
\begin{equation}
\begin{aligned}
&\prod_{j=1}^{n}\eta_{k_j}^\dagger\ket{G_{+}},\qquad n\;\text{is odd.}\qquad\quad(-,+),\\
&\prod_{j=1}^{n}\eta_{k_j}^\dagger\ket{G_{-}},\qquad n\;\text{is odd.}\qquad\quad(+,-),
\end{aligned}
\end{equation}
In the case of $\delta_{+}=-1$, we have the following eigenstates for each sector,
\begin{eqnarray}
\prod_{j=0}^{n}\eta_{k_j}^\dagger\ket{G_{+}},\qquad n\;\text{is even,}\qquad\quad(+,-),\\
\prod_{j=0}^{n}\eta_{k_j}^\dagger\ket{G_{-}},\qquad n\;\text{is even,}\quad\qquad(-,+),\\
\prod_{j=1}^{n}\eta_{k_j}^\dagger\ket{G_{+}},\qquad\; n\;\text{is odd,}\qquad\quad(-,-),\\
\prod_{j=1}^{n}\eta_{k_j}^\dagger\ket{G_{-}},\qquad\; n\;\text{is odd,}\qquad\quad(+,+).
\end{eqnarray}
As an example in the $(+,+)$ sector, we can write:
\begin{equation}\label{eq:BigSt-SmallSt}
    \ket{\phi_k}=\ket{+}_{_{0}}\otimes\ket{\varphi_k}\otimes\ket{+}_{_{L+1}},
\end{equation}
where $\ket{\phi_k}$ is an eigenstate of Hamiltonian \eqref{eq:FermionHamil}, $\ket{+}_{_{0,L+1}}$ are eigenstates of $\sigma^x$ at the position $0$ and $L+1$. The $\ket{\varphi_k}$ is an eigenstate of the BMF model. Knowing $\ket{\phi_k}$, we can obtain the $\ket{\varphi_k}$. 

The above argument means that to know the sector $(+,+)$ we need to figure out the value of $\delta_+$. The ground state of the Hamiltonian with boundary magnetic field is going to be one of the following two states:
\begin{eqnarray}
\ket{G_+}\qquad\delta_+=+1,\label{eq:SelRul_dp_GS}\\
\eta_{_{min}}^\dagger\ket{G_-}\qquad\delta_+=-1.\label{eq:SelRul_dm_GS}
\end{eqnarray}
%To find the right ground state we need to calculate $\delta_{\pm}$ which can be found by a bit of manipulations. The detail of the calculation is presented in \cite{XavierRajabpour2020}.

It is interesting to mention that exist transformations which change the sign of boundary couplings in the Hamiltonian \eqref{eq:FermionHamil}. Such transformations could be $T_{b_L}=\sigma_{L+1}^z$ and $T_{b_1}=\sigma_0^z$ where they change the sign of $x$ and $y$ component of boundary couplings ($\overrightarrow{b}_1,\:\overrightarrow{b}_L$ or equivalently $\alpha_1^0,\:\alpha_L^{L+1}$), without affecting the energy spectrum of the Hamiltonian:\small
\begin{equation}
    T_{_{b_1}}^\dagger T_{_{b_L}}^\dagger H(b_{1,x},b_{1,y},b_{1,z},b_{L,x},b_{L,y},b_{L,z}) T_{_{b_L}}T_{_{b_1}}=H(-b_{1,x},-b_{1,y},b_{1,z},-b_{L,x},-b_{L,y},b_{L,z}).
\end{equation}\normalsize
As a consequence, if the eigenstates of the BMF Hamiltonian \eqref{eq:Ham_intro} is found in one of the sectors of the Hilbert space of the Hamiltonian \eqref{eq:FermionHamil}, then other sectors are related to the eigenstates of BMF Hamiltonians with different signs of boundary couplings. For example, if a typical eigenstate of the BMF Hamiltonian like $\ket{\varphi_k}$, is in the $(+,+)$ sector, then from \eqref{eq:BigSt-SmallSt} we have $\ket{+}_{_{0}}\otimes\ket{\varphi_k}\otimes\ket{+}_{_{L+1}}$ for the eigenstates of Hamiltonian \eqref{eq:FermionHamil}. The action of $T_{b_1}T_{b_L}$ on a state like $\ket{\phi_k}$ would be
\begin{equation}
    T_{b_1}T_{b_L}\ket{\phi_k}=\ket{-}_{_{0}}\otimes\ket{\varphi_k}\otimes\ket{-}_{_{L+1}},
\end{equation}
which is equal to changing the sign of boundary couplings. Therefore, $\ket{\varphi_k}$ would be an eigenstate of BMF Hamiltonian with boundary couplings: $-b_{1,x},\:-b_{1,y},\:-b_{L,x}$ and $-b_{L,y}$. Equivalently, spectrum of of BMFH with negative couplings at boundary can be found in the $(-,-)$ sector of Hamiltonian \eqref{eq:FermionHamil}.

\section{Correlation functions}\label{sec:Correlation}
In this section, we would like to calculate the correlation matrix for different eigenstates of the system. It is more convenient to calculate the correlation matrices for Majorana fermions. For instance, one can use Majorana fermion correlations to calculate the entanglement in the system (for particular eigenstates). We introduce Majorana fermions as $\gamma_i=c_i+c_i^\dagger$ and $\bar{\gamma}_i=i(c_i^\dagger-c_i)$. We symbolize the correlation matrices as:
\begin{equation}\label{eq:GMatrixElement}
\langle\bar{\gamma}_j\gamma_k\rangle=iG_{jk},
\qquad
\langle\gamma_j\gamma_k\rangle=K_{jk},
\qquad
\langle\bar{\gamma}_j\bar{\gamma}_k\rangle=\bar{K}_{jk}.
\end{equation}
It is useful to write the later two point correlation in a block matrix form denoted by $\mathbf{\Gamma}$ as:
\begin{equation}\label{eq:GammaMatrixElement}
\boldsymbol{\Gamma}=
\left(\begin{array}{c c}
\boldsymbol{K}-\boldsymbol{\mathbf{I}} & -i\boldsymbol{G}^T\\
i\boldsymbol{G} & \boldsymbol{\bar{K}}-\boldsymbol{\mathbf{I}}
\end{array}\right).
\end{equation}
 One can easily find all the different elements of the $\boldsymbol{\Gamma}$ matrix. It is possible to write $\mathbf{G}$, $\mathbf{K}$ and $\mathbf{\bar{K}}$ in terms of correlation matrices of c-fermions
\begin{equation}\label{eq:KGmatrixCFmatrix}
\begin{aligned}
\mathbf{K}&=\mathbf{F}^\dagger+\mathbf{F}+\mathbf{C}-\mathbf{C}^T+\mathbb{I},\\
\mathbf{\bar{K}}&=-\mathbf{F}^\dagger-\mathbf{F}+\mathbf{C}-\mathbf{C}^T+\mathbb{I},\\
\mathbf{G}&=-\mathbf{F}^\dagger+\mathbf{F}+\mathbf{C}+\mathbf{C}^T-\mathbb{I},
\end{aligned}
\end{equation}
where $C_{ij}=\langle c_i^\dagger c_j\rangle$ and $F_{ij}=\langle c_i^\dagger c_j^\dagger\rangle$. The $\mathbf{C}$ is a Hermitian matrix and $\mathbf{F}$ is antisymmetric. Therefore, $\mathbf{K}$ and $\bar{\mathbf{K}}$ are Hermitian, and we can conclude that $\mathbf{G}$ is real. Knowing these properties, we can prove that the $\mathbf{\Gamma}$ correlation matrix is Hermitian too. All the analysis so far are valid for arbitrary eigenstates of the Hamiltonian \eqref{eq:FermionHamil}. In the rest, the correlations for vacuum and zero mode excited eigenstate (ZME state or $\eta_0^\dagger\ket{0}_\eta$) will be presented in details. The calculation of correlations of excited quasi-particle eigenstates is presented in appendix \ref{sec:AppExcitedCorrelations}.

\subsection{Correlations for vacuum state}\label{subsec:Correlation_Vacuum}

Using the notation introduced in \eqref{eq:HamDiago}, we calculate the correlations for the vacuum of $\eta$'s. In this case, we can write $\mathbf{C}^{^0}=\mathbf{h}^\dagger.\mathbf{h}$ and $\mathbf{F}^{^0}=\mathbf{h}^\dagger.\mathbf{g}$, where superscript zero stands for the expectation values calculated in the vacuum state. Putting these relations in \eqref{eq:KGmatrixCFmatrix}, we get
\begin{equation}\label{eq:KG matrix gh matrix}
\begin{aligned}
\mathbf{K}^{^0}=&(\mathbf{h}^\dagger+\mathbf{g}^\dagger).(\mathbf{h}+\mathbf{g}),\\
\mathbf{\bar{K}}^{^0}=&(\mathbf{h}^\dagger-\mathbf{g}^\dagger).(\mathbf{h}-\mathbf{g}),\\
\mathbf{G}^{^0}=&(\mathbf{h}^\dagger-\mathbf{g}^\dagger).(\mathbf{h}+\mathbf{g}).
\end{aligned}
\end{equation}
Therefore, for the vacuum of $\eta$-operators, we can find the correlations in terms of the elements of $\mathbf{U}$ matrix, which means correlation matrix calculations are straightforward. The above-mentioned correlation matrices have the form ($i,j\neq 0,L+1$):
{\arraycolsep=4.0pt\def\arraystretch{2.0}
	\begin{eqnarray}
	\mathbf{C}^{^{0}}&=&\left(\begin{array}{c|c|c}
	\frac{1}{2} &-\sum_{k=1}^{k=L+1}g_{k,0}^*h_{k,j}  &\tfrac{-1}{4}-\sum_{k=1}^{k=L+1}g_{k,0}^*g_{k,L+1}\\\hline
	-\sum_{k=1}^{k=L+1}h_{k,j}^*g_{k,0}&(\mathbf{h}^\dagger.\mathbf{h})_{i,j} &\sum_{k=1}^{k=L+1}h_{k,j}^*g_{k,L+1} \\\hline
	\tfrac{-1}{4}-\sum_{k=1}^{k=L+1}g_{k,L+1}^*g_{k,0}&\sum_{k=1}^{k=L+1}g_{k,L+1}^*h_{k,j}  &\frac{1}{2} \\
	\end{array}\right)\\
	\mathbf{F}^{^{0}}&=&\left(\begin{array}{c|c|c}
	0 &-\sum_{k=1}^{k=L+1}g_{k,0}^*g_{k,j}  &\tfrac{1}{4}-\sum_{k=1}^{k=L+1}g_{k,0}^*g_{k,L+1}\\\hline
	\sum_{k=1}^{k=L+1}g_{k,j}g_{k,0}^*&(\mathbf{h}^\dagger.\mathbf{g})_{i,j} &-\sum_{k=1}^{k=L+1}g_{k,j}g_{k,L+1}^* \\\hline
	-\tfrac{1}{4}+\sum_{k=1}^{k=L+1}g_{k,L+1}^*g_{k,0}&\sum_{k=1}^{k=L+1}g_{k,L+1}^*g_{k,j}  &0 \\
	\end{array}\right).
	\end{eqnarray}}
Using the relation \eqref{eq:KGmatrixCFmatrix}, for the correlation of Majorana fermions, we can write:
{\arraycolsep=4.0pt\def\arraystretch{2.2}
\begin{eqnarray}
	\mathbf{K}^{^{0}}&=&\left(\begin{array}{c|c|c}
	1 &0 &0\\\hline
	0&(\mathbf{h}^\dagger+\mathbf{g}^\dagger).(\mathbf{h}+\mathbf{g})_{i,j} &2\sum_{k=1}^{k=L+1}(h_{k,j}^*+g_{k,j}^*)g_{k,L+1} \\\hline
	0&2\sum_{k=1}^{k=L+1}g_{k,L+1}^*(h_{k,j}+g_{k,j})  &1 \\
	\end{array}\right),\\
	\mathbf{\bar{K}}^{^{0}}&=&\left(\begin{array}{c|c|c}
	1 &-2\sum_{k=1}^{k=L+1}g_{k,0}^*(h_{k,j}-g_{k,j})&0\\\hline
	-2\sum_{k=1}^{k=L+1}(h_{k,j}^*-g_{k,j}^*)g_{k,0}&(\mathbf{h}^\dagger-\mathbf{g}^\dagger).(\mathbf{h}-\mathbf{g})_{i,j} &0 \\\hline
	0&0  &1 \\
	\end{array}\right),\\
	\mathbf{G}^{^{0}}&=&\left(\begin{array}{c|c|c}
	0&-2\sum_{k=1}^{k=L+1}g_{k,0}^*(h_{k,j}+g_{k,j})  &-4\sum_{k=1}^{k=L+1}g_{k,0}^*g_{k,L+1}\\\hline
	0&(\mathbf{h}^\dagger-\mathbf{g}^\dagger).(\mathbf{h}+\mathbf{g})_{i,j} &2\sum_{k=1}^{k=L+1}(h_{k,j}^*-g_{k,j}^*)g_{k,L+1} \\\hline
	-1&0&0 \\
	\end{array}\right)
\end{eqnarray}}
To calculate the higher point correlation functions, one can use the Wick theorem, which is computationally favorable.  

\subsection{Correlations for ZME state}\label{subsec:Correlation_ZME}

From now on (for the sake of simplicity), we are going to indicate the ZME state by $\ket{\emptyset}=\eta_{0}^\dagger\ket{0}$. This state is degenerate with the vacuum, which is crucial for later studies. For this eigenstate, we get ($i,j\neq 0,L+1$):{\arraycolsep=4.0pt\def\arraystretch{2.2}
	\begin{equation}
	\mathbf{C}^{^\emptyset}=\left(\begin{array}{ccc}
	\frac{1}{2} &-\sum_{k=1}^{k=L+1}g_{k,0}^*h_{k,j}  &\tfrac{1}{4}-\sum_{k=1}^{k=L+1}g_{k,0}^*g_{k,L+1}\\
	-\sum_{k=1}^{k=L+1}h_{k,j}^*g_{k,0}&(\mathbf{h}^\dagger.\mathbf{h})_{i,j} &\sum_{k=1}^{k=L+1}h_{k,j}^*g_{k,L+1} \\
	\frac{1}{4}-\sum_{k=1}^{k=L+1}g_{k,L+1}^*g_{k,0}&\sum_{k=1}^{k=L+1}g_{k,L+1}^*h_{k,j}  &\frac{1}{2} \\
	\end{array}\right)
	\end{equation}}{\arraycolsep=4.0pt\def\arraystretch{2.2}
	\begin{equation}
	\mathbf{F}^{^\emptyset}=\left(\begin{array}{ccc}
	0 &-\sum_{k=1}^{k=L+1}g_{k,0}^*g_{k,j}  &\tfrac{-1}{4}-\sum_{k=1}^{k=L+1}g_{k,0}^*g_{k,L+1}\\
	\sum_{k=1}^{k=L+1}g_{k,j}g_{k,0}^*&(\mathbf{h}^\dagger.\mathbf{g})_{i,j} &-\sum_{k=1}^{k=L+1}g_{k,j}g_{k,L+1}^* \\
	\frac{1}{4}+\sum_{k=1}^{k=L+1}g_{k,L+1}^*g_{k,0}&\sum_{k=1}^{k=L+1}g_{k,L+1}^*g_{k,j}  &0 \\
	\end{array}\right).
	\end{equation}}
As a result, the correlation matrices $\mathbf{C}$ and $\mathbf{F}$ are only different in only two elements from state $\ket{0}$ to the state $\ket{\emptyset}$. It can be observed that the correlation matrices $\mathbf{K},\;\mathbf{\bar{K}}$ does not change from state $\ket{0}$ to $\ket{\emptyset}$. Form of the Majorana correlation matrices are presented below as:
{\arraycolsep=4.0pt\def\arraystretch{2.2}
\begin{eqnarray}
	\mathbf{K}^{^\emptyset}&=&\mathbf{K}^{^0},\quad\qquad\mathbf{\bar{K}}^{^\emptyset}=\mathbf{\bar{K}}^{^0},\\
%	\left(\begin{array}{ccc}
%	\mathbf{1} &0 &0\\
%	0&(\mathbf{h}^\dagger+\mathbf{g}^\dagger).(\mathbf{h}+\mathbf{g})_{i,j} &2\sum_{k=1}^{k=L+1}(h_{k,j}^*+g_{k,j}^*)g_{k,L+1} \\
%	0&2\sum_{k=1}^{k=L+1}g_{k,L+1}^*(h_{k,j}+g_{k,j})  &\mathbf{1} \\
%	\end{array}\right),\\
%	\mathbf{\bar{K}}^{^\emptyset}&=&\left(\begin{array}{ccc}
%	\mathbf{1} &-2\sum_{k=1}^{k=L+1}g_{k,0}^*(h_{k,j}-g_{k,j})&0\\
%	-2\sum_{k=1}^{k=L+1}(h_{k,j}^*-g_{k,j}^*)g_{k,0}&(\mathbf{h}^\dagger-\mathbf{g}^\dagger).(\mathbf{h}-\mathbf{g})_{i,j} &0 \\
%	0&0  &\mathbf{1}\\
%	\end{array}\right),\\
    \mathbf{G}^{^\emptyset}&=&\left(\begin{array}{ccc}
	0&-2\sum_{k=1}^{k=L+1}g_{k,0}^*(h_{k,j}+g_{k,j})  &-4\sum_{k=1}^{k=L+1}g_{k,0}^*g_{k,L+1}\\
	0&(\mathbf{h}^\dagger-\mathbf{g}^\dagger).(\mathbf{h}+\mathbf{g})_{i,j} &2\sum_{k=1}^{k=L+1}(h_{k,j}^*-g_{k,j}^*)g_{k,L+1} \\
	1&0&0 \\
	\end{array}\right).
\end{eqnarray}}
Similar to the result of previous subsection, for higher point functions, one make use of the Wick theorem to calculate the quantity of interest.

\subsection{Correlations for the general parity broken state}\label{subsec:Correlation_Beta}

In this subsection we study the correlation function of general states that break the parity, such as $\ket{\beta}$ defined in \eqref{eq:BetaParityBrokenState}. Calculating correlations (or any expectation value) with respect to the state $\ket{\beta}$, is not as trivial as the calculations for eigenstates, since we are not able to use the Wick theorem. However, there could be many subtleties when we come across quantities which are evaluated with respect to $\ket{\beta}$.

For instance, such a subtlety could be calculation of one point function with respect to the $\ket{\beta}$: 
%\begin{equation}\label{eq:BPBS_OnePoint_Corrl}
%\bra{\beta}c_j\ket{\beta}=\bra{\beta}c_j^{\dagger}\ket{\beta}=\frac{\mathrm{Re}[\beta]}{1+|\beta|^2}\delta_{j,0}
%\end{equation}
\begin{equation}\label{eq:BPBS_OnePoint_Corrl}
\bra{\beta}c_j\ket{\beta}=\frac{\mathrm{Re}[\beta]}{1+|\beta|^2}\delta_{j,0}+\frac{i\mathrm{Im}[\beta]}{1+|\beta|^2}\delta_{j,L+1}
\end{equation}
The above means that one point correlation functions are not necessarily zero for the state $\ket{\beta}$. In general, we can say that $\bra{\beta}\hat{\mathcal{O}}\ket{\beta}$ is not necessarily zero, if operator $\hat{\mathcal{O}}$ has odd number of fermionic operators. Nonetheless, if $\hat{\mathcal{O}}$ does not depend on $c_0$, $c_0^\dagger$, $c_{L+1}$ and $c_{L+1}^\dagger$ then we can write $\bra{\beta}\hat{\mathcal{O}}\ket{\beta}=\bra{0}\hat{\mathcal{O}}\ket{0}$. With this condition, we can assume that the state $\ket{\beta}$ obeys the Wick theorem. In the following, we denote the correlation matrices by superscript $\beta$ for the state $\ket{\beta}$. These correlations have the form: 
{\arraycolsep=4.0pt\def\arraystretch{2.0}
\begin{eqnarray}
	\mathbf{C}^{^{\beta}}&=&\left(\begin{array}{c|c|c}
	\frac{1}{2} &-\!\!\underset{k=1}{\overset{k=L+1}{\sum}}g_{k,0}^*h_{k,j}  &\tfrac{-1}{4}\frac{1-|\beta|^2}{1+|\beta|^2}-\!\!\underset{k=1}{\overset{k=L+1}{\sum}}g_{k,0}^*g_{k,L+1}\\\hline
	-\underset{k=1}{\overset{k=L+1}{\sum}}h_{k,j}^*g_{k,0}&(\mathbf{h}^\dagger.\mathbf{h})_{i,j} &\underset{k=1}{\overset{k=L+1}{\sum}}h_{k,j}^*g_{k,L+1} \\\hline
	\tfrac{-1}{4}\frac{1-|\beta|^2}{1+|\beta|^2}-\!\!\underset{k=1}{\overset{k=L+1}{\sum}}g_{k,L+1}^*g_{k,0}&\!\underset{k=1}{\overset{k=L+1}{\sum}}g_{k,L+1}^*h_{k,j}  &\frac{1}{2} \\
	\end{array}\right),\\
	\mathbf{F}^{^{\beta}}&=&\left(\begin{array}{c|c|c}
	0 &-\!\!\underset{k=1}{\overset{k=L+1}{\sum}}g_{k,0}^*g_{k,j}  &\tfrac{1}{4}\frac{1-|\beta|^2}{1+|\beta|^2}-\!\!\underset{k=1}{\overset{k=L+1}{\sum}}g_{k,0}^*g_{k,L+1}\\\hline
	\!\!\underset{k=1}{\overset{k=L+1}{\sum}}g_{k,j}g_{k,0}^*&(\mathbf{h}^\dagger.\mathbf{g})_{i,j} &-\!\!\underset{k=1}{\overset{k=L+1}{\sum}}g_{k,j}g_{k,L+1}^*\\\hline
	-\tfrac{1}{4}\frac{1-|\beta|^2}{1+|\beta|^2}+\!\!\underset{k=1}{\overset{k=L+1}{\sum}}g_{k,L+1}^*g_{k,0}&\!\underset{k=1}{\overset{k=L+1}{\sum}}g_{k,L+1}^*g_{k,j}  &0 \\
	\end{array}\right).
\end{eqnarray}	
Based on the above calculations, the correlations for $\ket{\beta}$ can be written in terms of correlations of vacuum and ZME state. % For instance, $\mathbf{K}^\beta=\frac{1}{2}\mathbf{K}^0+\frac{1}{2}\mathbf{K}^\emptyset$. 
  For rest of the correlation matrices, we have:
\begin{eqnarray}
    \label{eq:Beta_KMat_form}
	\mathbf{K}^{^\beta}&=&\mathbf{K}^{^0},\quad\qquad\mathbf{\bar{K}}^{^\beta}=\mathbf{\bar{K}}^{^0},\\
    \label{eq:Beta_GMat_form}
    \mathbf{G}^{^\beta}&=&\left(\begin{array}{c|c|c}
	0&-2\!\!\underset{k=1}{\overset{k=L+1}{\sum}}g_{k,0}^*(h_{k,j}+g_{k,j})  &-4\!\!\underset{k=1}{\overset{k=L+1}{\sum}}g_{k,0}^*g_{k,L+1}\\\hline
	0&(\mathbf{h}^\dagger-\mathbf{g}^\dagger).(\mathbf{h}+\mathbf{g})_{i,j} &2\!\!\underset{k=1}{\overset{k=L+1}{\sum}}(h_{k,j}^*-g_{k,j}^*)g_{k,L+1} \\\hline
	\frac{1-|\beta|^2}{1+|\beta|^2}&0&0 \\
	\end{array}\right).
\end{eqnarray}}\noindent
For a higher point correlation like $\hat{\mathcal{O}}$, when operator $\hat{\mathcal{O}}$ contains fermionic creation and annihilation operators at position $0$ or $L+1$, then the relation between $\langle\hat{\mathcal{O}}\rangle_0$ and $\langle\hat{\mathcal{O}}\rangle_{\beta}$ would not be trivial. For example, in some cases, one can have Wick theorem for the $\ket{\beta}$ too. Some of the interesting cases are listed in table \ref{tbl:expctationofOperator}.

Results of this part can be extended to the zero parity state. Form of the correlation matrices for the state $\ket{G_{\pm}}$ \eqref{eq:GPMstate}, can simply be obtained by putting $\beta\rightarrow\pm 1$.

\renewcommand{\arraystretch}{1.7}
\begin{table}[H]
	\centering
	\caption{\footnotesize Terms ``even" and ``odd" mean that the operator $\hat{\mathcal{O}}$ contains even or odd products of  $c$-operators. Also, $k,l\neq0,L+1$ and the Einstein summation rule is assumed. The notation $\langle\cdots\rangle_0$ stands for the expectation value calculated in the vacuum state.}
	\label{tbl:expctationofOperator}
	\begin{tabular}{|c|c|c|}
		\hline
		{}&\multicolumn{2}{|c|}{$\hat{\mathcal{O}}$ has no $c_{0}^{(\dagger)}$ and $c_{L+1}^{(\dagger)}$}\\
		\cline{2-3}
		{}&even&odd\\
		\hline
		$\bra{\beta}c_0\hat{\mathcal{O}}\ket{\beta}$&$\tfrac{\mathrm{Re}[\beta]}{1+|\beta|^2}\langle\hat{\mathcal{O}}\rangle_{_0}$&$g_{k,0}^*\bra{0}\eta_k\hat{\mathcal{O}}\ket{0}$\\ \hline
		$\bra{\beta}c_0^\dagger\hat{\mathcal{O}}\ket{\beta}$&$\tfrac{\mathrm{Re}[\beta]}{1+|\beta|^2}\langle\hat{\mathcal{O}}\rangle_{_0}$&$-g^*_{l,0}\bra{0}\eta_l\hat{\mathcal{O}}\ket{0}$\\ \hline
		$\bra{\beta}c_{L+1}\hat{\mathcal{O}}\ket{\beta}$&$\tfrac{i\mathrm{Im}[\beta]}{1+|\beta|^2}\langle\hat{\mathcal{O}}\rangle_{_0}$&$g_{k,L+1}^*\bra{0}\eta_k\hat{\mathcal{O}}\ket{0}$\\ \hline
		$\bra{\beta}c_{L+1}^\dagger\hat{\mathcal{O}}\ket{\beta}$&$\tfrac{-i\mathrm{Im}[\beta]}{1+|\beta|^2}\langle\hat{\mathcal{O}}\rangle_{_0}$&$g^*_{l,L+1}\bra{0}\eta_l\hat{\mathcal{O}}\ket{0}$\\ \hline
		$\bra{\beta}c_{0}^\dagger c_{0}\hat{\mathcal{O}}\ket{\beta}$&$\tfrac{1}{2}\langle\hat{\mathcal{O}}\rangle_{_{0}}$&$\tfrac{2\mathrm{Re}[\beta]}{1+|\beta|^2}\langle(g_{l,0}^*\eta_l-g_{l,0}\eta_l^\dagger)\hat{\mathcal{O}}\rangle_{_0}$\\ \hline
		$\bra{\beta}c_{L+1}^\dagger c_{L+1}\hat{\mathcal{O}}\ket{\beta}$&$\tfrac{1}{2}\bra{0}\hat{\mathcal{O}}\ket{0}$&$\tfrac{-2i\mathrm{Im}[\beta]}{1+|\beta|^2}\langle(g_{_{l,L+1}}^*\eta_l+g_{_{l,L+1}}\eta_l^\dagger)\hat{\mathcal{O}}\rangle_{_0}$\\ \hline
		$\bra{\beta}c_{0}^\dagger c_{L+1}\hat{\mathcal{O}}\ket{\beta}$&$\tfrac{1}{4}\tfrac{|\beta|^2-1}{|\beta|^2+1}\langle\hat{\mathcal{O}}\rangle_0-g_{_{l,0}}^* g_{_{k,L+1}}\langle\eta_l\eta_k^\dagger\hat{\mathcal{O}}\rangle_0$&$\tfrac{\mathrm{Re}[\beta]}{1+|\beta|^2}\langle g_{_{l,L+1}}^*\eta_l\hat{\mathcal{O}}\rangle_{_0}+\tfrac{i\mathrm{Im}[\beta]}{1+|\beta|^2}\langle g_{_{l,0}}^*\eta_l\hat{\mathcal{O}}\rangle_{_0}$\\ \hline
		$\bra{\beta}c_{0}^\dagger c_{L+1}^\dagger\hat{\mathcal{O}}\ket{\beta}$&$\tfrac{1}{4}\tfrac{1-|\beta|^2}{1+|\beta|^2}\langle\hat{\mathcal{O}}\rangle_0-g_{_{l,0}}^* g_{_{k,L+1}}\langle\eta_l\eta_k^\dagger\hat{\mathcal{O}}\rangle_0$&$\tfrac{\mathrm{Re}[\beta]}{1+|\beta|^2}\langle g_{_{l,L+1}}^*\eta_l\hat{\mathcal{O}}\rangle_{_0}-\tfrac{i\mathrm{Im}[\beta]}{1+|\beta|^2}\langle g_{_{l,0}}^*\eta_l\hat{\mathcal{O}}\rangle_{_0}$\\ \hline
		{}&\multicolumn{2}{|c|}{$\hat{\mathcal{O}}$ has no $\bar{\gamma}_{0}$ and $\gamma_{L+1}$}\\ \cline{2-3}
		%{}&even&odd\\ \hline
		$\bra{\beta}\gamma_{0}\hat{\mathcal{O}}\ket{\beta}$&$\tfrac{2\mathrm{Re}[\beta]}{1+|\beta|^2}\bra{0}\hat{\mathcal{O}}\ket{0}$&$0$\\ \hline
		$\bra{\beta}\bar{\gamma}_{0}\hat{\mathcal{O}}\ket{\beta}$&$0$&$-2ig^*_{k,0}\bra{0}\eta_k\hat{\mathcal{O}}\ket{0}$\\ \hline
	\end{tabular}
\end{table}
\pagebreak

\section{Reduced Density matrix}\label{sec:RDM}

In this section, we calculate the density matrix and reduced density matrix (RDM) of the particular states introduced previously. We are going to use the configurational basis result of section \ref{subsec:Configur_Basis} and coherent basis formulation to calculate the density matrix and RDM. The RDM will be presented in both coherent basis and operator form. The operator form of RDM is useful to calculate the entanglement content of the states, while the coherent basis form of the RDM can be used to study the formation probabilities. In subsection \ref{subsec:RDM_BetaState}, we start with the $\beta$-broken parity state \eqref{eq:BetaParityBrokenState}. We calculate the total density matrix and then the RDM in coherent basis and operator format. In subsections \ref{subsec:RDM_Vacuum}, \ref{subsec:RDM_ZeroBar} and \ref{subsec:RDM_Gpm} we present the same calculations for the states $\ket{0}_\eta$, $\ket{\emptyset}$ and $\ket{G_{\pm}}$, respectively.

\subsection{\texorpdfstring{$\ensuremath{\boldsymbol{\beta}}$}{} parity broken state}\label{subsec:RDM_BetaState}

We start by calculating the reduced density matrix for the state $\ket{\beta}$ defined in \eqref{eq:BetaParityBrokenState}. The total density matrix of a particular state $\ket{\psi}$ is defined as $\rho=\ket{\psi}\bra{\psi}$. We prefer to calculate the density matrix for the state $\ket{\beta}$ explicitly in an exponential form  using the definition \eqref{eq:VacEtaVacC}. With the $+1$ parity for the state $\ket{0}_\eta$, the density matrix has the form
\begin{equation}
\rho^\beta=\ket{\beta}\!\bra{\beta}=|C^\beta|^2e^{{}^{\tfrac{1}{2}R_{ij}c^\dagger_ic_j^\dagger}}(1+\beta\mathfrak{M}_{0k}c_k^\dagger)\ket{0}_c{}^{}_c\!\bra{0}(1+\beta^*\mathfrak{M}_{0l}^*c_l) e^{{}^{-\tfrac{1}{2}R^*_{ij}c_ic_j}},
\end{equation}
where $|C^\beta|^2=\Big((1+|\beta|^2)\sqrt{\text{det}[\mathbf{I}+\mathbf{R}^\dagger\mathbf{R}]}\Big)^{-1}$ and $\boldsymbol{\mathfrak{M}}=\mathbf{h}^*.\mathbf{R}+\mathbf{g}^*$.  To proceed, we use the Fermionic coherent state defined as
\begin{equation}\label{eq:FermiCoherentState}
\ket{\boldsymbol{\xi}}=\ket{\xi_1\xi_2\cdots\xi_N}=e^{-\sum_{k=1}^{N}\xi_kc_k^\dagger}\ket{0}_c,
\end{equation}
where $\xi_k$ are Grassmann variables. Therefore, we can write (following similar procedure as \cite{Peschel_2001})
\begin{equation}\label{eq:DensityMatGrass}
\bra{\boldsymbol{\xi}}\rho^\beta\ket{\boldsymbol{\xi}^\prime}=\rho^\beta\!(\boldsymbol{\xi},\boldsymbol{\xi}^\prime)=|C^\beta|^2 e^{{}^{\tfrac{1}{2}R_{ij}\bar{\xi}_i\bar{\xi}_j}}(1+\beta\mathfrak{M}_{0k}\bar{\xi}_k)(\beta^*\mathfrak{M}_{0l}^*\xi_l^\prime+1)  e^{{}^{-\tfrac{1}{2}R^*_{nm}\xi_n^\prime \xi_m^\prime}}.
\end{equation}

To obtain the reduced density matrix (RDM), we divide the system into two parts (subsystem) $\mathbf{1}$ and $\mathbf{2}$. Here, we denote parts of any matrix that correspond to the subsystem $\mathbf{1}$ ($\mathbf{2}$) with the subscript $1$ ($2$)\footnote{For instance, the $\mathbf{A}_{12}$ stands for the sub-matrix of $\mathbf{A}$ that rows and columns belong to subsystem $1$ and $2$ respectively}
. We trace out the subsystem $\mathbf{2}$ to find the RDM for subsystem $\mathbf{1}$, $\rho_{{}_\mathbf{1}}=\text{tr}_{{}_{\mathbf{2}}}\rho$. In order to do so, we use the trace formula for operators in the coherent basis. Therefore, we have:
\begin{equation}\label{eq:Partial_Trace_Def}
\rho_{\mathbf{1}}^\beta\!(\boldsymbol{\xi},\boldsymbol{\xi}^\prime)=\int\prod_{l\in\mathbf{2}}\!\text{d}\bar{\xi_l}\text{d}\xi_l\;e^{-\underset{n\in\mathbf{2}}{\sum}\bar{\xi}_n\xi_n}\bra{\xi_1,\cdots,\xi_k,-\xi_{k+1},\cdots,-\xi_{L}}\rho^\beta\ket{\xi_1^\prime,\cdots,\xi_k^\prime,\xi_{k+1},\cdots,\xi_{L}},
\end{equation}
where $\xi_1,\cdots,\xi_k$ belong to the subsystem $\mathbf{1}$ and $\xi_{k+1},\cdots,\xi_{L}$ belong to the subsystem $2$. For the details of calculation, see appendix \ref{sec:AppExcitedCorrelations}. The final result after partial tracing the \eqref{eq:DensityMatGrass} is:
\begin{equation}\label{eq:BetaRDMGrassFin}
    \rho^\beta_{\mathbf{1}}(\boldsymbol{\xi},\boldsymbol{\xi}^\prime)=\mathcal{C}^\beta\Big[\big(\frac{1}{\beta}+\boldsymbol{\mathcal{L}}_1.\bar{\boldsymbol{\xi}}+\boldsymbol{\mathcal{L}}_2.\boldsymbol{\xi}^\prime\big)\big(\frac{1}{\beta^*}+\boldsymbol{\mathcal{L}}_3.\bar{\boldsymbol{\xi}}+\boldsymbol{\mathcal{L}}_4.\boldsymbol{\xi}^\prime\big)-\text{Pf}[\boldsymbol{\mathcal{W}}]\Big]e^{\tfrac{1}{2}
	\left(\begin{matrix}
	\bar{\xi}&\!\xi^\prime\\
	\end{matrix}\right)\begin{matrix}
	\boldsymbol{\Omega}
	\end{matrix}\left(\begin{matrix}
	\bar{\xi}\\\xi^\prime
	\end{matrix}\right)},
\end{equation}
where the $\mathcal{C}^\beta$ and the introduced matrices are given by:
\begin{subequations}\label{subeq:Beta_Consts_Matrixs}
\begin{equation}\label{eq:BetaCcalConst}
    \mathcal{C}^\beta=\frac{|\beta|^2\sqrt{\det\big[\mathbf{I}+{\mathbf{R}_{22}}^\dagger\mathbf{R}_{22}\big]}}{(1+|\beta|^2)\sqrt{\det\big[\mathbf{I}+{\mathbf{R}}^\dagger\mathbf{R}\big]}},
\end{equation}
\begin{equation}\label{eq:BetaOmegaMatSimple}
\boldsymbol{\Omega}=\left(\begin{matrix}\mathbf{R}_{11}&0\\0&-{\mathbf{R}_{11}}^*\\\end{matrix}\right)+\left(\begin{matrix}\mathbf{R}_{12}&0\\0&{\mathbf{R}_{12}}^*\\\end{matrix}\right)\boldsymbol{\mathcal{A}}^{-1}\left(\begin{matrix}{\mathbf{R}_{12}}^T&0\\0&{\mathbf{R}_{12}}^\dagger\\\end{matrix}\right),
\end{equation}
\begin{equation}
\left(\begin{matrix}\boldsymbol{\mathcal{L}}_1&\boldsymbol{\mathcal{L}}_2\\ \boldsymbol{\mathcal{L}}_3&\boldsymbol{\mathcal{L}}_4\end{matrix}\right)=\left(\begin{matrix}\boldsymbol{\mathfrak{M}}_1&\boldsymbol{0}\\\boldsymbol{0}&\boldsymbol{\mathfrak{M}}_1^*\end{matrix}\right)+\left(\begin{matrix}\boldsymbol{\mathfrak{M}}_2&\boldsymbol{0}\\\boldsymbol{0}&-\boldsymbol{\mathfrak{M}}_2^*\end{matrix}\right)\boldsymbol{\mathcal{A}}^{-1}\left(\begin{matrix}\mathbf{R}_{12}^T&\mathbf{0}\\\mathbf{0}&\mathbf{R}_{12}^\dagger\end{matrix}\right),
\end{equation}
\begin{equation}
    \boldsymbol{\mathcal{W}}=\left(\begin{matrix}
    \boldsymbol{\mathfrak{M}}_2&\mathbf{0}\\ \mathbf{0}&-\boldsymbol{\mathfrak{M}}_2^*\\
    \end{matrix}\right)\boldsymbol{\mathcal{A}}^{-T}\left(\begin{matrix}
    \boldsymbol{\mathfrak{M}}_2^T&\mathbf{0}\\ \mathbf{0}&-\boldsymbol{\mathfrak{M}}_2^\dagger\\
    \end{matrix}\right),
\end{equation}
\begin{equation}\label{eq:BetaAMatGrass}
\boldsymbol{\mathcal{A}}=\left(\begin{matrix}
\mathbf{R}_{22}&-\mathbf{I}\\\mathbf{I}&-\mathbf{R}_{22}^*\\
\end{matrix}\right).
\end{equation}
\end{subequations}
It is useful also to have the RDM in the operator format (for example to calculate the R\'enyi entanglement entropy). To derive the operator form for $\rho_{{}_{\mathbf{1}}}$ from equation \eqref{eq:BetaRDMGrassFin}, we rewrite the exponential term as
\begin{equation}
    \text{Exp}\big[\tfrac{1}{2}
	\left(\begin{matrix}
	\bar{\xi}&\!\xi^\prime\\
	\end{matrix}\right)\begin{matrix}
	\boldsymbol{\Omega}
	\end{matrix}\left(\begin{matrix}
	\bar{\xi}\\\xi^\prime
	\end{matrix}\right)\big]=\text{Exp}\big[\tfrac{1}{2}\mathcal{X}_{ij}\bar{\xi_i}\bar{\xi_j}\big]\text{Exp}\big[\mathcal{Y}_{ij}\bar{\xi_i}\xi_j^\prime\big]\;\text{Exp}\big[\tfrac{1}{2}\mathcal{Z}_{ij}\xi_i^\prime\xi_j^\prime\big]
\end{equation}
with properly defined matrices $\mathcal{X}$, $\mathcal{Y}$ and $\mathcal{Z}$.
Using the relations $c_ic_j\ket{\boldsymbol{\xi}}=\xi_i\xi_j\ket{\boldsymbol{\xi}}$ and $\bra{\boldsymbol{\xi}}c_i^\dagger c_j^\dagger=\bra{\boldsymbol{\xi}}\bar{\xi_i}\bar{\xi_j},$ one can replace $\bar{\xi_i}\bar{\xi_j}$ with $c_i^\dagger c_j^\dagger$ and $\xi_i^\prime\xi_j^\prime$ with $c_ic_j$ in the left and right exponentials. The cross term can be rewritten $\mathcal{Y}_{ij}\bar{\xi_i}\xi_j^\prime\;\rightarrow\;\ln(\boldsymbol{\mathcal{Y}})_{ij}c^\dagger_ic_j$. The final operator form of RDM \eqref{eq:BetaRDMGrassFin} is given as:
\begin{equation}\label{eq:BetaRDMFinalVersion2}
\begin{aligned}
    \rho_{\mathbf{1}}^\beta(c,c^\dagger)=&\:\mathcal{C}^\beta\:
	 e^{^{\frac{1}{2}
	\left(\begin{matrix}
	\mathbf{c}^\dagger&\!\mathbf{c}\\
	\end{matrix}\right)\begin{matrix}
	\boldsymbol{\mathcal{M}}
	\end{matrix}\left(\begin{matrix}
	\!\mathbf{c}\\\mathbf{c}^\dagger
	\end{matrix}\right)}} e^{^{\tfrac{1}{2}\text{tr}\ln(\tfrac{1}{2}\boldsymbol{\Omega}_{12}-\tfrac{1}{2}\boldsymbol{\Omega}_{21}^T)}}\Big[\text{Pf}[\boldsymbol{\mathcal{W}}]-\boldsymbol{\mathcal{L}}_{3}\boldsymbol{\mathcal{T}}_{22}\boldsymbol{\mathcal{L}}_{2}\\
	&+\big(\boldsymbol{\mathcal{L}}_{1}\boldsymbol{\mathcal{T}}_{22}\boldsymbol{c}^\dagger+(\boldsymbol{\mathcal{L}}_{1}\boldsymbol{\mathcal{T}}_{21}+\boldsymbol{\mathcal{L}}_{2})\boldsymbol{c}+\frac{1}{\beta}\big)\big(\boldsymbol{\mathcal{L}}_{3}\boldsymbol{\mathcal{T}}_{22}\boldsymbol{c}^\dagger+(\boldsymbol{\mathcal{L}}_{4}+\boldsymbol{\mathcal{L}}_{3}\boldsymbol{\mathcal{T}}_{21})\boldsymbol{c}+\frac{1}{\beta^*}\big)\Big]
\end{aligned}
\end{equation}
where
%\begin{subequations}
\begin{equation}\label{BetaMTmatrixdef}
    \boldsymbol{\mathcal{M}}=\ln\boldsymbol{\mathcal{T}};\quad\boldsymbol{\mathcal{T}}={\left(\begin{matrix} 
    \tfrac{1}{2}\boldsymbol{\Omega}_{12}-\tfrac{1}{2}{\boldsymbol{\Omega}_{21}}^T+2\boldsymbol{\Omega}_{11}(\boldsymbol{\Omega}_{12}^T-{\boldsymbol{\Omega}_{21}})^{-T}\boldsymbol{\Omega}_{22}&\;2\boldsymbol{\Omega}_{11}(\boldsymbol{\Omega}_{12}^T-{\boldsymbol{\Omega}_{21}})^{-T}\\2(\boldsymbol{\Omega}_{12}^T-\boldsymbol{\Omega}_{21})^{-T}\boldsymbol{\Omega}_{22}&2(\boldsymbol{\Omega}_{12}^T-{\boldsymbol{\Omega}_{21}})^{-T}
    \end{matrix}\right)},
\end{equation}    
%\begin{equation}
%\boldsymbol{\Omega}=\left(\begin{matrix}\mathbf{R}_{11}&0\\0&-{\mathbf{R}_{11}}^*\\\end{matrix}\right)+\left(\begin{matrix}\mathbf{R}_{12}&0\\0&{\mathbf{R}_{12}}^*\\\end{matrix}\right){{\boldsymbol{\mathcal{A}}}}^{-1}\left(\begin{matrix}{\mathbf{R}_{12}}^T&0\\0&{\mathbf{R}_{12}}^\dagger\\\end{matrix}\right);\qquad
%\boldsymbol{\mathcal{A}}=\left(\begin{matrix}
%\mathbf{R}_{22}&-\mathbf{I}\\\mathbf{I}&-\mathbf{R}_{22}^*\\
%\end{matrix}\right)
%\end{equation}
%begin{equation}
 %   \boldsymbol{\mathcal{W}}=\left(\begin{matrix}
    %\boldsymbol{\mathfrak{M}}_2&\mathbf{0}\\ %\mathbf{0}&-\boldsymbol{\mathfrak{M}}_2^*\\
    %\end{matrix}\right)\boldsymbol{\mathcal{A}}^{-T}\left(\begin{matrix}
%    \boldsymbol{\mathfrak{M}}_2^T&\mathbf{0}\\ \mathbf{0}&-\boldsymbol{\mathfrak{M}}_2^\dagger\\
 %   \end{matrix}\right)
%\end{equation}
%\begin{equation}
%\left(\begin{matrix}\boldsymbol{\mathcal{L}}_1&\boldsymbol{\mathcal{L}}_2\\ \boldsymbol{\mathcal{L}}_3&\boldsymbol{\mathcal{L}}_4\end{matrix}\right)=\left(\begin{matrix}\boldsymbol{\mathfrak{M}}_1&\boldsymbol{0}\\\boldsymbol{0}&\boldsymbol{\mathfrak{M}}_1^*\end{matrix}\right)+\left(\begin{matrix}\boldsymbol{\mathfrak{M}}_2&\boldsymbol{0}\\\boldsymbol{0}&-\boldsymbol{\mathfrak{M}}_2^*\end{matrix}\right)\boldsymbol{\mathcal{A}}^{-1}\left(\begin{matrix}\mathbf{R}_{12}^T&\mathbf{0}\\\mathbf{0}&\mathbf{R}_{12}^\dagger\end{matrix}\right).
%\end{equation}
%\end{subequations}
and the matrix $\boldsymbol{\Omega}$ and rest of the matrices are defined in \eqref{subeq:Beta_Consts_Matrixs}. Also, $\boldsymbol{\mathcal{T}}_{11}$, $\boldsymbol{\mathcal{T}}_{12}$, $\boldsymbol{\mathcal{T}}_{21}$ and $\boldsymbol{\mathcal{T}}_{22}$ stand for the sub-matrices (blocks) of matrix $\boldsymbol{\mathcal{T}}$. In the above expression, to move the exponential with fermionic operators, we have used the following relation, coming from Baker-Hausdorff formula,
\begin{equation}\label{eq:BeckHausd_Relation}
    F^{-1}\left(\begin{matrix}\mathbf{c}\\\mathbf{c}^\dagger \end{matrix}\right)F=\boldsymbol{\mathcal{T}}\left(\begin{matrix}\mathbf{c}\\\mathbf{c}^\dagger \end{matrix}\right)\;\;\Rightarrow\;\;\left(\begin{matrix}\mathbf{c}\\\mathbf{c}^\dagger \end{matrix}\right)F=F\boldsymbol{\mathcal{T}}\left(\begin{matrix}\mathbf{c}\\\mathbf{c}^\dagger \end{matrix}\right);\qquad F=e^{{\tfrac{1}{2}
	\left(\begin{matrix}
	\mathbf{c}^\dagger&\!\mathbf{c}\\
	\end{matrix}\right)\begin{matrix}
	\boldsymbol{\mathcal{M}}
	\end{matrix}\left(\begin{matrix}
	\!\mathbf{c}\\\mathbf{c}^\dagger
	\end{matrix}\right)}}.
\end{equation}
In the expression of RDM, having all the creation and annihilation operators in the argument of exponential is preferred (some of calculations would be simplified). For this reason, we present another calculation of $\rho_1^\beta$ in the appendix \ref{sec:AppRDMcalculations}, where using a trick, we managed to get the RDM with two exponentials. The above equations are valid for the RDM of $\ket{\beta}$-state, which also means, for any given bi partition, one can use \eqref{eq:BetaRDMFinalVersion2}. However, it should be noted that the spin and fermion representations for solvable quantum chains lead to different RDMs. For more details see section \ref{subsec:RDM_SpinFermion}.

We can propose an ansatz to write the RDM for $\ket{\beta}$ in term of correlation matrix. Although computationally favorable, the down side of such an ansatz is that we could only use it for a particular type of bipartition. while we can not apply Wick theorem to $\ket{\beta}$, one can use the $\mathbf{\Gamma^{^\beta}}$ matrix to calculate the RDM of a subsystem which starts from one boundary. One has to make an adjustment to the Peschel method. Such a procedure for $\beta=\pm1$ has been shown in \cite{XavierRajabpour2020}; here we are extending that result. 

Since the $\boldsymbol{\Gamma^{^\beta}}$ matrix is a skew symmetric matrix it can be written in a block form using an
orthogonal matrix $\mathbf{V}$ as:
\begin{equation}\label{eq:Beta_GamaCorr_Diag1}
\mathbf{V}\boldsymbol{\Gamma^{^\beta}}\mathbf{V}^T=\begin{pmatrix}
\mathbf{0}&i\boldsymbol{\nu}\\
-i\boldsymbol{\nu}&\mathbf{0}
\end{pmatrix},
\end{equation}
where $\boldsymbol{\nu}$ is a diagonal matrix. Then, we can define the following fermionic operators:
\begin{equation}\label{eq:Beta_GamaCorr_Diag12}
\begin{pmatrix}\boldsymbol{d}^\dagger\\ \boldsymbol{d}\end{pmatrix}
=\frac{1}{\sqrt{2}}\mathbf{W}\begin{pmatrix}\boldsymbol{\gamma}\\\boldsymbol{\bar{\gamma}}\end{pmatrix}=\frac{1}{2}\begin{pmatrix}\mathbf{I}&i\mathbf{I}\\ \mathbf{I}& -i\mathbf{I}\end{pmatrix}
\mathbf{V}\begin{pmatrix}\boldsymbol{\gamma}\\\boldsymbol{\bar{\gamma}}\end{pmatrix}.
\end{equation}
Similar to the results in \cite{XavierRajabpour2020}, one can make an ansatz for the
RDM of the subsystem $\mathbf{1}$. To be precise, we are assuming that the subsystem $\mathbf{1}$ is a connected bipartite of the system starting from site $0$ to $\ell$. The ansatz should have a form like below with respect to the operators that diagonalize the $\boldsymbol{\Gamma^\beta}$ matrix.
\begin{equation*}%\label{eq:RDM_Beta_DiagonalForm}
\rho_{\mathbf{1}}^{^\beta}(d,d^\dagger)=\mathrm{g}(d_0,d_0^\dagger,d_0^\dagger d_0)\times\prod_{k=1}^{\ell}\Big(\frac{1+\nu_k}{2}d_k^\dagger d_k+\frac{1-\nu_k}{2}d_k d_k^\dagger\Big),
\end{equation*}
where $\mathrm{g}$ is an arbitrary function to be determined. From correlation matrices (section \ref{sec:Correlation}), we can realize that 
\begin{equation}%\label{eq:RDM_Beta_DiagonalForm}
\mathrm{g}(c_0,c_0^\dagger,c_0^\dagger c_0)=\frac{\mathrm{Re}[\beta]}{1+|\beta|^2}(c^\dagger_0+c_0)+\frac{1}{2}\mathbf{I}.
\end{equation}
This ansatz satisfies the expectations of the state $\ket{\beta}$ including one point functions \eqref{eq:BPBS_OnePoint_Corrl}. It is easy to show that $\boldsymbol{c}^\dagger+\boldsymbol{c}=\sqrt{2}\big(\boldsymbol{W}_{11}^{\dagger}\boldsymbol{d}^\dagger+\boldsymbol{W}_{11}^{T}\boldsymbol{d}\big)$, where $\boldsymbol{W}$ is the unitary transformation which diagonalizes the matrix $\boldsymbol{\Gamma}^\beta$. Then, in terms of $d$ and $d^\dagger$ operators we have:
\begin{equation}\label{eq:RDM_Beta_DiagonalForm}
\rho_{\mathbf{1}}^{^\beta}(d,d^\dagger)=\frac{\mathrm{Re}[\beta]}{1+|\beta|^2}\big(\tfrac{\sqrt{2}}{2}d_0+\tfrac{\sqrt{2}}{2}d_0^\dagger+\tfrac{1}{2}\mathbf{I}\big)\times\prod_{k=1}^{\ell}\Big(\frac{1+\nu_k}{2}d_k^\dagger d_k+\frac{1-\nu_k}{2}d_k d_k^\dagger\Big),
\end{equation}
%where the matrix $\mathbf{T}$ is related to the unitary transformation which diagonalizes $\boldsymbol{\Gamma^\beta}$:
%\begin{equation}\label{eq:Beta_TMat_Def}
%\mathbf{T}=\frac{\sqrt{2}\Re[\beta]}{1+|\beta|^2}\begin{pmatrix}\mathbf{I}&-i\mathbf{I}\\ \mathbf{I}& i\mathbf{I}\end{pmatrix}
%\boldsymbol{W}^\dagger
%\end{equation}
The ansatz \eqref{eq:RDM_Beta_DiagonalForm} respects the generalized Wick’s theorem, which means that this RDM produces all the correlation functions correctly. As an example, one point function is not necessarily zero for the $\ket{\beta}$:
\begin{equation}
\langle d_k^\dagger\rangle_\beta=\frac{1}{\sqrt{2}}\frac{\mathrm{Re}[\beta]}{1+|\beta|^2}\delta_{k,0},
\end{equation}
\begin{equation}
\langle d_k\rangle_\beta=\frac{1}{\sqrt{2}}\frac{\mathrm{Re}[\beta]}{1+|\beta|^2}\delta_{k,0}.
\end{equation}
%The prior claim has been verified numerically \cite{XavierRajabpour2020}, which can also be proved analytically. 
Based on \eqref{eq:Beta_KMat_form}, %\eqref{eq:Gpm_KbMat_form},
\eqref{eq:Beta_GMat_form} and \eqref{eq:GammaMatrixElement}, the first row and column of  matrix $\boldsymbol{\Gamma}^\beta$ is zero which evidently means it has a zero eigenvalue, $\nu_0=0$. From the earlier statement it can be inferred that the matrix which diagonalizes the $\boldsymbol{\Gamma}^\beta$ -called $\boldsymbol{W}$- has the following form:
\begin{equation}\label{eq:Beta_WMat_Form}
    \boldsymbol{W}=\left(\begin{matrix}
    \boldsymbol{W}_{11}&\boldsymbol{W}_{12}\\
    {\boldsymbol{W}_{11}}^*&{\boldsymbol{W}_{12}}^*\\
    \end{matrix}\right);\qquad\boldsymbol{W}_{11}=\left(\begin{matrix}
    1&0&\cdots&0\\
    \underset{\vdots}{0}&{}&{\begingroup\Large\boldsymbol{\omega}\endgroup}&{}\\
    {0}&{}&&{}\\
    \end{matrix}\right).
\end{equation}
The reason we have defined the $\boldsymbol{W}$ as above is that the $\boldsymbol{d}^\dagger$ and $\boldsymbol{d}$ should be related with a conjugate transpose transformation. From \eqref{eq:Beta_GamaCorr_Diag12}, The $\boldsymbol{W}$ and $\boldsymbol{V}$ matrices are related by $\boldsymbol{W}\propto{\begingroup\footnotesize\begin{pmatrix}\mathbf{I}&i\mathbf{I}\\ \mathbf{I}& -i\mathbf{I}\end{pmatrix}\endgroup}\mathbf{V}$, which means we can write:
\begin{equation}
    \begin{pmatrix}\boldsymbol{d}^\dagger\\ \boldsymbol{d}\end{pmatrix}
    =\frac{1}{\sqrt{2}}\begin{pmatrix}
    \boldsymbol{W}_{11}\boldsymbol{\gamma}+\boldsymbol{W}_{12}\bar{\boldsymbol{\gamma}}\\
    \boldsymbol{W}_{11}^*\boldsymbol{\gamma}+\boldsymbol{W}_{12}^*\bar{\boldsymbol{\gamma}}\\
    \end{pmatrix}.
\end{equation}
We know that $\langle\bar{\gamma}_i\rangle_\beta=0$ and $\langle\gamma_i\rangle_\beta=\delta_{i,0}$, then we can write $\langle d_k^\dagger\rangle_\beta=(\boldsymbol{W}_{11})_{k,0}$. Making use of the relation \eqref{eq:Beta_WMat_Form}, we have $\langle d_k^\dagger\rangle_\beta\propto\delta_{k,0}$. For the $\langle d_k\rangle_\beta$ case, we use the fact that $\boldsymbol{d}$ and $\boldsymbol{d}^\dagger$ are related by a conjugation, which means that $\langle d_k\rangle_\beta\propto\delta_{k,0}$ and the proof is complete.

Note that, the above calculation is also valid for states $\ket{G_+}$, $\ket{G_-}$ and any state created from the action of $\eta^\dagger$'s on these states. Consequently, with a proper $\boldsymbol{\Gamma}$ matrix, a similar ansatz to  \eqref{eq:RDM_Beta_DiagonalForm} can be used to study the entanglement for all the states belonging to the four sectors created with $\ket{G_\pm}$ as introduced in subsection \ref{subsec:SelectionRule}.

\subsection{Vacuum state}\label{subsec:RDM_Vacuum}

As it was mentioned before the limit $\beta\rightarrow0$ in the $\ket{\beta}$, gives the $\ket{0}_\eta$. Equivalently, $\underset{\beta\rightarrow0}{\text{lim}}\:\rho^\beta=\rho^0=\ket{0}_{_\eta} {}_{_{\eta}}\!\!\bra{0}$ . Simply, we are going to apply this limit on the results of subsection \ref{subsec:RDM_BetaState} to find the required quantities in this subsection. For instance, in the limit $\beta\rightarrow0$, the only nonzero term in \eqref{eq:BetaRDMGrassFin} is $|\mathcal{C}^\beta|^2\frac{1}{|\beta|^2}$. Therefore for the RDM of vacuum state in the coherent basis we have:
\begin{equation}\label{eq:VacRDMGrassFin}
    \rho^{0}_{\mathbf{1}}(\boldsymbol{\xi},\boldsymbol{\xi}^\prime)=\frac{\sqrt{\det\big[\mathbf{I}+{\mathbf{R}_{22}}^\dagger\mathbf{R}_{22}\big]}}{\sqrt{\det\big[\mathbf{I}+{\mathbf{R}}^\dagger\mathbf{R}\big]}}e^{\tfrac{1}{2}
	\left(\begin{matrix}
	\bar{\xi}&\!\xi^\prime\\
	\end{matrix}\right)\begin{matrix}
	\boldsymbol{\Omega}
	\end{matrix}\left(\begin{matrix}
	\bar{\xi}\\\xi^\prime
	\end{matrix}\right)}.
\end{equation}
\iffalse where:
\begin{subequations}
\begin{equation}\
\boldsymbol{\mathcal{A}}=\left(\begin{matrix}
\mathbf{R}_{22}&-\mathbf{I}\\\mathbf{I}&-\mathbf{R}_{22}^*\\
\end{matrix}\right),
\end{equation}
\begin{equation}\label{eq:VacOmegaMatSimple}
\boldsymbol{\Omega}=\left(\begin{matrix}\mathbf{R}_{11}&0\\0&-{\mathbf{R}_{11}}^*\\\end{matrix}\right)+\left(\begin{matrix}\mathbf{R}_{12}&0\\0&{\mathbf{R}_{12}}^*\\\end{matrix}\right)\boldsymbol{\mathcal{A}}^{-1}\left(\begin{matrix}{\mathbf{R}_{12}}^T&0\\0&{\mathbf{R}_{12}}^\dagger\\\end{matrix}\right).
\end{equation}
\end{subequations}\fi 
The operator form of RDM reads as
\begin{equation}\label{eq:VacRDMFinalVersion2}
\begin{aligned}
    \rho_{\mathbf{1}}^0(c,c^\dagger)=&\frac{\sqrt{\det\big[\mathbf{I}+{\mathbf{R}_{22}}^\dagger\mathbf{R}_{22}\big]}}{\sqrt{\det\big[\mathbf{I}+{\mathbf{R}}^\dagger\mathbf{R}\big]}}
	 e^{^{\frac{1}{2}
	\left(\begin{matrix}
	\mathbf{c}^\dagger&\!\mathbf{c}\\
	\end{matrix}\right)\begin{matrix}
	\boldsymbol{\mathcal{M}}
	\end{matrix}\left(\begin{matrix}
	\!\mathbf{c}\\\mathbf{c}^\dagger
	\end{matrix}\right)}} e^{^{\tfrac{1}{2}\text{tr}\ln(\tfrac{1}{2}\boldsymbol{\Omega}_{12}-\tfrac{1}{2}\boldsymbol{\Omega}_{21}^T)}}.
\end{aligned}
\end{equation}
\iffalse where
\begin{subequations}\label{eq:RDMVacMatDef}
\begin{equation}\label{eq:RDMVacMatDef1}
    \boldsymbol{\mathcal{M}}=\ln\boldsymbol{\mathcal{T}};\quad\boldsymbol{\mathcal{T}}={\left(\begin{matrix} 
    \tfrac{1}{2}\boldsymbol{\Omega}_{12}-\tfrac{1}{2}{\boldsymbol{\Omega}_{21}}^T+2\boldsymbol{\Omega}_{11}(\boldsymbol{\Omega}_{12}^T-{\boldsymbol{\Omega}_{21}})^{-T}\boldsymbol{\Omega}_{22}&\;2\boldsymbol{\Omega}_{11}(\boldsymbol{\Omega}_{12}^T-{\boldsymbol{\Omega}_{21}})^{-T}\\2(\boldsymbol{\Omega}_{12}^T-\boldsymbol{\Omega}_{21})^{-T}\boldsymbol{\Omega}_{22}&2(\boldsymbol{\Omega}_{12}^T-{\boldsymbol{\Omega}_{21}})^{-T}
    \end{matrix}\right)},
\end{equation}    
\begin{equation}\label{eq:RDMVacMatDef2}
\boldsymbol{\Omega}=\left(\begin{matrix}\mathbf{R}_{11}&0\\0&-\mathbf{R}_{11}^*\\\end{matrix}\right)+\left(\begin{matrix}\mathbf{R}_{12}&0\\0&\mathbf{R}_{12}^*\\\end{matrix}\right){{\boldsymbol{\mathcal{A}}}}^{-1}\left(\begin{matrix}\mathbf{R}_{12}^T&0\\0&\mathbf{R}_{12}^\dagger\\\end{matrix}\right);\qquad
\boldsymbol{\mathcal{A}}=\left(\begin{matrix}
\mathbf{R}_{22}&-\mathbf{I}\\\mathbf{I}&-\mathbf{R}_{22}^*\\
\end{matrix}\right).
\end{equation}
\end{subequations}\fi

Since one can use Wick theorem for the state $\ket{0}_{\eta}$, the RDM can be written in terms of Majorana fermions and their correlations (introduced in section \ref{sec:Correlation}) in subsystem $\mathbf{1}$. The reduced density matrix can be written as \cite{Peschel_2003}:
\begin{equation}\label{eq:RDM_Vac_Majorana}
\rho_\mathbf{1}^0(\boldsymbol{\gamma},\boldsymbol{\bar{\gamma}})=[\text{det}\frac{\mathbf{I}-\boldsymbol{\Gamma_1^0}}{2}]^{\frac{1}{2}}e^{\frac{1}{4}
\begin{pmatrix}\boldsymbol{\gamma}&\boldsymbol{\bar{\gamma}}\end{pmatrix} 
\text{ln}\frac{\mathbf{I}+\boldsymbol{\Gamma_1^0}}{\mathbf{I}-\boldsymbol{\Gamma_1^0}}
  \begin{pmatrix}
  \boldsymbol{\gamma}\\ \boldsymbol{\bar{\gamma}}
  \end{pmatrix}},
\end{equation}
where the $\begin{pmatrix}\boldsymbol{\gamma}&\boldsymbol{\bar{\gamma}}\end{pmatrix}$ contains all the Majorana fermions of subsystem $\mathbf{1}$. The $\boldsymbol{\Gamma_1^0}$ stands for the correlation matrix defined in section \ref{sec:Correlation} calculated for the vacuum state and subsystem $\mathbf{1}$. The constant in \eqref{eq:RDM_Vac_Majorana} can be simplified as:
\begin{equation}
\text{det}[\frac{\mathbf{I}-\boldsymbol{\Gamma_1^0}}{2}]=\text{det}[\mathbf{K_1^0}]\text{det}[\frac{\bar{\mathbf{K}}^0_1-\mathbf{G_1^0}.\mathbf{K_1^0}^{-1}.\mathbf{G_1^0}^T}{4}].
\end{equation} 

\subsection{ZME state}\label{subsec:RDM_ZeroBar}

The ZME state can be obtained by the limit $\ket{\emptyset}=\underset{\beta\rightarrow\infty}{\text{lim}}\ket{\beta}$. Therefore, the RDM $\rho^\emptyset=\ket{\emptyset}\!\bra{\emptyset}$ in the operator form will be the large $\beta$ limit of the results of the subsection \ref{subsec:RDM_BetaState}. 
\begin{equation}\label{eq:RDM_ZME_C_Operator}
\begin{aligned}
    \rho_{\mathbf{1}}^\beta(c,c^\dagger)=&\:\mathcal{C}^\emptyset\:
	 e^{^{\frac{1}{2}
	\left(\begin{matrix}
	\mathbf{c}^\dagger&\!\mathbf{c}\\
		\end{matrix}\right)\begin{matrix}
	\boldsymbol{\mathcal{M}}
	\end{matrix}\left(\begin{matrix}
	\!\mathbf{c}\\\mathbf{c}^\dagger
	\end{matrix}\right)}} e^{^{\tfrac{1}{2}\text{tr}\ln(\tfrac{1}{2}\boldsymbol{\Omega}_{12}-\tfrac{1}{2}\boldsymbol{\Omega}_{21}^T)}}\Big[\text{Pf}[\boldsymbol{\mathcal{W}}]-\boldsymbol{\mathcal{L}}_{3}\boldsymbol{\mathcal{T}}_{22}\boldsymbol{\mathcal{L}}_{2}\\
	&+\big(\boldsymbol{\mathcal{L}}_{1}\boldsymbol{\mathcal{T}}_{22}\boldsymbol{c}^\dagger+(\boldsymbol{\mathcal{L}}_{1}\boldsymbol{\mathcal{T}}_{21}+\boldsymbol{\mathcal{L}}_{2})\boldsymbol{c}\big)\big(\boldsymbol{\mathcal{L}}_{3}\boldsymbol{\mathcal{T}}_{22}\boldsymbol{c}^\dagger+(\boldsymbol{\mathcal{L}}_{4}+\boldsymbol{\mathcal{L}}_{3}\boldsymbol{\mathcal{T}}_{21})\boldsymbol{c}\big)\Big]
\end{aligned}
\end{equation}
where the definition of $\boldsymbol{\mathcal{M}}$, $\boldsymbol{\mathcal{W}}$, $\boldsymbol{\Omega}$, $\boldsymbol{\mathcal{T}}$ and $\boldsymbol{\mathcal{L}}$ can be found in \eqref{subeq:Beta_Consts_Matrixs} and \eqref{BetaMTmatrixdef}.

The above equation is lengthy and calculation of entanglement seem difficult with the above RDM. However, since the ZME state has an opposite parity with respect to $\ket{0}_{\!\eta}$. One can use the \textit{Tilda transformation} introduced in the subsection \ref{subsec:Configur_Basis} to write the ZME state as:
\begin{equation}\label{eq:ZME_to_VacC}
\ket{\emptyset}=C^{\emptyset}e^{\tfrac{1}{2}\sum_{i,j}R_{ij}^\emptyset \tilde{c}^\dagger_i\tilde{c}_j^\dagger}\tilde{\ket{0}}_c
\end{equation}
with properly defined $\mathbf{R}^\emptyset$ matrix and constant $C^\emptyset$. The above Gaussian form for the ZME state will simplify some of the calculations exceedingly. For instance, using the \eqref{eq:ZME_to_VacC}, it is possible to write a shorter notation for RDM as:
\begin{equation}\label{eq:RDM_ZME_Gaussian}
    \rho_{\mathbf{1}}^\emptyset(c,c^\dagger)={C^\emptyset}^\prime
	 e^{\tfrac{1}{2}
	\left(\begin{matrix}
	\mathbf{c}^\dagger&\!\mathbf{c}\\
	\end{matrix}\right)
	\boldsymbol{\mathcal{M}^\emptyset}
	\left(\begin{matrix}
	\!\mathbf{c}\\\mathbf{c}^\dagger\!
	\end{matrix}\right)},
\end{equation}
where ${C^\emptyset}^\prime$ is the normalization factor and $\boldsymbol{\mathcal{M}^\emptyset}=\ln\boldsymbol{\mathcal{T}^{\emptyset}}$ which we have:\small
\begin{subequations}\label{eq:ZME_Tilda_Omega&Mmat}
\begin{equation}
\boldsymbol{\mathcal{T}^\emptyset}={\left(\begin{matrix} 
    \tfrac{1}{2}\boldsymbol{\Omega}_{12}^\emptyset-\tfrac{1}{2}{{\boldsymbol{\Omega}_{21}^\emptyset}}^T+2\boldsymbol{\Omega}_{11}^\emptyset({\boldsymbol{\Omega}_{12}^\emptyset}^T-{\boldsymbol{\Omega}_{21}^\emptyset})^{-T}\boldsymbol{\Omega}_{22}^\emptyset&\;2\boldsymbol{\Omega}_{11}^\emptyset({\boldsymbol{\Omega}_{12}^\emptyset}^T-{\boldsymbol{\Omega}_{21}^\emptyset})^{-T}\\2({\boldsymbol{\Omega}_{12}^\emptyset}^T-\boldsymbol{\Omega}_{21}^\emptyset)^{-T}\boldsymbol{\Omega}_{22}^\emptyset&2({\boldsymbol{\Omega}_{12}^\emptyset}^T-{\boldsymbol{\Omega}_{21}^\emptyset})^{-T}
    \end{matrix}\right)},
\end{equation}
\begin{equation}
\boldsymbol{\Omega^\emptyset}=\left(\begin{matrix}\mathbf{R}_{11}^\emptyset&0\\0&-{\mathbf{R}_{11}^\emptyset}^*\\\end{matrix}\right)+\left(\begin{matrix}\mathbf{R}_{12}^\emptyset&0\\0&{\mathbf{R}_{12}^\emptyset}^*\\\end{matrix}\right){{\boldsymbol{\mathcal{A}^\emptyset}}}^{-1}\left(\begin{matrix}{\mathbf{R}_{12}^\emptyset}^T&0\\0&{\mathbf{R}_{12}^\emptyset}^\dagger\\\end{matrix}\right);\qquad
\boldsymbol{\mathcal{A}^\emptyset}=\left(\begin{matrix}
\mathbf{R}_{22}^\emptyset&-\mathbf{I}\\\mathbf{I}&-{\mathbf{R}_{22}^\emptyset}^*\\
\end{matrix}\right).
\end{equation}
\end{subequations}
\normalsize

The Wick theorem can also be applied to ZME state. Likewise, it is possible to write the RDM in terms of correlation matrices and Majorana fermions of subsystem $\mathbf{1}$. 
%Using the definition of Majorana fermions with respect to the $c$-fermions:
%\begin{equation}
%\begin{pmatrix}\boldsymbol{\gamma}\\\boldsymbol{\bar{\gamma}}\end{pmatrix}
%=\begin{pmatrix}\mathbf{I}&\mathbf{I}\\ i\mathbf{I}&-i\mathbf{I} \end{pmatrix}
%\begin{pmatrix}\mathbf{c}^\dagger\\ \mathbf{c}\end{pmatrix},
%\end{equation}
We can write the RDM as:
\begin{equation}\label{eq:RDM_ZME_Majorana}
\begin{aligned}
\rho_\mathbf{1}^\emptyset(\boldsymbol{\gamma},\boldsymbol{\bar{\gamma}})=&[\text{det}\frac{\mathbf{I}-\boldsymbol{\Gamma_1^\emptyset}}{2}]^{\frac{1}{2}}e^{\frac{1}{4}
\begin{pmatrix}\boldsymbol{\gamma}&\boldsymbol{\bar{\gamma}}\end{pmatrix}
\text{ln}\frac{\mathbf{I}+\boldsymbol{\Gamma_1^\emptyset}}{\mathbf{I}-\boldsymbol{\Gamma_1^\emptyset}}
\begin{pmatrix} \boldsymbol{\gamma}\\ \boldsymbol{\bar{\gamma}}\end{pmatrix}}.
\end{aligned}
\end{equation}
In the above expression, the matrix $\boldsymbol{\Gamma}^\emptyset_{_\mathbf{1}}$ is defined in subsection \ref{subsec:Correlation_ZME}, and the subscript stands for the correlation matrix for the subsystem. %Rest of the matrices are defined in \eqref{eq:RDM_ZME_C_Operator_Constants}.

\subsection{Zero parity eigenstates}\label{subsec:RDM_Gpm}

The zero parity eigenstates $\ket{G_\pm}$ are the cases where $\beta=\pm1$. Using the results of \eqref{eq:BetaRDMGrassFin} and \eqref{eq:BetaRDMFinalVersion2}, the RDM matrix in coherent basis and operator form for $\ket{G_\pm}$ are given by setting $\beta=\pm1$.
\subsection{Spin versus fermion representation}\label{subsec:RDM_SpinFermion}

\begin{figure}[!htb]
	\centering
	\includegraphics[width=\textwidth]{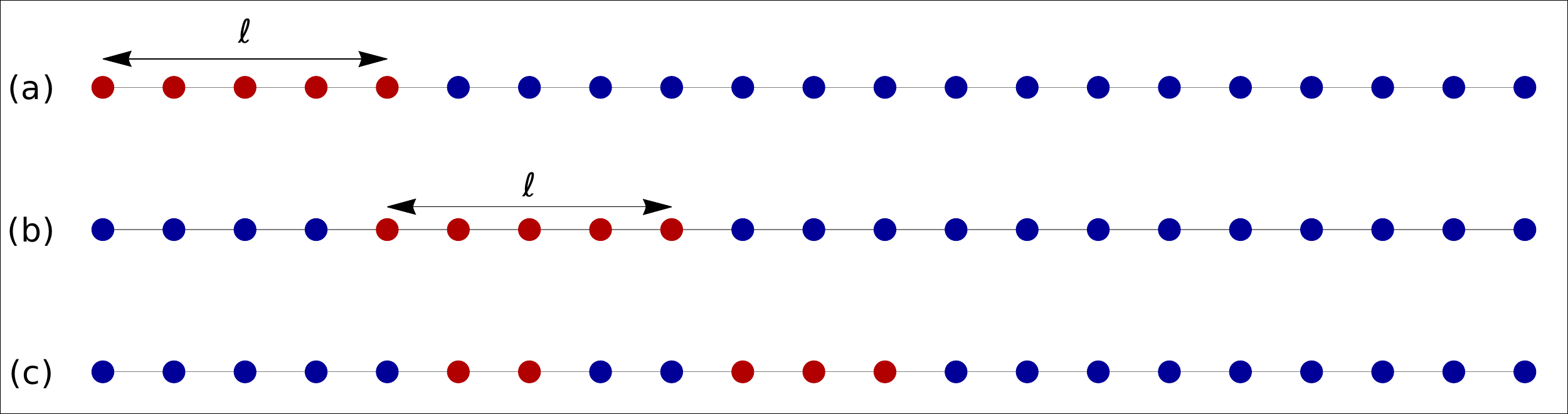}
	\caption{\footnotesize The three subsystem block configurations we consider here. (a) a block of length $\ell$ starting from the boundary. (b) is a block of length $\ell$ at a distance from the boundary. The case (c) shows a disconnected subsystem which does not start from any of boundary points.}\label{fig:SpinFermionBiPart}
\end{figure}

Although the previous considerations are advantageous numerically to study RDM's, we have to point out that the RDM for the spin representation and the fermionic representation (of the Hamiltonian) are not identical, necessarily. Correspondingly, the entanglement entropies could end up to be different. Based on the way of selecting the subsystem, RDM's (of spin and fermion representations) could be different or equal. This difference can be expected due to the non-local structure of the Jordan-Wigner transformation \cite{Igloi2010}. In general, we are interested in two scenarios for subsystem bipartition as demonstrated in the figure \ref{fig:SpinFermionBiPart}.

For start, if our desired state (like $\ket{0}$ and $\ket{\emptyset}$) is an eigenstate of parity operator, then RDM has equal form in spin and fermion representations for types (a) and (b) of subsystems in figure \ref{fig:SpinFermionBiPart}. Since the effect of the Jordan-Wigner strings disappears in these cases. This statement is true for any boundary as long as boundary terms do not break the parity symmetry. If subsystem is not connected, then starting from fermionic representation, to obtain a spin correlation function (like $\langle\sigma^x_i\sigma^x_j\rangle$), one needs information about the string of sites between two blocks of subsystem in the fermionic picture, while, this is not necessary if one asks only for fermionic correlations.

In the related case to our study, which we are dealing with an open boundary case and a state which does not respects the parity (like $\ket{\beta}$), then the relation between spin and fermion version of RDM is more peculiar. Essentially, the string of $\sigma^z$'s from Jordan-Wigner transformation would
break off the correspondence between spins and fermions for a subsystem separated from the boundary, similar to (b) in figure \ref{fig:SpinFermionBiPart}. It is an anomaly of parity broken state that the RDM of any interval starting from the boundary is the same for spins and fermions. While not starting from one of the boundaries of the chain, other techniques should be used to find the RDM of spin representation \cite{Fagotti_2011}. The arguments that are presented in this part can be summarized in the table \ref{tab:RDM_SpinVsFerm}.

\begin{table}[!htb]
    \centering
    \caption{\footnotesize A handy representation of the statements in the section \ref{subsec:RDM_SpinFermion}. In the below, $\rho_{_\mathrm{spin}}$ stands for the RDM in the spin representation and $\rho_{_\mathrm{fermion}}$ is the RDM in the fermionic representation of the Hamiltonian. The subsystem types are demonstrated in figure \ref{fig:SpinFermionBiPart}. In addition, we have assumed the restriction on $\beta\neq0,\infty$.}
    \label{tab:RDM_SpinVsFerm}
    \begin{tabular}{|c|c|c|c|}
        \cline{1-4}
        \diagbox{State}{Subsystem} & type (a) & type (b) & type (c) \\\hline
        \multicolumn{1}{|c|}{$\ket{0}$, $\ket{\emptyset}$}& $\rho_{_\mathrm{spin}}=\rho_{_\mathrm{fermion}}$ & $\rho_{_\mathrm{spin}}=\rho_{_\mathrm{fermion}}$ & $\rho_{_\mathrm{spin}}\neq\rho_{_\mathrm{fermion}}$ \\\hline
        \multicolumn{1}{|c|}{$\ket{\beta}$}& $\rho_{_\mathrm{spin}}=\rho_{_\mathrm{fermion}}$& $\rho_{_\mathrm{spin}}\neq\rho_{_\mathrm{fermion}}$ &$\rho_{_\mathrm{spin}}\neq\rho_{_\mathrm{fermion}}$\\\hline
    \end{tabular}
\end{table}

\section{Entanglement Entropy}\label{sec:Entanglement}

%The entanglement entropy is a measure of the degree of quantum entanglement between two subsystems constituting a bipart composite quantum system. Exists famous entropy measures of entanglement for a subsystem $A$ such as R\'enyi, Tsallis or Von Neumann.
Given the reduced density matrix $\rho_{_A}$ describing knowledge of the state of a subsystem $A$, the R\'enyi entanglement entropy is given by:
\begin{equation}\label{eq:RenyiEntGeneral}
    S_n(A)=\frac{1}{1-n}\log(\text{tr}_{\!_A}\rho_{\!_A\!}^n).
\end{equation}
The R\'enyi entanglement entropy (EE) can be seen as the generalized version of von Neumann entanglement measure. In the limit $n\rightarrow1$, R\'enyi EE produces the Von Neumann EE ($S_{vN}$),  
\begin{equation}
    S_{vN}(A)=-\text{tr}_{\!_{A}}[\rho_{_{A}}\log\rho_{_A}].
\end{equation}
%The Tsallis entanglement measure is not as famous as previous ones, still it is a good measure of entanglement. This entropy in terms of RDM is given by:
%\begin{equation}
%    S_{n}^{\text{Tsallis}}(A)=\frac{1-\text{tr}_{_A}[\rho_{_A}^n]}{n-1}.
%\end{equation}
The difficulty of calculation of the entanglement grows exponentially with size of the subsystem. One can find a  basis which RDM is diagonal, however, it would still be computationally disadvantageous. 
%In this basis The R\'enyi entropy will be:
%\begin{equation}
 %   S_\alpha(A)=\frac{1}{1-\alpha}\log(\sum_i\lambda_i^\alpha),
%\end{equation}
%where $\lambda_i$'s are eigenvalues of RDM. 

It is possible to have special structure for the RDM to simplify the entanglement calculation. For instance having a Gaussian form for RDM simplifies the calculation, or being able to write the RDM in terms of correlation matrices. In the rest of this section, we first present the result of entanglement for the vacuum state and the ZME state which are basically the results in \cite{Vidal2001,jin2004quantum,Keating2004,Peschel_Eisler_2009}. We are going to give the entanglement in terms of the eigenvalues of the correlation matrices and Gaussian form of the RDM. Next, we will discuss %entanglement for the case of ZME state which also can be expressed in different methods. Finally, we hand out 
the entanglement calculation for a general parity broken state such as $\ket{\beta}$ and the limiting cases of $\ket{G_\pm}$ which are new.

\subsection{Vacuum and ZME state}\label{subsec:EntVacuum}

Since the Wick theorem can be applied to the vacuum state we were able to write the RDM as \eqref{eq:VacRDMFinalVersion2}. 
%For a correctly defined matrix $\boldsymbol{\mathcal{M}}$ we have:
%\begin{equation}\label{eq:TraceFormula1}
 %   \text{tr}\:e^{\begingroup\tiny
  %  \frac{1}{2}
%	\left(\begin{matrix}
%	\mathbf{c}^\dagger&\!\!\!\mathbf{c}
%	\end{matrix}\right)\begin{matrix}
%	\boldsymbol{\mathcal{M}}
%	\end{matrix}\left(\begin{matrix}
%	\!\mathbf{c}\!\\\mathbf{c}^\dagger\!
%	\end{matrix}\right)\endgroup
%	}=\det\big(\mathbf{I}+e^{\boldsymbol{\mathcal{M}}}\big)^{\tfrac{1}{2}}.
%\end{equation}
Using the equations \eqref{eq:VacRDMFinalVersion2} and \eqref{eq:RenyiEntGeneral}, we can trace the RDM and write the R\'enyi entanglement for this state as
\begin{equation}\label{eq:EntVac}
    S_n^0(A)=\frac{1}{1-n}\log\Big(\mathrm{C}^{\alpha} \text{det}\big[ \mathbf{I}+e^{n\boldsymbol{\mathcal{M}}} \big]^{\tfrac{1}{2}} \Big),
\end{equation}
where the constant $\mathrm{C}$ is given by
\begin{equation}
    \mathrm{C}=\tfrac{\sqrt{\det\big[\mathbf{I}+{\mathbf{R}_{22}}^\dagger\mathbf{R}_{22}\big]}}{\sqrt{\det\big[\mathbf{I}+{\mathbf{R}}^\dagger\mathbf{R}\big]}}e^{
    \begingroup\footnotesize
    \tfrac{1}{2}\text{tr}\ln(\tfrac{1}{2}\boldsymbol{\Omega}_{12}-\tfrac{1}{2}\boldsymbol{\Omega}_{21}^T)
    \endgroup}.
\end{equation}
The matrices $\boldsymbol{\mathcal{M}}$ and $\boldsymbol{\Omega}$ are already defined in in \eqref{subeq:Beta_Consts_Matrixs} and \eqref{BetaMTmatrixdef}%\eqref{eq:RDMVacMatDef}
. This relation is computationally favorable, since all we need to calculate is a determinant. 

As it was written in section \ref{subsec:RDM_Vacuum}, we could have also express the RDM in terms of correlation matrices of Majorana fermions \eqref{eq:RDM_Vac_Majorana}. Therefore, it is possible to find the entanglement in terms of the correlation matrix $\boldsymbol{\Gamma}^0$  (of the vacuum state) for the subsystem $A$. Therefore, using the RDM \eqref{eq:RDM_Vac_Majorana} and the trace formula %\eqref{eq:TraceFormula1}
for Gaussian function of fermions, the R\'enyi EE is given by
\begin{equation}\label{eq:Ent_Vac_Correl}
    S_{n}^0(A)=\frac{1}{2(1-n)}\log\text{det}\big[(\frac{1+\boldsymbol{\Gamma}^0_{\!\!_A}}{2})^n+(\frac{1-\boldsymbol{\Gamma}_{\!\!_A}^0}{2})^n\big]=\frac{1}{(1-n)}\sum_{j}\log\big[(\frac{1+\nu^0_{\!\!_j}}{2})^n+(\frac{1-\nu_{\!\!_j}^0}{2})^n\big]
\end{equation}
where the set of $\{\pm\nu_k\}$ are the eigenvalues of $\boldsymbol{\Gamma}_A^0$ for the subsystem $A$. The above relation is favorable because it is straightforward to find the correlation matrix for states that obeys Wick theorem.

%\subsection{ZME state}\label{subsec:EntZME}
%In this subsection we are going to calculate the entanglement for the state $\ket{\emptyset}=\eta_0^\dagger\ket{0}$ in different methods.
The entanglement for the ZME state can have a similar form.
For instance, in subsection \ref{subsec:RDM_ZeroBar}, it was mentioned that through some canonical transformations it is possible to write the RDM in Gaussian form \eqref{eq:RDM_ZME_Gaussian}. Similar to previous state, using the Gaussian form of RDM and tracing that, we can get the R\'enyi entanglement for subsystem $A$ same as \eqref{eq:EntVac} but we have to use $\mathbf{R}^\emptyset$ instead of $\mathbf{R}^0$.
%\begin{equation}\label{eq:Reny_Ent_ZME}
 %   S_\alpha^\emptyset(A)=\frac{1}{1-\alpha}\log\Big(\mathrm{C^\emptyset}^{\alpha} \text{det}\big[ \mathbf{I}+e^{\alpha\boldsymbol{\mathcal{M}}^\emptyset} \big]^{\tfrac{1}{2}} \Big),
%\end{equation}
%The matrix $\boldsymbol{\mathcal{M}}^\emptyset$ was already defined in \eqref{eq:ZME_Tilda_Omega&Mmat} and the constant $\mathrm{C}^\emptyset$ is given by
%\begin{equation}
 %   \mathrm{C}^\emptyset=\tfrac{\sqrt{\det\big[\mathbf{I}+{\mathbf{R}_{22}^\emptyset}^\dagger\mathbf{R}_{22}^\emptyset\big]}}{\sqrt{\det\big[\mathbf{I}+{\mathbf{R}^\emptyset}^\dagger\mathbf{R}^\emptyset\big]}}e^{
  %  \begingroup\footnotesize
%    \tfrac{1}{2}\text{tr}\ln(\tfrac{1}{2}\boldsymbol{\Omega}_{12}^\emptyset-\tfrac{1}{2}{\boldsymbol{\Omega}_{21}^\emptyset}^T)
 %   \endgroup}.
%\end{equation}
On the other hand, the Wick theorem can be applied to the ZME state. Hence, it is possible to relate the EE to the correlation matrices, similar to \eqref{eq:Ent_Vac_Correl}. We only need to use the $\boldsymbol{\Gamma}_A^\emptyset$ given in section \ref{subsec:Correlation_ZME}.
%where the $\boldsymbol{\Gamma}_A^\emptyset$ can be defined using the results of section \ref{subsec:Correlation_ZME}. Therefore, for R\'enyi EE we get:   
%\begin{equation}\label{eq:Ent_ZME_Correl}
 %   S_{\alpha}^\emptyset(A)=\frac{1}{1-\alpha}\sum_{l}\log\big((\frac{1+\upsilon}{2})^\alpha+(\frac{1-\upsilon}{2})^\alpha\big),
%\end{equation}
%where the set of $\{\pm\upsilon_k\}$ are the eigenvalues of $\boldsymbol{\Gamma}_A^\emptyset$ for subsystem $A$.

Based on the form of $\boldsymbol{\Gamma}_A^\emptyset$ and $\boldsymbol{\Gamma}_A^0$, one can deduce that for any subsystem $A$ which does not include the last lattice point $L+1$, they have equal set of eigenvalues ($\{\nu^0\}=\{\nu^\emptyset\}$). It also means that the entanglement $S_{\alpha}^0$ and $S_{\alpha}^\emptyset$ are equal. It means that the states $\ket{0}$ and $\ket{\emptyset}$ have the same entanglement properties for any given model ($\mathbf{A}$ and $\mathbf{B}$ matrices which depend on the model).

\subsection{General \texorpdfstring{$\beta$}{}-parity broken state}\label{subsec:EntBetaState}

Unlike the vacuum and ZME states, there are not many computationally easy methods to study the entanglement entropy for a state such as $\ket{\beta}$. Namely, we can not use the correlation matrix blindly for this state. In this subsection, we first focus on 
a computationally favorable method to study the EE for such state. Then, we offer a relation for special type of bipartition which relies on correlation matrix. For start, when $n\in \mathbb{N}$ we can use Berezin integrations to find the ${\rho_{A}^\beta}^n$. The steps of calculation are written in appendix \ref{sec:App_Ent_alpha_Beta_calculation}. The R\'enyi entanglement entropy for $n=2$ is given by:
\begin{equation}\label{eq:Trace_Beta_Final}
\begin{aligned}
    \text{tr}({\rho_{A}^\beta}^2)=&{\mathcal{C}^\beta}^2\text{Pf}[\boldsymbol{\mathcal{B}}]\Big\{\big(\frac{1}{|\beta|^2}+\text{Pf}[\boldsymbol{\mathcal{W}}]\big)^2+\text{Pf}[\boldsymbol{\mathcal{W}}]\big(\mathbb{C}\boldsymbol{\mathcal{B}}^{^{\!-T}}\mathbb{C}^T|_{_{1,2}}+\mathbb{C}\boldsymbol{\mathcal{B}}^{-T}\mathbb{C}^T|_{_{3,4}}\big)\\
    &-\frac{\mathbb{C}\boldsymbol{\mathcal{B}}^{^{\!-T}}\mathbb{C}^T|_{_{2,4}}}{\beta^2}-\frac{\mathbb{C}\boldsymbol{\mathcal{B}}^{^{\!-T}}\mathbb{C}^T|_{_{1,3}}}{{\beta^*}^2}+\frac{1}{|\beta|^2}\big(\mathbb{C}\boldsymbol{\mathcal{B}}^{^{\!-T}}\mathbb{C}^T|_{_{1,2}}+\mathbb{C}\boldsymbol{\mathcal{B}}^{^{\!-T}}\mathbb{C}^T|_{_{3,4}}-\mathbb{C}\boldsymbol{\mathcal{B}}^{^{\!-T}}\mathbb{C}^T|_{_{2,3}}\\
    &-\mathbb{C}\boldsymbol{\mathcal{B}}^{^{\!-T}}\mathbb{C}^T|_{_{1,4}}\big)+\text{Pf}[\mathbb{C}.\boldsymbol{\mathcal{B}}^{^{\!-T}}\!\!\!.\mathbb{C}^T]\Big\}.
\end{aligned}
\end{equation}
Where,
\begin{equation}
    \boldsymbol{\mathcal{B}}=\left(\begin{matrix}
    \boldsymbol{\Omega}_{_{11}}&\boldsymbol{0}&-\boldsymbol{\Omega}_{_{12}}&\mathbf{I}\\
    \boldsymbol{0}&\boldsymbol{\Omega}_{_{11}}&\mathbf{I}&\boldsymbol{\Omega}_{_{12}}\\
    -\boldsymbol{\Omega}_{_{21}}&-\mathbf{I}&\boldsymbol{\Omega}_{_{22}}&\boldsymbol{0}\\
    -\mathbf{I}&\boldsymbol{\Omega}_{_{21}}&\boldsymbol{0}&\boldsymbol{\Omega}_{_{22}}\\
    \end{matrix} \right),\qquad
    \mathbb{C}=\left(\begin{matrix}
    -\boldsymbol{\mathcal{L}}_1&\boldsymbol{0}&\boldsymbol{\mathcal{L}}_2&\boldsymbol{0}\\
    -\boldsymbol{\mathcal{L}}_3&\boldsymbol{0}&\boldsymbol{\mathcal{L}}_4&\boldsymbol{0}\\
    \boldsymbol{0}&\boldsymbol{\mathcal{L}}_1&\boldsymbol{0}&\boldsymbol{\mathcal{L}}_2\\
    \boldsymbol{0}&\boldsymbol{\mathcal{L}}_3&\boldsymbol{0}&\boldsymbol{\mathcal{L}}_4\\
    \end{matrix} \right).
\end{equation}
The $\boldsymbol{\Omega}$ and $\boldsymbol{\mathcal{L}}$ are given by \eqref{subeq:Beta_Consts_Matrixs}. Also, the notation $\mathbb{X}|_{r,s}$ stand for the element of the matrix $\mathbb{X}$ at the position ($r,s$). The above expression can be used for any bipartition of the system. However, we are restricted to the $n=2$. From section \ref{subsec:Correlation_Beta}, we expect to see no $\beta$-dependence in the entanglement entropy when the subsystem does not contain boundary points \cite{Igloi2010}. Despite that, the entanglement content would not be the same from spin perspective to the fermion one. For the case of general parity broken state, if subsystem is not a connected bipartition starting from boundary, then spin entanglement and fermion entanglement do not agree.  

In section \ref{subsec:RDM_BetaState}, we proved that it is possible to have an ansatz to write the RDM in the diagonal form of \eqref{eq:RDM_Beta_DiagonalForm}. Using this form of RDM one can write the R\'enyi EE as:
\begin{equation}\label{entanglement-beta-satat}
    S_{n}^{\beta}(A)=\frac{1}{1-n}\sum_{j=1}^{l}\log\big((\frac{1+\nu_j}{2})^n+(\frac{1-\nu_j}{2})^n\big)+\frac{1}{1-n}\log[(\tfrac{1}{2}+\mathrm{f}_\beta)^n+(\tfrac{1}{2}-\mathrm{f}_\beta)^n],
\end{equation}
where $\mathrm{f}_\beta=\frac{\mathrm{Re}[\beta]}{1+|\beta|^2}$, and it is the eigenvalue of the zeroth part of RDM \eqref{eq:RDM_Beta_DiagonalForm}. In the above $\nu_j$'s are the eigenvalues of correlation matrix $\boldsymbol{\Gamma}^\beta$. The formula above simplifies the entanglement studies, however, it is valid for a certain type of the subsystems. The set $A$ should be connected subsystem of the system containing the site 0 (first site). Otherwise, we would not be able to get entanglement from correlations of the system.

It is crucial to mention that form of correlation $\boldsymbol{\Gamma}^\beta$, in \eqref{eq:Beta_KMat_form} and \eqref{eq:Beta_GMat_form}, indicates that the eigenvalues of $\boldsymbol{\Gamma}^\beta_A$ for any subsystem $0\in A$ and $L+1\notin A$ are the same as $\boldsymbol{\Gamma}^0_A$. This statement means that there is a relation between entanglements $S_{n}^{0}(A)$, $S_{n}^{\emptyset}(A)$ and $S_{n}^{\beta}(A)$ (only when $n\neq1$);
\begin{equation}\label{entanglement-beta-satate}
    S_{n}^{0}(A)=S_{n}^{\emptyset}(A)=S_{n}^{\beta}(A)+\frac{1}{1-n}\log[\frac{2^{1-n}}{(\tfrac{1}{2}+\mathrm{f}_\beta)^n+(\tfrac{1}{2}-\mathrm{f}_\beta)^n}].
\end{equation}

\subsection{Zero parity eigenstates}\label{subsec:Ent_Gpm}

The R\'enyi entanglement for states $\ket{G_\pm}$ can be computed in different ways. One can use the \eqref{eq:Trace_Beta_Final} in the limit $\beta\rightarrow1$. However, similar to section \ref{subsec:RDM_Gpm},  it is possible to have an ansatz to write the RDM in the diagonal form of \eqref{eq:RDM_Beta_DiagonalForm}. Using this form of RDM one can write the R\'enyi EE as \cite{XavierRajabpour2020}:
\begin{equation}
    S_{n}^{\pm}(A)=\frac{1}{1-n}\sum_{j=0}^{l}\log\big((\frac{1+\nu_j}{2})^n+(\frac{1-\nu_j}{2})^n\big)-\log2.
\end{equation}
In the above $\nu_j$ are the eigenvalues of correlation matrix $\boldsymbol{\Gamma}^\pm$. The above formula simplifies the entanglement studies, however, it is valid for a certain type of the subsystems. The set $A$ should be connected subsystem of the system containing the site 0 (first site). Otherwise, we would not be able to get entanglement from correlations of the system.

It is crucial to mention that form of correlation $\boldsymbol{\Gamma}^\pm$, in \eqref{eq:Beta_KMat_form} and \eqref{eq:Beta_GMat_form}, indicates that the eigenvalues of $\boldsymbol{\Gamma}^\pm_A$ for any subsystem $0\in A$ and $L+1\notin A$ are the same as $\boldsymbol{\Gamma}^0_A$. This statement means that there is a relation between entanglements $S_{n}^{0}(A)$, $S_{n}^{\emptyset}(A)$ and $S_{n}^{\pm}(A)$;
\begin{equation}
    S_{n}^{0}(A)=S_{n}^{\emptyset}(A)=S_{n}^{\pm}(A)+\log2.
\end{equation}
And it is correct for any valid choices of $\mathbf{A}$ and $\mathbf{B}$ matrices. In fact, one can extend this argument to any excited state created from $\ket{0}$, $\ket{\emptyset}$ and $\ket{G_\pm}$ with excitation of same modes, as in
\begin{equation*}
    \ket{\psi}=\prod_{k_j\in \mathbb{E}}\eta_{k_j}^\dagger\ket{0},\qquad
    \ket{\phi}=\prod_{k_j\in \mathbb{E}}\eta_{k_j}^\dagger\ket{\emptyset},\qquad
    \ket{\chi_\pm}=\prod_{k_j\in \mathbb{E}}\eta_{k_j}^\dagger\ket{G_\pm}.
\end{equation*}
where set $\mathbb{E}$ does not contain mode zero. The general form of correlations for states above can be found in appendix \ref{sec:AppExcitedCorrelations}. Based on the results of that appendix, we conclude that 
\begin{equation}
    S_{n}^{\psi}(A)=S_{n}^{\phi}(A)=S_{n}^{\chi_{_\pm}}(A)+\log2.
\end{equation}

\section{Physical interpretation of the parity-broken state}\label{sec:betastate}

In this section, we give a simple interpretation of the $\ket{\beta}$ state which helps to understand some simple cases of the results that we presented so far. This sate can be written as follows:
\begin{equation}\label{beta-Gpm-form}
  \ket{\beta}=\frac{1}{\sqrt{2(1+|\beta|^2)}} \Big{(} (1+\beta)\ket{G_+}+(1-\beta)\ket{G_-}\Big{)}.
\end{equation}
In the spin representation the above state can be written in an interesting form. Consider $\delta_+=1$ then we can write
\begin{equation}\label{beta-threeQubit}
  \ket{\beta}=\frac{1}{\sqrt{2(1+|\beta|^2)}} \Big{(} (1+\beta)\ket{+}_0\ket{\phi_{++}}\ket{+}_{L+1}+(1-\beta)\ket{-}_0\ket{\phi_{--}}\ket{-}_{L+1}\Big{)},
\end{equation}
where $\ket{\phi_{--}}$ and $\ket{\phi_{++}}$ are normalized states. The above form suggests that the whole system can be considered as two qubit with one qubit at site $0$ ($L+1$) and the other the rest of the system. Interestingly, the entanglement structure of these three parts is independent of the size of the system and one can easily calculate for example the entanglement of the site $0$ ($L+1$) with the rest, i.e. $S_n(0)$ ($S_n(L+1)$); 
\begin{equation}
    S_{n}(0)=S_{n}(L+1)=\frac{1}{1-n}\log[(\tfrac{1}{2}+\mathrm{f}_\beta)^n+(\tfrac{1}{2}-\mathrm{f}_\beta)^n],
\end{equation}
which is consistent with the equation (\ref{entanglement-beta-satat}). 
%The above result suggest that as far as $\beta\neq\pm1$ the two ancillary sites, although far from each other, are strongly entangled.
We note that one can generalize the above argument for all the $\ket{\beta}$ states that can be made out of the eigenstates. The extension to $\delta_+=-1$ is straightforward.
Finally, there is one extra piece that one can add to this story by considering the following mixed state:
\begin{equation}\label{beta-mixed}
  \rho^{\beta}=\frac{1}{2(1+|\beta|^2)} \Big{(} |1+\beta|^2\ket{G_+}\bra{G_+}+|1-\beta|^2\ket{G_-}\bra{G_-}\Big{)}.
\end{equation}
One can easily show that the reduced density matrix of the above state is exactly equal to the reduced density matrix of the $\ket{\beta}$ state which is guaranteed because of the especial form of the state. This makes the preparation of states with the desired reduced density matrix very easy. However, it is clear that in the mixed state scenario the von Neumann entropy does not have entanglement interpretation anymore.

\section{Examples}\label{sec:Examples}

In this section, we provide a couple of examples to show how the general results that we derived can be applied in specific cases. The first example which we are able to do the entire calculation analytically is the Hamiltonian  (\ref{eq:FermionHamil}) with $\boldsymbol{A}=\boldsymbol{B}=0$. There are a few good reasons to study this Hamiltonian. First of all, for this Hamiltonian one can follow all the calculations analytically and show the validity of all the presented results. Second, it is a Hamiltonian that can be used to diagonalize a Hermitian Hamiltonian with linear creation and annihilation operators which makes it worth studying. Finally interestingly the entanglement structure that emerges from this Hamiltonian is entirely universal. In other words the general Hamiltonians with generic parameters end up to have similar entanglement structure. This is shown in the example of XY chain with arbitrary boundary magnetic fields which is the second example of this section. The boundary conformal entanglement entropy at the critical point in this case has been studied already in \cite{XavierRajabpour2020}, however, here we are more concentrated on general aspects of entanglement with respect to the $\beta$ parameter and boundary magnetic fields.

\subsection{\texorpdfstring{$\boldsymbol{A}=\boldsymbol{B}=0$}{}}
We are going to focus on the entanglement properties here of the case where the $\mathbf{A}$ and $\mathbf{B}$ matrices are zero. The coupling of fermions in the system has been demonstrated in figure \ref{fig:A&BZero}. A detailed study is presented in Appendix \ref{sec:App_A_B_Zero} for diagonalization and correlations of this special case.
\begin{figure}[!tb]
	\centering
		\includegraphics[width=0.85\textwidth, height=4cm]{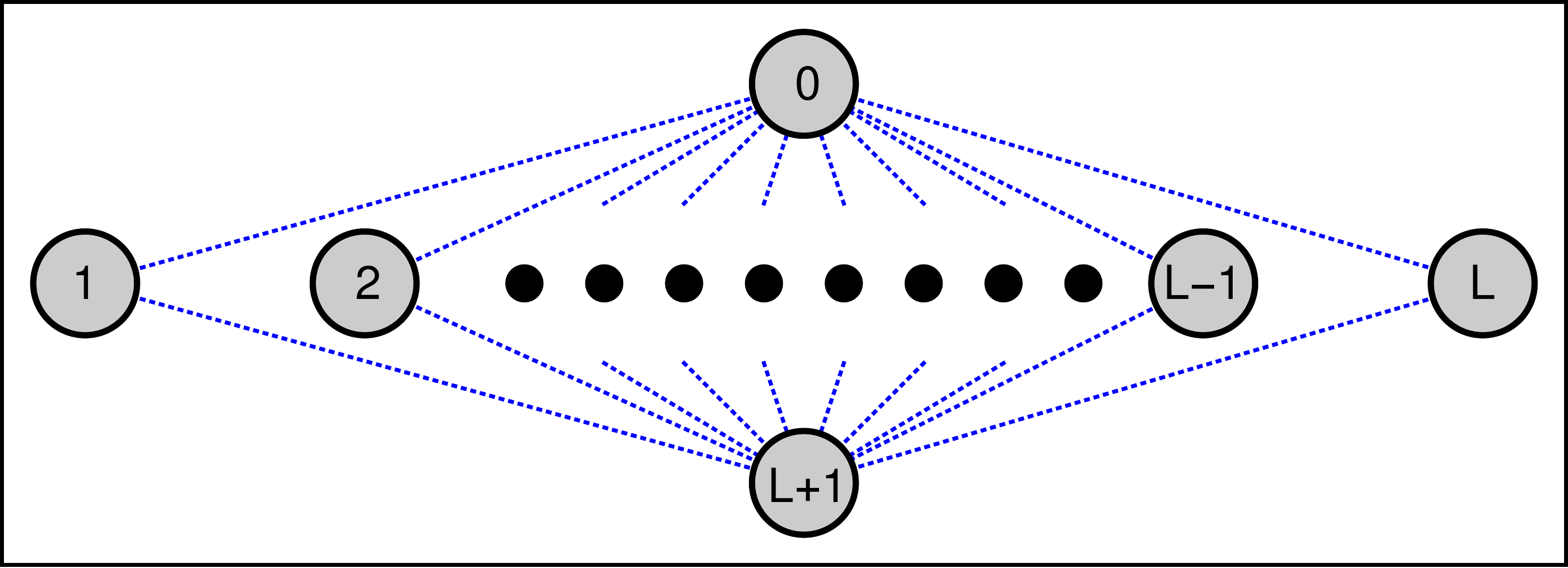}
	\caption{\footnotesize In this figure the type of interaction among fermions at different sites, for $\mathbf{A}=\mathbf{B}=0$ is demonstrated. As one can see, there is no particle hoping among sites $1$ to $L$ and the only allowed hoping is from sites $0$ or $L+1$ with the rest of sites.}\label{fig:A&BZero}
\end{figure}

For the vacuum state and ZME state, which Wick theorem is applicable as it was stated in section \ref{subsec:EntVacuum}, we can relate the R\'enyi entanglement to the eigenvalues of correlation matrix $\boldsymbol{\Gamma}$ introduced in \eqref{eq:GammaMatrixElement}. In a special type of system bipartition, we could also use the $\boldsymbol{\Gamma}^\pm$ to find entanglement for the $\ket{G_\pm}$ case. For any connected subsystem which contains the site $0$ (or site $L+1$), the (positive valued) eigenvalues of correlation matrices $\boldsymbol{\Gamma}^0$, $\boldsymbol{\Gamma}^\emptyset$ and $\boldsymbol{\Gamma}^\pm$ are given by the set:
\begin{equation}
\{\nu\}=\begin{cases}
\{0,\sqrt{1-\frac{(\sum_{j=1}^{\ell}|\alpha_j|^2)(\sum_{j=\ell+1}^{L}|\alpha_j|^2)}{(\sum_{j=1}^{L}|\alpha_j|^2)^2}},1,\cdots,1\} & \ell \leq L,\\
%\{0,1,\cdots,1\} & \ell=L\\
\{1,\cdots\cdots,1\} & \ell=L+1.
\end{cases}    
\end{equation}
The fact that we get the same eigenvalues for each of the states above was explained in section \ref{sec:Entanglement}. Based on the form of correlation matrices similar (up to a zero eigenvalue) eigenvalues was expected. Using the eigenvalues above the entanglement will be given by ($\ell\leq L$)
\begin{eqnarray}
S_n^{0}(\ell)&=&S_n^{\emptyset}(\ell)=\frac{1}{1-n}\log\Big((\frac{1}{2}+\frac{1}{2}\sqrt{1-\tfrac{(\sum_{j=1}^{\ell}|\alpha_j|^2)(\sum_{j=\ell+1}^{L}|\alpha_j|^2)}{(\sum_{j=1}^{L}|\alpha_j|^2)^2}})^n\nonumber\\
&&+(\frac{1}{2}-\frac{1}{2}\sqrt{1-\tfrac{(\sum_{j=1}^{\ell}|\alpha_j|^2)(\sum_{j=\ell+1}^{L}|\alpha_j|^2)}{(\sum_{j=1}^{L}|\alpha_j|^2)^2}})^n\Big)+\log2,\\
S_n^\pm(\ell)&=&S_n^{0}(\ell)-\log2.
\end{eqnarray}
In case where all the $\alpha_i$'s are constant, we would get
\begin{equation}
S_n^{0}(\ell)=\frac{1}{1-n}\log\Big((\tfrac{1}{2}+\tfrac{1}{2}\sqrt{\tfrac{L^2-\ell L+\ell^2}{L^2}})^n+(\tfrac{1}{2}-\tfrac{1}{2}\sqrt{\tfrac{L^2-\ell L+\ell^2}{L^2}})^n\Big)+\log2,
\end{equation}
In the thermodynamic limit ($L\to \infty$), we simply get $S_n^{0}(\ell)=S_n^{\emptyset}(\ell)=\log2$ and $S_n^{\pm}(\ell)=0$.

For any connected bipartition of system with length $\ell$ which does not contain the $0^\mathrm{th}$ and $L+1^{\mathrm{th}}$ sites, eigenvalues of $\boldsymbol{\Gamma}^0$ and $\boldsymbol{\Gamma}^\emptyset$ are given by
\begin{equation}%\label{eq:GMatEigenvalues2}
\{\nu\}=\begin{cases}
\{\frac{(\sum_{k=\ell+1}^{L}|\alpha_{k}|^2}{(\sum_{k=1}^{L}|\alpha_{k}|^2)^2},1,1,\cdots,1\}; & \ell\leq L,\\
\{0,1,1,\cdots,1\}; & \ell=L+1.
\end{cases}
\end{equation}
With this type of bipartition for the system, then the correlation matrix can be used to calculate the entanglement for states $\ket{0}$ and $\ket{\emptyset}$. While for the state $\ket{G_{\pm}}$ we can not, despite the fact that the $\boldsymbol{\Gamma}^\pm$ has the same eigenvalues.

\begin{figure}[H]
	\centering
	%\begin{subfigure}[b]{0.48\textwidth}
	\includegraphics[width=0.85\textwidth]{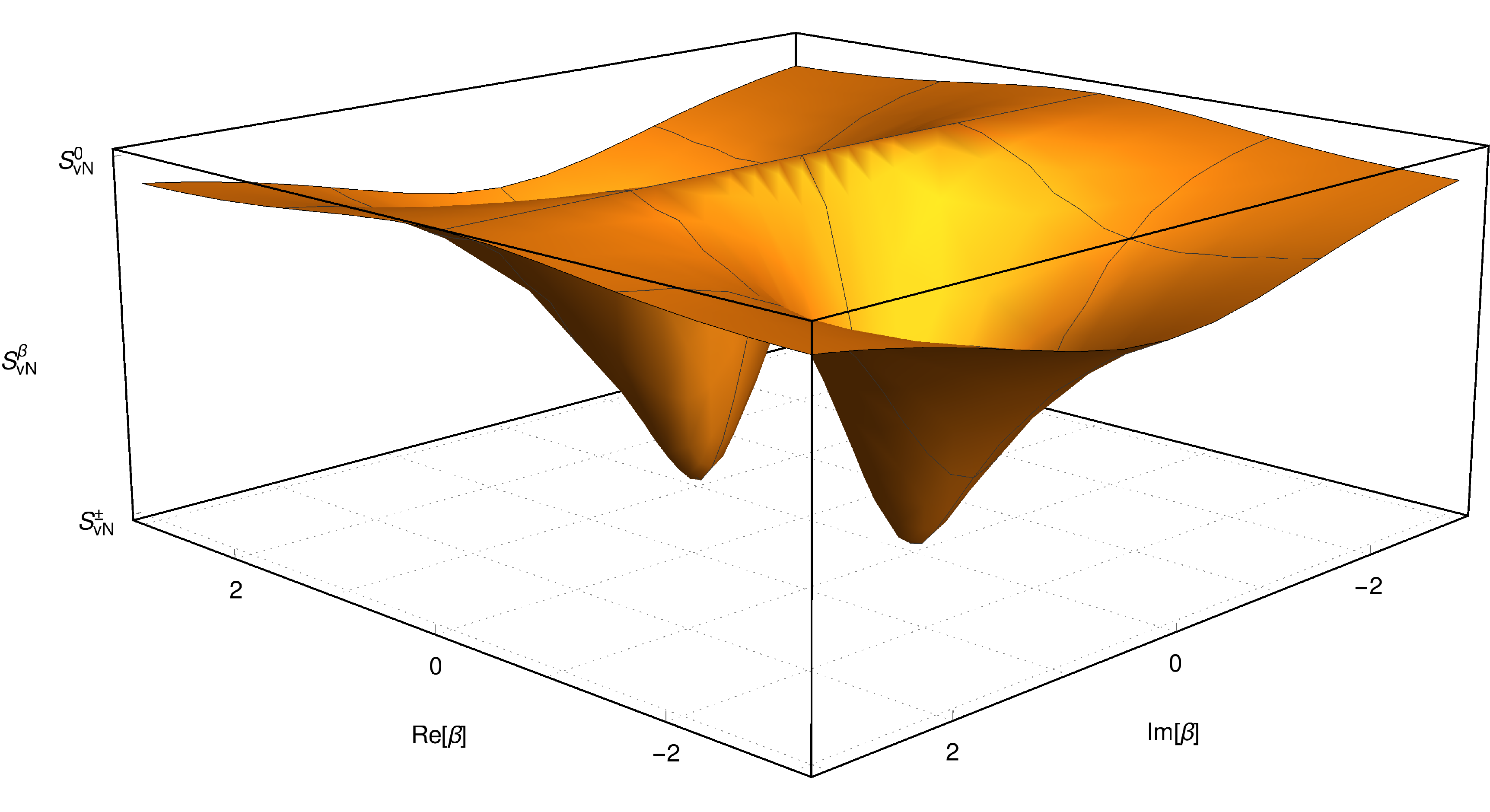}
	%	\caption{}\label{fig:gull}
	%\end{subfigure}
	%\hfill
	%\begin{subfigure}[b]{0.48\textwidth}
	%	\includegraphics[width=\textwidth]{ENTvsBeta.pdf}
	%	\caption{}\label{fig:tiger}
	%\end{subfigure}
	\caption{\footnotesize Plot of von Neumann entanglement entropy with respect to $\beta$. We already expected that the entanglement to be minimum for the case of $\beta=\pm1$ which is apparent in the figure. In the case where $\beta$ is purely imaginary, we notice that the EE is again maximum and would not change. This figure is valid for a connected subsystem starting from site $0$, for a general set of couplings in the Hamiltonian and size of the full system and subsystem.}\label{fig:EntVSBeta}
\end{figure}

%Now, for the partitions that we can use the eigenvalues of $\boldsymbol{\mathcal{G}}^T\boldsymbol{\mathcal{G}}$ (to calculate the entanglement), one can actually compare the eigenvalues and show that the eigenvalues \eqref{eq:GMatEigenvalues} are bigger than those in \eqref{eq:GMatEigenvalues2}.
%$$\nu_k-\nu_k^\prime>0,$$
%which also means that in this case one could conclude that
%$$S^{\pm}(A)<S^{0}(A),S^{\emptyset}(A).$$
In the case of general $\beta$-state, and when $\alpha_i\in\mathbb{R}$, the von Neumann entanglement will be:
\begin{equation}
\begin{aligned}
    S_{\!_{vN}}^\beta\!(A)=&-\tfrac{1}{2}\log\Big[(\frac{1}{4}-\mathrm{f}_\beta)\frac{\Sigma_{l}\Sigma_{L-l}}{4\Sigma_{L}^2}\Big] -\mathrm{f}_\beta\log\big(\frac{1+2\mathrm{f}_\beta}{1-2\mathrm{f}_\beta}\big)\\
    &-\tfrac{\sqrt{\Sigma_{L}^2-\Sigma_{l}\Sigma_{L}+\Sigma_{l}^2}}{2\Sigma_{L}}\log\Big[\frac{\Sigma_{L}+\sqrt{\Sigma_{L}^2-\Sigma_{l}\Sigma_{L}+\Sigma_{l}^2}}{\Sigma_{L}-\sqrt{\Sigma_{L}^2-\Sigma_{l}\Sigma_{L}+\Sigma_{l}^2}}\Big].
\end{aligned}
\end{equation}
Where in the above $\Sigma_l=\sum_{j=1}^{l}\alpha_j^2$ and $\mathrm{f}_\beta=\frac{\mathrm{Re}[\beta]}{1+|\beta|^2}$.
%\begin{equation}
%\begin{aligned}
 %   \text{tr}({\rho_{A}^\beta}^2)=&\frac{(\Sigma_l+2\Sigma_{L-l})|\beta|^2}{4\Sigma_l(1+|\beta|^2)}\Big\{\big(\frac{1+|\beta|^2}{|\beta|^2}\big)^2+\frac{8\Re[\beta^2]}{|\beta|^4}\big(\frac{2\Sigma_L^2-\Sigma_l\Sigma_{L-l}}{(\Sigma_l+2\Sigma_{L-l})^2}\big)+\frac{8}{|\beta|^2}\big(\frac{2\Sigma_L^2-\Sigma_l\Sigma_{L-l}}{(\Sigma_l+2\Sigma_{L-l})^2}\big)\Big\}.
%\end{aligned}
%\end{equation}
It is particularly interesting to see the behavior of entanglement with respect to the $\beta$. As it was mentioned previously, parameter $\beta$ can be thought as a parameter which breaks the parity continuously. In figure \ref{fig:EntVSBeta}, we have demonstrated a typical behavior of EE for different values of $\beta$ in complex plain when all $\alpha_i$'s are real constants. As you can see, for real values of $\beta$, the entanglement is maximum for $\beta=0,\infty$ and minimum for $\beta=\pm1$, which the first one corresponds to vacuum and ZME state and second one is the $\ket{G_\pm}$. On the other hand, for the purely imaginary $\beta$, entanglement is constant and equal to $S_{_{2}}^{0}(l)$ for any value of $\beta$ which is not a trivial observation.

\subsection{Modified \textbf{XY} chain with boundary magnetic field}\label{subsec:XYExample}

In this section, we consider the modified transverse field XY chain with arbitrary direction of the boundary magnetic field. We are interested in the following XY Hamiltonian\cite{XavierRajabpour2020} with open boundary conditions
\begin{equation}\label{eq:XY_Ham_Ghost}
\begin{aligned}
    \mathcal{H}^{^{XY}}\!=&J\sum_{i=1}^{L-1}\big[\frac{1+\gamma}{4}\sigma^x_i\sigma^x_{i+1}+\frac{1-\gamma}{4}\sigma^y_i\sigma^y_{i+1}\big]-\frac{h}{2}\sum_{i=1}^{L}\sigma^z_{i}\\
    &+\frac{1}{2}\big(b_{1,x}\sigma^x_{0}\sigma^x_{1}+b_{1,y}\sigma^x_{0}\sigma^y_{1}+b_{1,z}\sigma^z_{1}+b_{L,x}\sigma^x_{L}\sigma^x_{L+1}+b_{L,y}\sigma^y_{L}\sigma^x_{L+1}+b_{L,z}\sigma^z_{L}\big)
\end{aligned}
\end{equation}
where $\overrightarrow{b}_{\!\!1}=(b_1\sin\theta_1\cos\varphi_1,b_1\sin\theta_1\sin\varphi_1,	b_1\cos\theta_1)$ and $\overrightarrow{b}_{\!\! L}=(b_L\sin\theta_L\cos\varphi_L,b_L\sin\theta_L\sin\varphi_L,b_L\cos\theta_L)$ are constant vectors. Form of $\mathcal{H}^{^{XY}}$ suggests that we can diagonalize and find the eigenstates analytically. Using the Jordan-Wigner transformation
\begin{equation}\label{eq:XY_example_JWtransf}
    c_{l}^\dagger=\prod_{j=0}^{l-1}\sigma^z_{j}\sigma^+_{l},
\end{equation}
where $\sigma^{\pm}_{n}=\frac{\sigma^x_{n}\pm i\sigma^y_{n}}{2}$, we can map the Hamiltonian \eqref{eq:XY_Ham_Ghost} to the free fermion Hamiltonian \eqref{eq:Ham matrix form},
%\begin{equation}
 %   H^{^{XY}}_{f}=\sum_{i,j=0}^{L+1}\big[c_{i}^{\dagger}A^{^{\!XY}}_{ij}c_{j}+\tfrac{1}{2}c_{i}^{\dagger}B^{^{\!XY}}_{ij}c_{j}^{\dagger}+\frac{1}{2}c_{i}{B^{^{\!XY}}_{ji}}^{*}c_{j}\big]-\frac{1}{2}\text{tr}({\mathbf{A}^{^{\!XY}}}^*).
%\end{equation}
with properly chosen matrices $\mathbf{M}$ (see appendix \ref{sec:App_M_XY} for a demonstration of this matrix).
%$H^{^{XY}}_{f}$ commutes with $\sigma^x_0$ and $\sigma^x_{L+1}$. Due to this fact it is possible to block diagonalize H in four distinct sectors, labeled by the eigenvalues of $\sigma^x_{0}$ and $\sigma^x_{L+1}$. We are going to denote these sectors by ($s_0$, $s_{L+1}$), where $!$.
The diagonalization process has been explained in details in section \ref{subsec:QuadHamDiagonaliz}. The above Hamiltonian is related to an XY chain with boundary magnetic fields as:
\begin{equation}\label{eq:XY_Ham_BMF}
    H^{^{XY}}\!=J\sum_{i=1}^{L-1}\big[\frac{1+\gamma}{4}\sigma^x_i\sigma^x_{i+1}+\frac{1-\gamma}{4}\sigma^y_i\sigma^y_{i+1}\big]-\frac{h}{2}\sum_{i=1}^{L}\sigma^z_{i}+\frac{1}{2}(\overrightarrow{b}_{\!\!1}.\overrightarrow{\sigma}_{\!\!1}+\overrightarrow{b}_{\!\!L}.\overrightarrow{\sigma}_L).
\end{equation}
In the above expression, we can think of $\overrightarrow{b}_{\!\!1}$ and $\overrightarrow{b}_{\!\! L}$ as magnetic fields at boundaries of our spin chain. If we add two auxiliary spins at positions $0$ and $L+1$, if they can only interact with the $x$ component of spins at position $1$ and $L$, then we get \eqref{eq:XY_Ham_Ghost}. It is apparent that $\mathcal{H}^{^{XY}}$ commutes with $\sigma^x_0$ and $\sigma^x_{L+1}$. Due to this fact it is possible to divide the Hilbert space of $\mathcal{H}^{^{XY}}$ in four distinct sectors, labeled by the eigenvalues of $\sigma^x_{0}$ and $\sigma^x_{L+1}$. Therefore, the eigenstates of Hamiltonian \eqref{eq:XY_Ham_BMF} would be found in one of these sectors. This procedure has been explained in section \ref{subsec:SelectionRule}.

\subsubsection{Determination of BMF Hamiltonian eigenstates}

\begin{figure}[!htb]
	\centering
    \begin{subfigure}[H]{0.49\linewidth}
		\includegraphics[width=7.5cm, keepaspectratio]{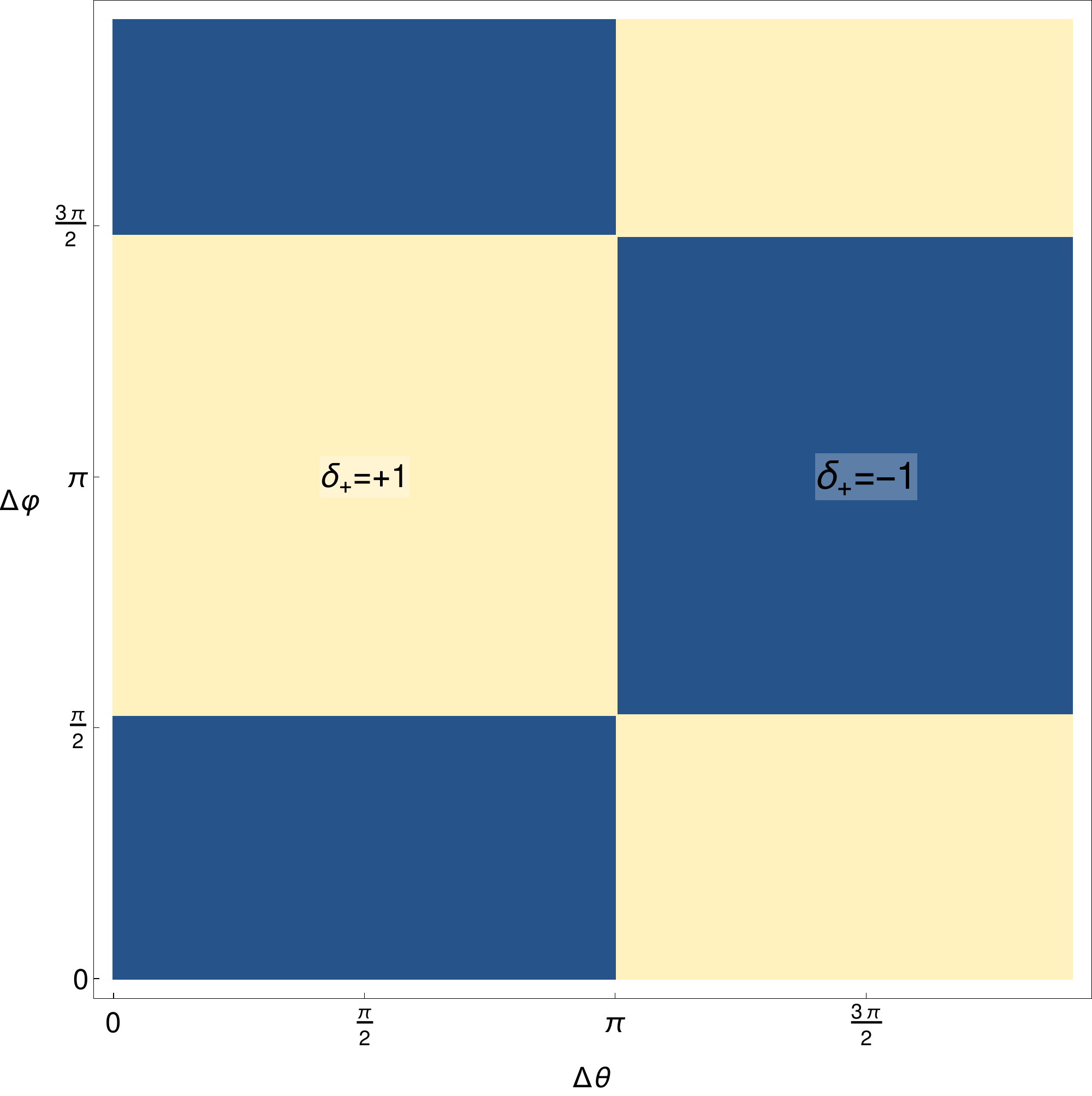}
	\end{subfigure}
	\hfill
	\begin{subfigure}[H]{0.49\linewidth}
		\includegraphics[width=7.5cm, keepaspectratio]{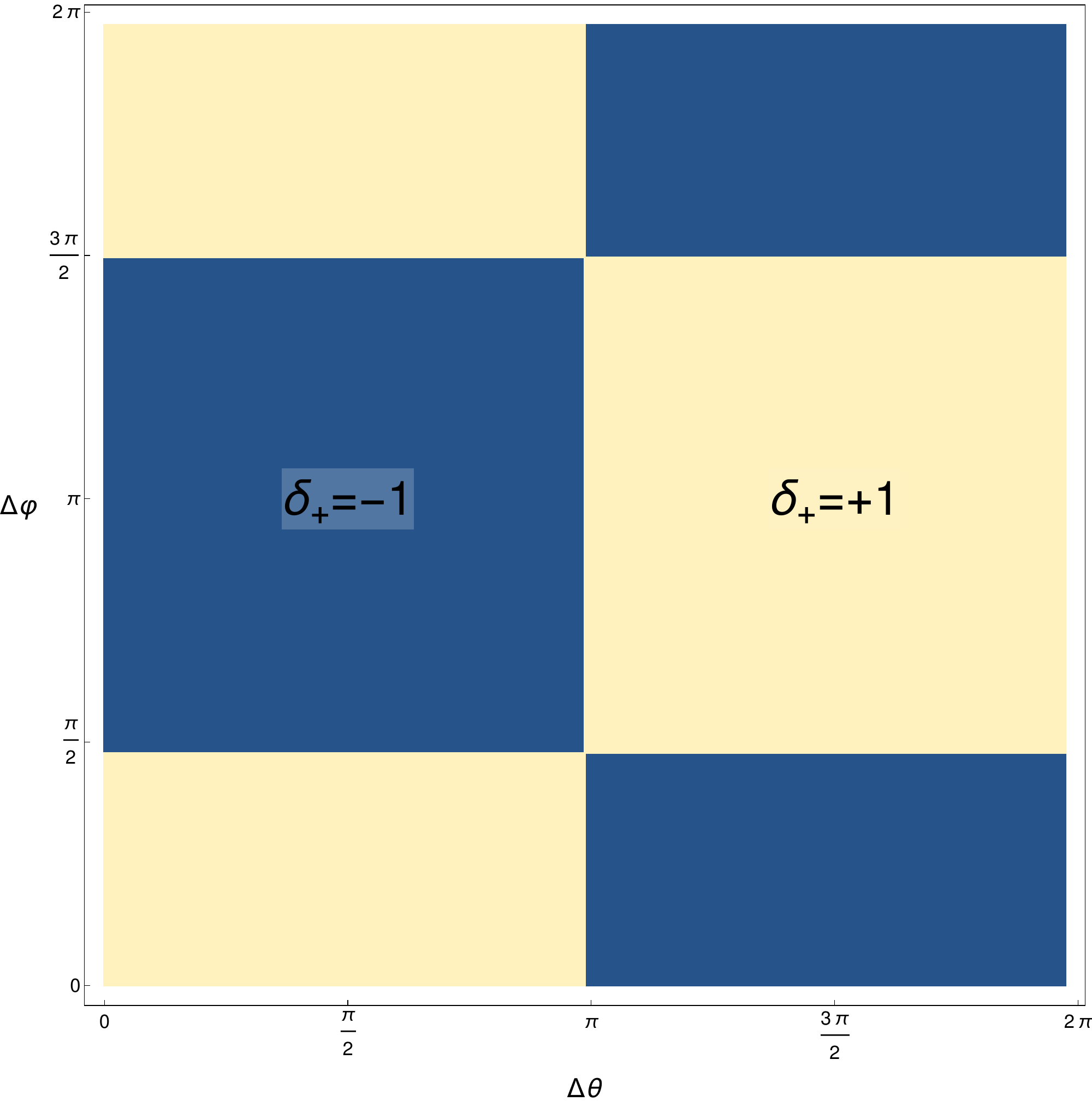}
	\end{subfigure}
	\caption{\footnotesize Different values of $\delta_+$ for difference in the directions of boundary fields $\Vec{b_1}=(\sin\theta_1\cos\varphi_1,\sin\theta_1\sin\varphi_1,\cos\theta_1)$ and $\Vec{b_L}=(\sin\theta_L\cos\varphi_L,\sin\theta_L\sin\varphi_L,\cos\theta_L)$. We have defined $\Delta\varphi=\varphi_L-\varphi_1$ and $\Delta\theta=\theta_L-\theta_1$. The direction of boundary field at the first site is fixed by $\theta_1=\varphi_1=\frac{\pi}{200}$ and we change the angles at end point of the chain. On the right, we have $L=9$ and left hand side corresponds to $L=10$. Rest of the parameters are: $J=-1$, $\gamma=1/2$ and $h=0.518$.}\label{fig:XYdeltapVsAngs}
\end{figure}

The eigenstates of Hamiltonian \eqref{eq:XY_Ham_BMF} can be found with a correct projection in the Hilbert space of \eqref{eq:XY_Ham_Ghost}, which was mentioned in the section \ref{subsec:SelectionRule}. Here, we focus on the XY-chain case and present (apparent) patterns to find the ground state of \eqref{eq:XY_Ham_BMF}. Based on results of \eqref{eq:SelRul_dp_GS} and \eqref{eq:SelRul_dm_GS}, one can identify the ground state of BMF spin chain in $\ket{G_+}$ or $\eta_{min}^\dagger\ket{G_-}$ by finding the $\delta_+$. Numerical investigations suggests that the value of $\delta_+$ does not depend on the parameters such as $\gamma$, $h$ (fixing $J=-1$) and strength of the magnetic field at boundaries. The size of the full system however changes the value of $\delta_+$; For equal directions of boundary fields ($\overrightarrow{b}_{\!\! 1}$ and $\overrightarrow{b}_{\!\! L}$) we get:
\begin{equation}
    \begin{cases}
    \delta_+=-1 & L\; \mathrm{even},\\
    \delta_+=+1 & L\; \mathrm{odd}.
    \end{cases}
\end{equation}
Nonetheless, the change in the directions of BMF's at ends of the chain can affect the value of $\delta_+$. Figure \ref{fig:XYdeltapVsAngs} shows the $\delta_+$ with respect to the change in the azimuthal and polar angle of direction of the magnetic field at boundaries. If one eliminates the BMF at one end of the system (either by putting $\theta_1=0,\pi$ or $b_1=0$), then changing the direction or strength of BMF at the other end would not affect the $\delta_+$ outcome. 

To be able to use the relation \eqref{eq:BetaRDMFinalVersion2}, one need to have the unitary transformation $\mathbf{U}$, which diagonalizes the Hamiltonian, in the form given by \eqref{eq:Diag_U_Mat}. Necessary condition is to make sure that eigenstates of $\mathbf{M}$ corresponding to zero modes are written correctly, they are orthogonal and satisfy the \eqref{eq:JOperator}. We have already introduced the zero mode of the system in section \ref{subsec:QuadHamZeroMode}, which explains the 2-fold degeneracy of the ground state. However, based on some values of coupling parameters (such as uniform external magnetic field or direction of boundary fields) there could arise more zero modes in the spectrum of $\mathbf{M}$ matrix. For instance, when $\gamma=1$ and $\varphi_1,\varphi_L=\frac{\pi}{2}$ (and general bulk and boundary magnetic fields $h$, $\overrightarrow{b}_{\!\! 1}$, $\overrightarrow{b}_{\!\! L}$), we would have extra zero mode and more degeneracy. For this values, the extra zero modes eigenvectors of $\mathbf{M}$ would be (besides \eqref{eq:zeromode evec 1})
\begin{equation}\label{eq:XY_zeromode_2}
\arraycolsep=1.4pt\def\arraystretch{1.2}
\ket{u_0^3}=\tfrac{1}{\sqrt{4+2\zeta_{_1}^2+2\zeta_{_L}^2}}\left(\begin{array}{c}
-i\zeta_{_1}\\ 1\\0\\ \vdots\\ 0\\-1\\ i\zeta_{_L}\\ i\zeta_{_1}\\1\\0 \\ \vdots \\0 \\1\\ i\zeta_{_L} \end{array}\right),\qquad \ket{u_0^4}=\tfrac{1}{\sqrt{4+2\zeta_{_1}^2+2\zeta_{_L}^2}}\left(\begin{array}{c}
-i\zeta_{_1}\\ 1\\0\\ \vdots\\ 0\\1\\ -i\zeta_{_L}\\ i\zeta_{_1}\\1\\0 \\ \vdots \\0 \\-1\\ -i\zeta_{_L} \end{array}\right);
\end{equation}
where $\zeta_{_{1,L}}=\frac{h-b_{_{1,L}}\cos{\theta_{_{1,L}}}}{b_{_{1,L}}\sin{\theta_{_{1,L}}}}$. This is an exact zero mode which comes from the fact that BMF's does not have any component in the $x$-direction. In general, there could be more zero modes for large sizes. It does not seem easy to identify analytically all the points where we face degeneracy more than the two mentioned. To study entanglement, one should be careful with the zero modes. As an example, numerical investigations suggests that for $L>>1$, we would expect one more zero mode for $h>1$ and independent from BMF values.

\subsubsection{Entanglement studies}

\begin{figure}[!htb]
	\centering
	%\begin{subfigure}[h]{0.49\linewidth}
		\includegraphics[width=12cm, keepaspectratio]{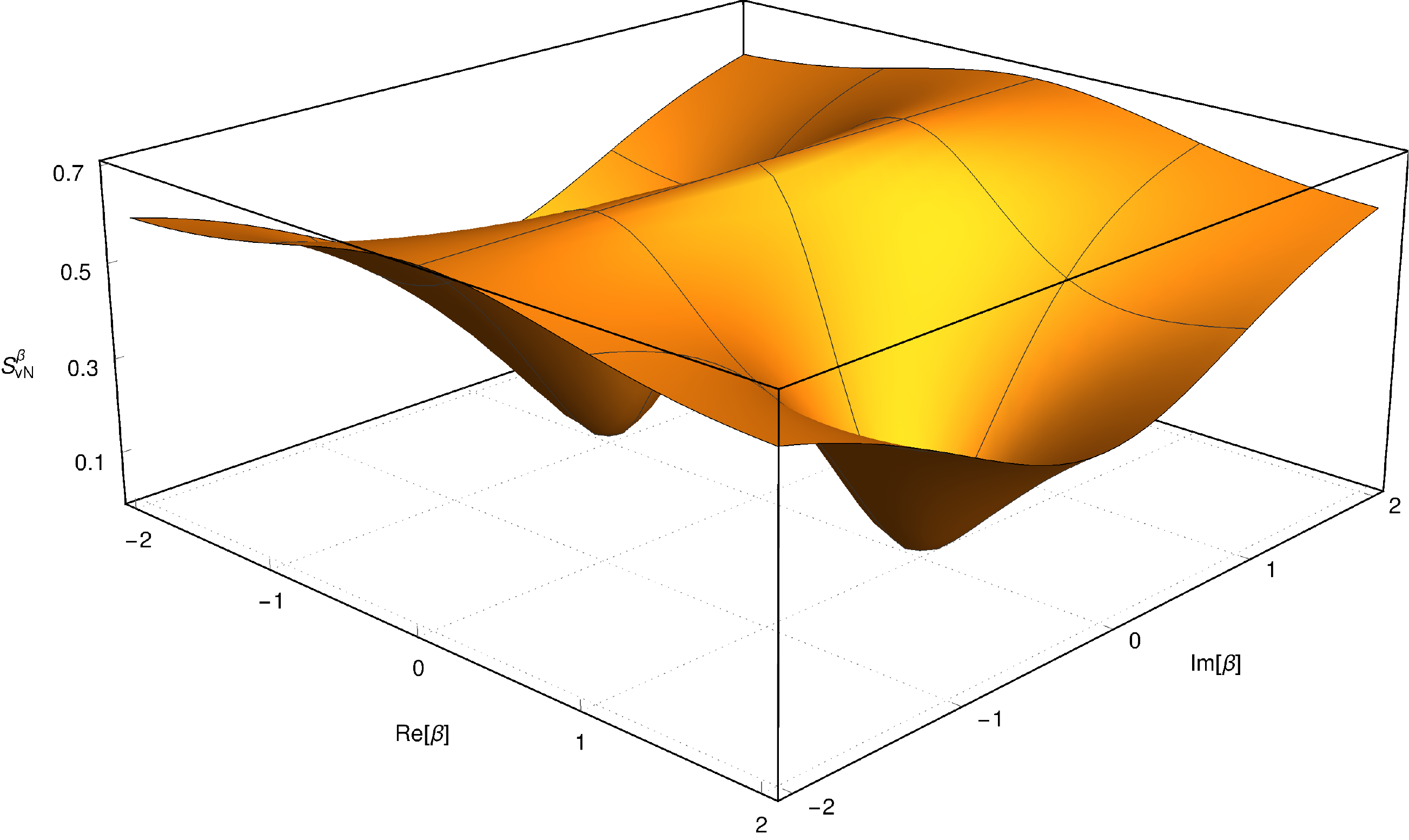}
	%	\caption{}\label{fig:gull}
	%\end{subfigure}
	%\hfill
	%\begin{subfigure}[h]{0.49\linewidth}
	%	\includegraphics[width=7.5cm, keepaspectratio]{XYFig1-2:EntVsBeta2ndBiP.pdf}
	%	\caption{}\label{fig:tiger}
	%\end{subfigure}
	\caption{\footnotesize In above, the typical behavior of von Neumann entropy with respect to parameter $\beta$ is plotted for a subsystem of length $\ell$ starting from boundary of the chain. Size of the full system is $L=30$ and subsystem is $l=14$. Rest of the parameters are: $J=\gamma=1$, $h=\frac{1}{2}$, $b_{_{L,x,y}}=1$, $b_{_{1,x,y}}=1$ and $b_{{1,L,z}}=0$.}\label{fig:XYEntVsBeta}
\end{figure}

Regarding the entanglement, we have looked into the entanglement properties of parity broken state $\ket{\beta}$. For instance, the behavior of von Neumann entanglement entropy is plotted with respect to the parameter $\beta$ in the figure \ref{fig:XYEntVsBeta} for fixed and equal angles of boundary fields. Looking to the discussion of section \ref{subsec:EntBetaState}, the entanglement entropy obeys the relation
\begin{equation}
    S_{n}^{0}(A)=S_{n}^{\emptyset}(A)=S_{n}^{\beta}(A)+\frac{1}{1-n}\log[\frac{2^{1-n}}{(\tfrac{1}{2}+\mathrm{f}_\beta)^n+(\tfrac{1}{2}-\mathrm{f}_\beta)^n}], \qquad \mathrm{f}_\beta=\frac{\mathrm{Re}[\beta]}{1+|\beta|^2}.
\end{equation}
Since the relation above is general and does not depend on parameters of the model, it is expected to see the same behavior of entanglement entropy with the change in the $\beta$ as in figure \ref{fig:EntVSBeta}.

\begin{figure}[!htb]
	\centering
	\begin{subfigure}[]{0.49\linewidth}
	    \caption{\footnotesize $h=\frac{1}{2}$}\label{fig:cat}
		\includegraphics[width=7.4cm, keepaspectratio]{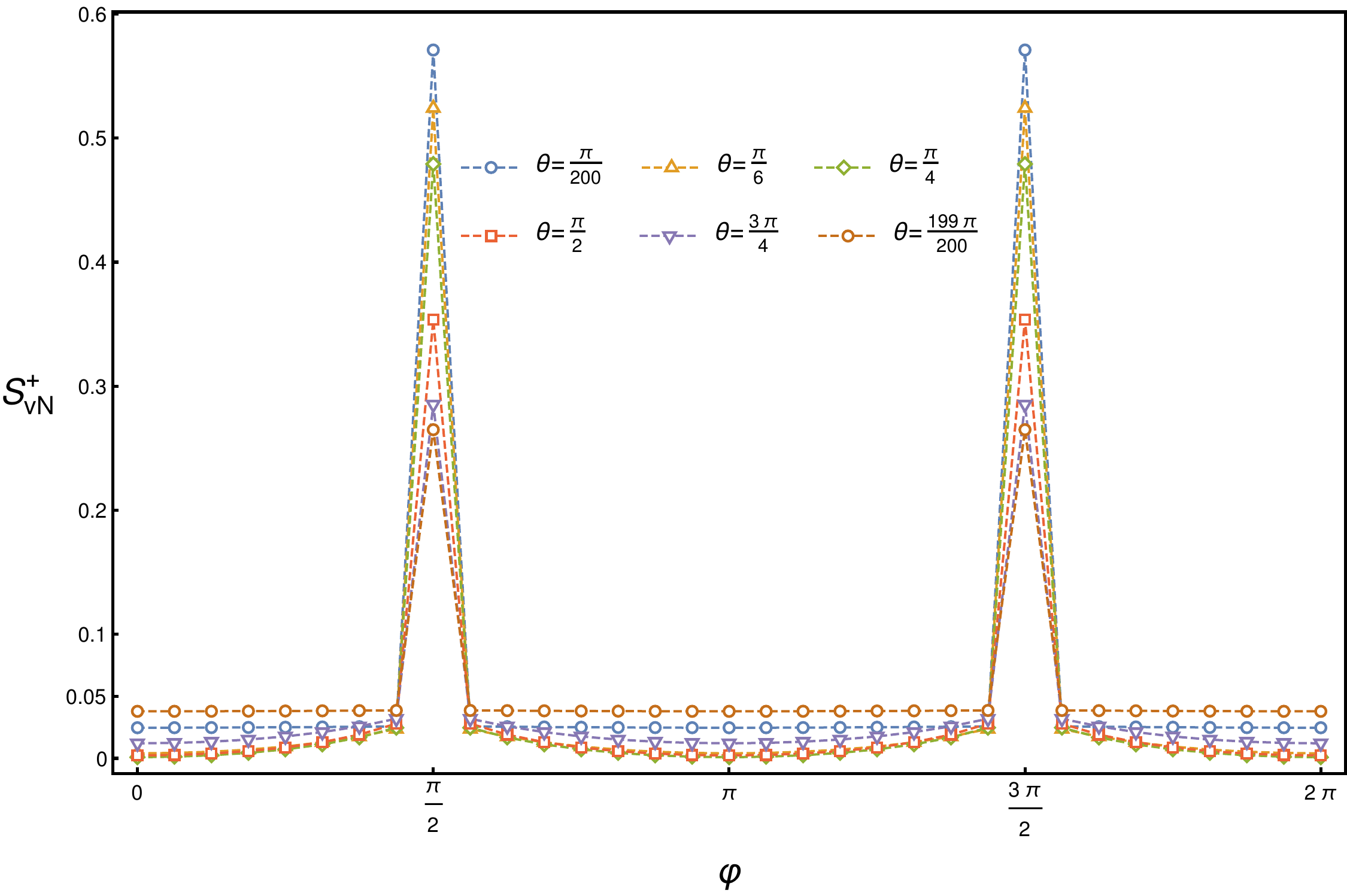}
	\end{subfigure}
	\hfill
	\begin{subfigure}[]{0.49\linewidth}
		\caption{\footnotesize $h=1$}\label{fig:dog}
		\includegraphics[width=7.4cm, keepaspectratio]{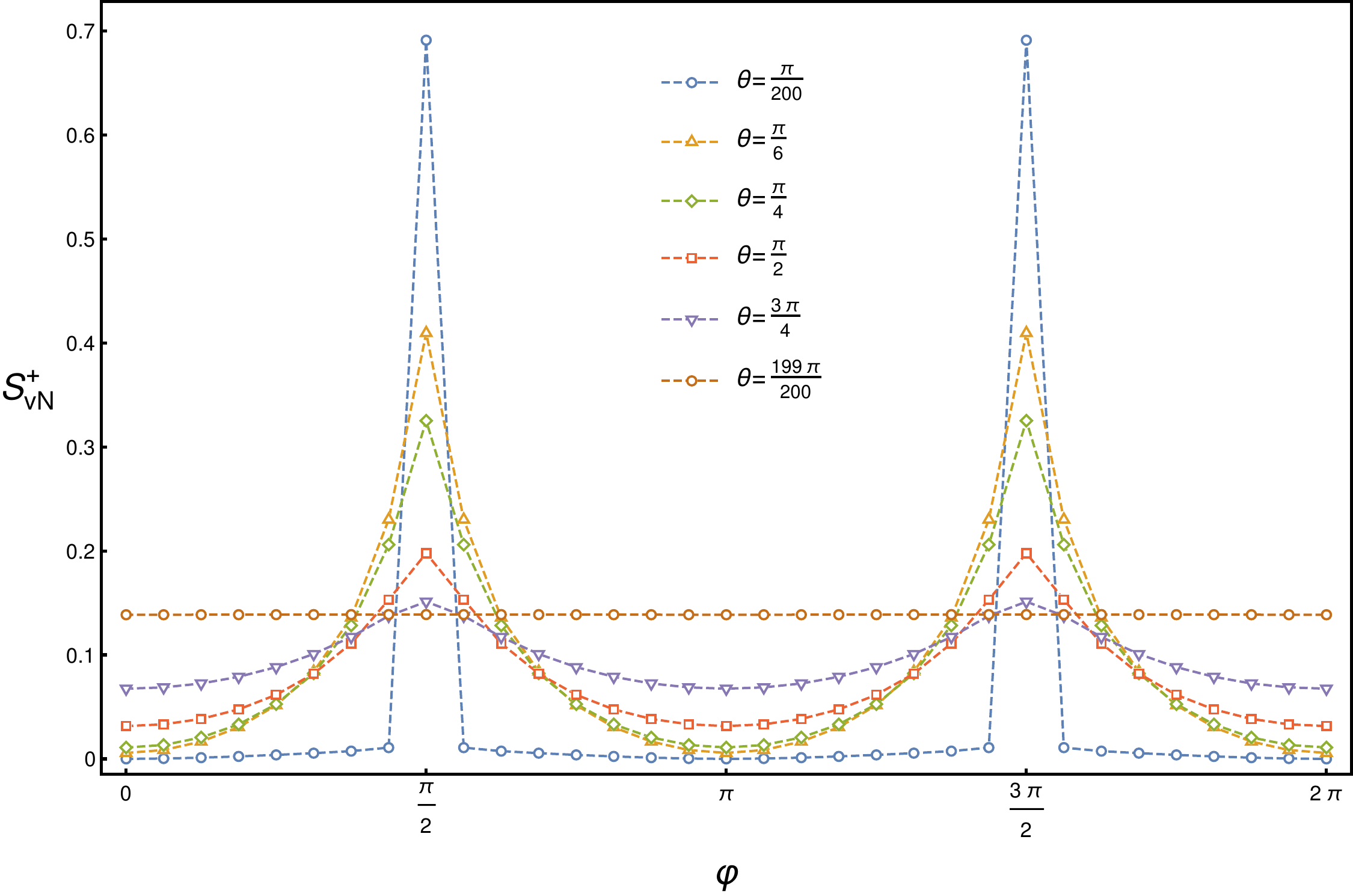}
	\end{subfigure}
	
	\begin{subfigure}[]{0.49\linewidth}
		\includegraphics[width=7.4cm, keepaspectratio]{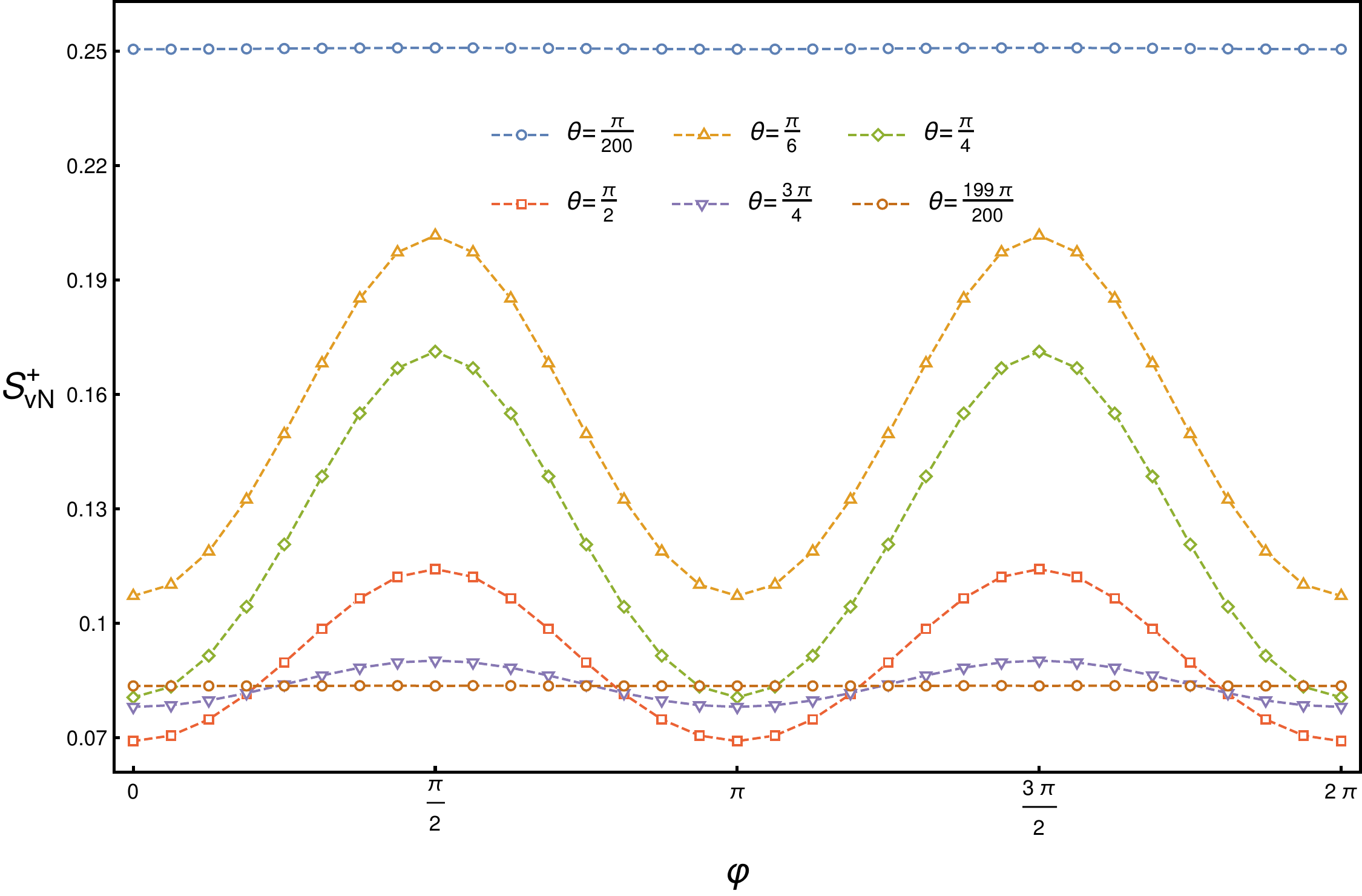}
		\caption{\footnotesize $h=\frac{3}{2}$}\label{fig:elephent}
	\end{subfigure}
	\caption{\footnotesize In this figure, the von Neumann entropy is plotted with respect to the angles of BMF, for three values of magnetic field, $h=\tfrac{1}{2},\:1,\:\tfrac{3}{2}$ respectively. We have put $b_{1,x}=b_{L,x}=\sin(\theta)\cos(\varphi)$, $b_{1,y}=b_{L,y}=\sin(\theta)\sin(\varphi)$ and $b_{1,z}=b_{L,z}=\cos(\theta)$. Rest of  parameters are: $L=30$, $l=2$, $\beta=+1$, $J=-1$ and $\gamma=1$. To better manifest the change in the entanglement entropy, size of the subsystem is selected to be close to the beginning of the chain. We observe that there is a change in the entanglement at $\varphi=\tfrac{\pi}{2},\tfrac{3\pi}{2}$. As mentioned before, there is (one) more zero mode in the eigenvalues of $\mathbf{M}$ matrix in \eqref{eq:M matrix} for $h>1$. However, this is not an exact zero mode and for size $L=30$, it is small but not zero. Therefore, observation above is intact.}\label{fig:XYEntVsAngl}
\end{figure}

For most parts of this article, we are interested in case where the boundary magnetic fields are the same in both edges, i. e., $\Vec{b}_1=\Vec{b}_L$. Another compelling observation would be the effect of the boundary field direction (same in both ends) on the entanglement of a particular state such as $\ket{G_+}$. Results have been demonstrated in figure \ref{fig:XYEntVsAngl}. For specific angles of BMF in the $xy$-plane, we observe a huge change in the value of entanglement entropy. %Notably, when magnetic fields at ends of system do not have any component in the $x$-direction ($b_x=0$), entanglement jumps our drops. 

For the small magnetic field ($h<J$ as in figure \ref{fig:cat}), there are two degenerate ground states for BMF Ising model: one can write $\frac{1}{2}(\ket{\rightarrow\rightarrow\cdots\rightarrow}+\ket{\leftarrow\leftarrow\cdots\leftarrow})$ as the ground state. When $\varphi\approx0$, the first spin prefers $\ket{\rightarrow}$ over the other possibility, $\ket{\leftarrow}$, which lowers the boundary entanglement
\footnote{Size of the subsystem has considered to be small so effect of boundary entanglement be more apparent}
. As $\varphi\rightarrow\frac{\pi}{2},\frac{3\pi}{2}$, both possible state of $x$-spin for the first spin would be equally probable, since the boundary magnetic field aligns the first spin in the (positive or negative) $y$ direction. This would result in increase in entanglement. As the angle $\theta$ increases, the intensity of BMF interaction $\Vec{b}_1.\Vec{S}_1$ opposes the uniform magnetic field $h$ in the system. So, we would expect smaller jumps in the entanglement for bigger $\theta$.

In the large magnetic fields ($h>J$ as in figure \ref{fig:elephent} or the paramagnetic state) the ground state is not degenerate, being all spins almost aligned in the direction of $h$. When BMF in the xy-plan is small ($\theta\approx0,\pi$), entanglement would not change much by change in $\varphi$. Although almost constant, entanglement is a  bit higher for small $\theta\sim0$ rather than $\theta\sim\pi$. %The reason is BMF forces the first spin to align anti-parallel to positive $z$-direction (opposing $h$), which allows the spin to be in a superposition of $\ket{\downarrow}$ and $\ket{\uparrow}$. 
As the intensity of BMF in xy-plan increases (for example $\theta=\frac{\pi}{4}$), first few spins at the beginning of the chain would get out of all parallel positioning and the entanglement changes with the angle of $\varphi$.

%\begin{figure}[!htb]
%	\centering		
%	\includegraphics[width=16cm, keepaspectratio]{XYFig2:EntVsAngs.pdf}
%	\caption{\footnotesize }\label{fig:XYEntVsAngl2}
%\end{figure}

\section{Conclusions}

In this paper we investigated a generic quantum spin chain Hamiltonian with arbitrary boundary magnetic fields. As far as the bulk Hamiltonian can be mapped to free fermions even though the boundary terms are not quadratic with respect to free fermions we were able to diagonalize the Hamiltonian exactly. This was done using two ancillary extra sites and later projection of the eigenstates. The extended Hamiltonian was studied in depth and many properties of the Hamiltonian was studied including eigenstates in configuration bases, the correlation function of eigenstates and the reduced density matrix. The extended Hamiltonian always has at least one zero mode which guaranties the presence of degeneracy. To get the eigenstates of the original spin chain one needs to go to the sector in which the parity number symmetry is broken. We studied comprehensively these eigenstates and found the correlation functions, reduced density matrix and the entanglement entropy. Interestingly the general features are independent of the parameters of the Hamiltonian and one can get universal results for the reduced density matrix and entanglement for generic eigenstates. The procedure used here can be extended for the other similar situations such as local breaking of the parity number symmetry in quantum spin chains due to local magnetic field impurity. It can be also useful to study local quantum quenches in quantum spin chains.

\section*{Acknowledgements}
MAR thanks CNPq and FAPERJ (grant number 210.354/2018) for partial support. MAR thanks ICTP for its hospitality during the visit which part of this work is done. The work done by AJ has been supported by CNPq.
\clearpage
\pagebreak

\begin{appendices}
\addtocontents{toc}{\protect\setcounter{tocdepth}{1}}
\renewcommand{\theequation}{\thesection.\arabic{equation}}

\section{Correlations for excited quasiparticle eigenstates}\label{sec:AppExcitedCorrelations}
\setcounter{equation}{0}

In section \ref{sec:Correlation}, we introduced the correlation functions and the method to produce these matrices for the ground state. In this part, we are going to present the method to calculate the correlations for excited quasiparticles \eqref{eq:ExcitedState1} using the unitary transformation $\mathbf{U}$ which diagonalize the Hamiltonian.   

The quasiparticle excited state such as $\ket{\psi}$ is defined by
\begin{equation}
\ket{\psi}=\ket{k_1,k_2,\cdots,k_N}=\prod_{k_j\in \mathbb{E}}\eta_{k_j}^\dagger\ket{0}_{_\eta}
\end{equation}   
where set $\mathbb{E}$ could be any subset of modes. For this state, we start by calculating the $\bra{\psi} c^\dagger_i c_j\ket{\psi}$ and $\bra{\psi} c^\dagger_i c_j^\dagger\ket{\psi}$ 
\begin{equation}
\langle c^\dagger_i c_j\rangle_{\scalebox{0.7}{$\psi$}}=\bra{0}\prod_{k_j\in \mathbb{E}}\eta_{k_j}\sum_{k,l}(h_{li}^*\eta_l+g_{ki}\eta_k^\dagger)\sum_{n,m}(g_{mj}^*\eta_m+h_{nj}\eta_n^\dagger)\prod_{k_j\in \mathbb{E}}\eta_{k_j}^\dagger\ket{0},
\end{equation}
\begin{equation}
\langle c^\dagger_i c_j^\dagger\rangle_{\scalebox{0.8}{$\psi$}}=\bra{0}\prod_{k_j\in \mathbb{E}}\eta_{k_j}\sum_{k,l}(h_{li}^*\eta_l+g_{ki}\eta_k^\dagger)\sum_{n,m}(h_{mj}^*\eta_m+g_{nj}\eta_n^\dagger)\prod_{k_j\in \mathbb{E}}\eta_{k_j}^\dagger\ket{0}.
\end{equation}
Using the Wick theorem we can simplify these expressions to get:
\begin{equation}
\begin{aligned}
\langle c^\dagger_i c_j\rangle_{\scalebox{0.7}{$\psi$}}&=(\mathbf{h}^\dagger.\mathbf{h})_{i,j}+\sum_{k_j\in\mathbb{E}}(g_{k_j,i}g_{k_j,j}^*-h_{k_j,i}^*h_{k_j,j}),\\
\langle c^\dagger_i c_j^\dagger\rangle_{\scalebox{0.7}{$\psi$}}&=(\mathbf{h}^\dagger.\mathbf{g})_{i,j}+\sum_{k_j\in\mathbb{E}}(g_{k_j,i}h_{k_j,j}^*-h_{k_j,i}^*g_{k_j,j}).
\end{aligned}
\end{equation}
Therefore, using the definitions \eqref{eq:KGmatrixCFmatrix}, we can write the $\mathbf{K}$, $\bar{\mathbf{K}}$ and $\mathbf{G}$ for the state $\ket{\psi}$ in a short version as
\begin{equation}\label{eq:Vac_KCorrl_Exc}
\mathbf{K}_{ij}^{\scalebox{0.7}{$\psi$}}=(\mathbf{h}^\dagger+\mathbf{g}^\dagger).(\mathbf{h}+\mathbf{g})_{ij}-2i\Im\Big[\sum_{k_j\in\mathbb{E}}(h_{k_j,i}^*+g_{k_j,i}^*)(g_{k_j,j}+h_{k_j,j})\Big],
\end{equation}
\begin{equation}\label{eq:Vac_KbCorrl_Exc}
\mathbf{\bar{K}}_{ij}^{\scalebox{0.7}{$\psi$}}=(\mathbf{h}^\dagger-\mathbf{g}^\dagger).(\mathbf{h}-\mathbf{g})_{ij}-2i\Im\Big[\sum_{k_j\in\mathbb{E}}(h_{k_j,i}^*-g_{k_j,i}^*)(h_{k_j,j}-g_{k_j,j})\Big],
\end{equation}
\begin{equation}\label{eq:Vac_GCorrl_Exc}
\begin{aligned}
\mathbf{G}_{ij}^{^\psi}=&(\mathbf{h}^\dagger-\mathbf{g}^\dagger).(\mathbf{h}+\mathbf{g})_{ij}+2\Re\Big[\sum_{k_j\in\mathbb{E}}(g_{k_j,i}^*-h_{k_j,i}^*)(h_{k_j,j}+g_{k_j,j})\Big].
\end{aligned}
\end{equation}

The prior expressions are useful for the study of entanglement in excited quasiparticle states. One can use the $\boldsymbol{\Gamma}^{^\psi}$ to calculate the RDM for a given subsystem $A$ using the relation
\begin{equation}%\label{eq:RDM_ZME_Majorana}
\rho_{_A}^{^{\psi}}(\boldsymbol{\gamma},\boldsymbol{\bar{\gamma}})=[\text{det}\frac{\mathbf{I}-\boldsymbol{\Gamma_{\!_A}^{^{\psi}}}}{2}]^{\frac{1}{2}}e^{\frac{1}{4}
\begin{pmatrix}\boldsymbol{\gamma}&\boldsymbol{\bar{\gamma}}\end{pmatrix}
\text{ln}\frac{\mathbf{I}+\boldsymbol{\Gamma_{_A}^{^\psi}}}{\mathbf{I}-\boldsymbol{\Gamma_{_A}^{^\psi}}}
\begin{pmatrix} \boldsymbol{\gamma}\\ \boldsymbol{\bar{\gamma}}\end{pmatrix}},
\end{equation}
where $\gamma$ and $\bar{\gamma}$ are the Majorana fermions defined in section \ref{sec:Correlation}. Equivalently, it is possible to use the preceding correlations to find the $\mathbf{R}^{^\psi}$ matrix (as in \eqref{eq:ExcStRMat}), $\mathbf{R}^{^\psi}={\mathbf{F}^{^\psi}}^*(\mathbf{I}-\mathbf{C}^{^\psi})^{-1}$, which could lead to finding the RDM.

For excited states created from ZME state, we can do the same calculations. These quasiparticle excited state are created as:
\begin{equation}
\ket{\phi_{_{\!\emptyset}}}=\ket{n_1,n_2,\cdots,n_N}=\prod_{n_j\in \mathbb{E}}\eta_{n_j}^\dagger\ket{\emptyset},
\end{equation}   
where $n_j\neq0$. Having the correlation matrices for the above states allows us to study the excited state entanglement for these states likewise. To write such matrices, we can use the result of section \ref{subsec:Correlation_ZME}. Correlations for the family of ZME excited states are demonstrated in \pageref{eq:ZME_Exc_Corr}. The matrices $\mathbf{K}^{\phi_{\emptyset}}$ and $\bar{\mathbf{K}}^{\phi_{\emptyset}}$ have the same form as \eqref{eq:Vac_KCorrl_Exc} and \eqref{eq:Vac_KbCorrl_Exc}, therefore, we have not included their forms.

As it was explained in subsection \ref{subsec:SelectionRule}, one could divide the Hilbert space of Hamiltonian \eqref{eq:FermionHamil} into four different towers which one of the these towers or sectors corresponds to the eigenstates of boundary magnetic field Hamiltonian \eqref{eq:Ham_intro}. We devote this part to find the correlation functions for states in a given tower, given the ground state for the tower (sector), can be found with $\ket{G_{\pm}}$. The base of calculation is similar to the above cases, therefore here we only hand out the final result. For start, an excited state in one of the sectors has the form
\begin{equation}
\ket{\chi_{_\pm}\!}=\prod_{m_j\in\mathbb{E}}\eta^\dagger_{m_j}\ket{G_{_\pm}\!}.
\end{equation}
In the preceding expression, set $\mathbb{E}$ contains excited modes acting on $\ket{G_{_\pm\!}}$ and $0\notin \mathbb{E}$. For a general case of $\mathbb{E}$ we can write the correlations as in page \pageref{eq:PM_Exc_Corrl}. Similar to the ZME and the $\ket{G_\pm}$ case, the matrices $\mathbf{K}^{\chi_{\pm}}$ and $\bar{\mathbf{K}}^{\chi_{\pm}}$ have the same form as $\mathbf{K}^{\psi}$ and $\bar{\mathbf{K}}^{\psi}$ in \eqref{eq:Vac_KCorrl_Exc} and \eqref{eq:Vac_KbCorrl_Exc}. 

%\newgeometry{lmargin=0.5in, tmargin=0.5in, bmargin=0.9in}%{top=100mm, bottom=0mm, left=11mm}  
\begin{landscape}
\vspace*{0.8cm}

{\arraycolsep=4.0pt\def\arraystretch{2.2}
\begin{eqnarray}\label{eq:ZME_Exc_Corr}
	\mathbf{C}^{^{\phi_{_{\emptyset}}}\!}&=&\left(\begin{array}{ccc}
	\frac{1}{2}%+\sum_{m\in\mathbb{E}}(g_{m,0}g_{m,0}^*-g_{m,0}^*g_{m,0})
	&C^{^\emptyset}_{\!\!_{0,j}}+\underset{n\in\mathbb{E}}{\sum}(g_{n,0}g_{n,j}^*+g_{n,0}^*h_{n,j})  &C^{^\emptyset}_{\!\!_{0,L+1}}+\underset{n\in\mathbb{E}}{\sum}(g_{n,0}g_{n,L+1}^*+g_{n,0}^*g_{n,L+1})\\
	C^{\emptyset}_{\!\!_{i,0}}+\underset{n\in\mathbb{E}}{\sum}(g_{n,i}g_{n,0}^*+h_{n,i}^*g_{n,0})&C^{\emptyset}_{i,j}+\underset{n\in\mathbb{E}}{\sum}(g_{n,i}g_{n,j}^*-h_{n,i}^*h_{n,j}) &C^{^\emptyset}_{\!\!_{i,L+1}}+\underset{n\in\mathbb{E}}{\sum}(g_{n,i}g_{n,L+1}^*-h_{n,i}^*g_{n,L+1}) \\
	C^{^\emptyset}_{\!\!_{L+1,0}}+\underset{n\in\mathbb{E}}{\sum}(g_{_{n,L+1}}g_{_{n,0}}^*+g_{_{n,L+1}}^*g_{_{n,0}}) &C^{^\emptyset}_{\!\!_{L+1,j}}+\underset{n\in\mathbb{E}}{\sum}(g_{n,L+1}g_{n,j}^*-g_{n,L+1}^*h_{n,j})  &\frac{1}{2}\\
    \end{array}\right),\\
    {}\nonumber\\
	\mathbf{F}^{\phi_{_{\!\emptyset}}}&=&\left(\begin{array}{ccc}
	0 &F^{^{\emptyset}}_{\!\!_{0,j}}+\underset{n\in\mathbb{E}}{\sum}(g_{_{n,0}\!}h_{_{n,j}\!}^*+g_{_{n,0}\!}^*g_{_{n,j}\!})  &F^{^{\emptyset}}_{\!\!_{0,L+1}}+\underset{n\in\mathbb{E}}{\sum}(g_{_{n,0}\!}g_{_{n,L+1}\!}^*+g_{_{n,0}\!}^*g_{_{n,L+1}\!})\\
	F^{^{\emptyset}}_{\!\!_{i,0}}-\underset{n\in\mathbb{E}}{\sum}(g_{_{n,i}\!}g_{_{n,0}\!}^*+h_{_{n,i}\!}^*g_{_{n,0}\!}) &F^{^{\pm}}_{\!\!_{i,j}}+\underset{n\in\mathbb{E}}{\sum}(g_{_{n,i}\!}h_{_{n,j}\!}^*-h_{_{n,i}\!}^*g_{_{n,j}\!}) &F^{^{\emptyset}}_{\!\!_{i,L+1}} +\underset{n\in\mathbb{E}}{\sum}(g_{_{n,i}\!}g_{_{n,L+1}\!}^*-g_{_{n,i}\!}^*g_{_{n,L+1}\!})\\
	F^{^{\emptyset}}_{\!\!_{L+1,0}}-\underset{n\in\mathbb{E}}{\sum}(g_{_{n,L+1}\!}g_{_{n,0}\!}^*+g_{_{n,L+1}\!}^*g_{_{n,0}\!})&\quad F^{^{\emptyset}}_{\!\!_{L+1,j}} +\underset{n\in\mathbb{E}}{\sum}(g_{_{n,L+1}\!}h_{_{n,j}\!}^*-g_{_{n,L+1}\!}^*g_{_{n,j}\!})&0 \\
	\end{array}\right)\\
	{}\nonumber\\
	\mathbf{G}^{\phi_{_{\!\emptyset}}}&=&\left(\begin{array}{ccc}
    0\qquad&G^{^{\emptyset}}_{0,j}+4\Re\big[\underset{n\in\mathbb{E}}{\sum}g_{n,0}^*(h_{n,j}+g_{n,j})\big]&G^{^{\emptyset}}_{_{\!0,L+1}}+8\Re\big[\underset{n\in\mathbb{E}}{\sum}g_{n,0}^*g_{n,L+1}\big]\\
	0\qquad&G^{^{\emptyset}}_{i,j}+2\Re\big[\underset{n\in\mathbb{E}}{\sum}(g_{n,0}^*-h^*_{n,0})(h_{n,j}+g_{n,j})\big]&\quad G^{^{\emptyset}}_{i,L+1}+4\Re[\underset{n\in\mathbb{E}}{\sum}(g_{n,i}^*-h^*_{n,i})g_{n,L+1}] \\
	\mathbf{1}\qquad&0&0 \\
	\end{array}\right).
\end{eqnarray}}

\end{landscape}
%\restoregeometry    

%\section{Correlations for parity broken sectors}\label{sec:AppSectorsCorrelations}

%\newgeometry{lmargin=0.5in, tmargin=0.5in, bmargin=0.9in}%{top=100mm, bottom=0mm, left=11mm}  
\begin{landscape}
\vspace*{0.8cm}

{\arraycolsep=1.0pt\def\arraystretch{2.0}
\begin{eqnarray}\label{eq:PM_Exc_Corrl}
	\mathbf{C}^{^{\chi_{_\pm\!}\!}\!}&=&\left(\begin{array}{ccc}
	\frac{1}{2}%+\sum_{m\in\mathbb{E}}(g_{m,0}g_{m,0}^*-g_{m,0}^*g_{m,0})
	&C^{^\pm}_{\!\!_{0,j}}+\underset{m\in\mathbb{E}}{\sum}(g_{m,0}g_{m,j}^*+g_{m,0}^*h_{m,j})  &C^{^\pm}_{\!\!_{0,L+1}}+\underset{m\in\mathbb{E}}{\sum}(g_{m,0}g_{m,L+1}^*+g_{m,0}^*g_{m,L+1})\\
	C^{\pm}_{\!\!_{i,0}}+\underset{m\in\mathbb{E}}{\sum}(g_{m,i}g_{m,0}^*+h_{m,i}^*g_{m,0})&C^{\pm}_{i,j}+\underset{m\in\mathbb{E}}{\sum}(g_{m,i}g_{m,j}^*-h_{m,i}^*h_{m,j}) &C^{^\pm}_{\!\!_{i,L+1}}+\underset{m\in\mathbb{E}}{\sum}(g_{m,i}g_{m,L+1}^*-h_{m,i}^*g_{m,L+1}) \\
	C^{^\pm}_{\!\!_{L+1,0}}+\underset{m\in\mathbb{E}}{\sum}(g_{_{m,L+1}}g_{_{m,0}}^*+g_{_{m,L+1}}^*g_{_{m,0}}) &C^{^\pm}_{\!\!_{L+1,j}}+\underset{m\in\mathbb{E}}{\sum}(g_{m,L+1}g_{m,j}^*-g_{m,L+1}^*h_{m,j})  &\frac{1}{2}%+\sum_{m\in\mathbb{E}}(g_{m,L+1}g_{m,L+1}^*-g_{m,L+1}^*g_{m,L+1}) 
	\\
    \end{array}\right),\\
    {}\nonumber\\
	\mathbf{F}^{^{\chi_{_\pm\!}\!}\!}&=&\left(\begin{array}{ccc}
	0 &F^{^{\pm}}_{\!\!_{0,j}}+\underset{m\in\mathbb{E}}{\sum}(g_{_{m,0}\!}h_{_{m,j}\!}^*+g_{_{m,0}\!}^*g_{_{m,j}\!})  &F^{^{\pm}}_{\!\!_{0,L+1}}+\underset{m\in\mathbb{E}}{\sum}(g_{_{m,0}\!}g_{_{m,L+1}\!}^*+g_{_{m,0}\!}^*g_{_{m,L+1}\!})\\
	F^{^{\pm}}_{\!\!_{i,0}}-\underset{m\in\mathbb{E}}{\sum}(g_{_{m,i}\!}g_{_{m,0}\!}^*+h_{_{m,i}\!}^*g_{_{m,0}\!})&F^{^{\pm}}_{\!\!_{i,j}}+\underset{m\in\mathbb{E}}{\sum}(g_{_{m,i}\!}h_{_{m,j}\!}^*-h_{_{m,i}\!}^*g_{_{m,j}\!}) &F^{^{\pm}}_{\!\!_{i,L+1}} +\underset{m\in\mathbb{E}}{\sum}(g_{_{m,i}\!}g_{_{m,L+1}\!}^*-g_{_{m,i}\!}^*g_{_{m,L+1}\!})\\
	F^{^{\pm}}_{\!\!_{L+1,0}}-\underset{m\in\mathbb{E}}{\sum}(g_{_{m,L+1}\!}g_{_{m,0}\!}^*+g_{_{m,L+1}\!}^*g_{_{m,0}\!})&F^{^{\pm}}_{\!\!_{L+1,j}} +\underset{m\in\mathbb{E}}{\sum}(g_{_{m,L+1}\!}h_{_{m,j}\!}^*-g_{_{m,L+1}\!}^*g_{_{m,j}\!})&0 \\
	\end{array}\right),\\
	{}\nonumber\\
	\mathbf{G}^{^{\chi_{_\pm\!}\!}\!}&=&\left(\begin{array}{ccc}
	0\qquad&G^{^{\pm}}_{0,j}+4\Re\big[\underset{m\in\mathbb{E}}{\sum}g_{m,0}^*(h_{m,j}+g_{m,j})\big]&G^{^{\pm}}_{_{\!0,L+1}}+8\Re\big[\underset{m\in\mathbb{E}}{\sum}g_{m,0}^*g_{m,L+1}\big]\\
	0\qquad&G^{^{\pm}}_{i,j}+2\Re\big[\underset{m\in\mathbb{E}}{\sum}(g_{m,0}^*-h^*_{m,0})(h_{m,j}+g_{m,j})\big]&\quad G^{^{\pm}}_{i,L+1}+4\Re[\underset{m\in\mathbb{E}}{\sum}(g_{m,i}^*-h^*_{m,i})g_{m,L+1}] \\
	0\qquad&0&0
	\end{array}\right).
\end{eqnarray}}

\end{landscape}%\restoregeometry    

\section{Reduced density matrix calculations}\label{sec:AppRDMcalculations}
\setcounter{equation}{0}
In this appendix, we are presenting the calculations of \eqref{eq:BetaRDMFinalVersion2} in details. Starting from \eqref{eq:Partial_Trace_Def}, we can write
\begin{equation}\label{eq:App_Trace_1}
\begin{aligned}
\rho_{\mathbf{1}}^\beta\!(\boldsymbol{\xi},\boldsymbol{\xi}^\prime)=&\int\prod_{l\in\mathbf{2}}\!\text{d}\bar{\xi_l}\text{d}\xi_l\;e^{-\underset{n\in\mathbf{2}}{\sum}\bar{\xi}_n\xi_n}\bra{\xi_1,\cdots,\xi_k,-\xi_{k+1},\cdots,-\xi_{L}}\rho^\beta\ket{\xi_1^\prime,\cdots,\xi_k^\prime,\xi_{k+1},\cdots,\xi_{L}}\\
=&|C^{\beta}|^2e^{\tfrac{1}{2}(\mathbf{R}_{11})_{ij}\bar{\xi_i}\bar{\xi}_j-\tfrac{1}{2}(\mathbf{R}_{11}^*)_{ji}\xi_j^\prime\xi_i^\prime}\int\prod_{l\in\mathbf{2}}\!\text{d}\bar{\xi_l}\text{d}\xi_l\;e^{-\bar{\xi}_n\xi_n}\mathcal{F}(\{\xi_n\},\{\xi_i^\prime\},\{\bar{\xi}_j\},\{\bar{\xi}_m\})\\
&\times\big[e^{-\tfrac{1}{2}(\mathbf{R}_{12})_{in}\bar{\xi}_i\bar{\xi}_n-\tfrac{1}{2}(\mathbf{R}_{21})_{ni}\bar{\xi}_n\bar{\xi}_i} e^{\tfrac{1}{2}(\mathbf{R}_{22})_{mn}\bar{\xi}_m\bar{\xi}_n}e^{-\tfrac{1}{2}(\mathbf{R}_{22}^*)_{nm}\xi_n\xi_m+\tfrac{1}{2}(\mathbf{R}_{12}^*)_{jm}\xi^\prime_j\xi_m+\tfrac{1}{2}(\mathbf{R}_{21}^*)_{mj}\xi_m\xi_j^\prime}\big],
\end{aligned}
\end{equation}
where in the above Einstein summation convention is used. To clarify the notation above, the indices $i,j$ belong to subsystem $\mathbf{1}$ and indices $n,m$ to subsystem $\mathbf{2}$. Eventually, the function $\mathcal{F}$ is given by
\begin{eqnarray}
\mathcal{F}(\{\xi_n\},\{\xi_i^\prime\}\!\!\!\!\!\!&&\!\!\!\!,\{\bar{\xi}_j\},\{\bar{\xi}_m\})=1+\beta(\boldsymbol{\mathfrak{M}}_1)_{0j}\bar{\xi}_j-\beta(\boldsymbol{\mathfrak{M}}_2)_{0m}\bar{\xi}_m+\beta^*(\boldsymbol{\mathfrak{M}}_1^*)_{0i}\xi_i^\prime\nonumber\\
&&+\beta^*(\boldsymbol{\mathfrak{M}}_2^*)_{0n}\xi_n+|\beta|^2(\boldsymbol{\mathfrak{M}}_1)_{0j}(\boldsymbol{\mathfrak{M}}_1^*)_{0i}\bar{\xi}_j\xi_i^\prime+|\beta|^2(\boldsymbol{\mathfrak{M}}_1)_{0j}(\boldsymbol{\mathfrak{M}}_2^*)_{0n}\bar{\xi}_j\xi_{n}\nonumber\\
&&-|\beta|^2(\boldsymbol{\mathfrak{M}}_2)_{0m}(\boldsymbol{\mathfrak{M}}_1^*)_{0i}\bar{\xi}_{m}\xi_i^\prime-|\beta|^2(\boldsymbol{\mathfrak{M}}_2)_{0m}(\boldsymbol{\mathfrak{M}}_2^*)_{0n}\bar{\xi}_{m}\xi_{n}.
\end{eqnarray}
In \eqref{eq:App_Trace_1} we have divided $\mathbf{R}$ into four submatrices $\mathbf{R_{11}}$, $\mathbf{R_{12}}$, $\mathbf{R_{21}}=-\mathbf{R_{12}}^T$ and $\mathbf{R_{22}}$, according to the part we are tracing out (or not). Although these submatrices do not need to have same size, $\mathbf{R_{11}}$ and $\mathbf{R_{22}}$ should be square matrices. We write the \eqref{eq:App_Trace_1} in the compact form:
\begin{equation}\label{eq:App_Trace_2}
\rho_{\mathbf{1}}(\boldsymbol{\xi},\boldsymbol{\xi}^\prime)=|C^{\beta}|^2e^{\tfrac{1}{2}
\left(\begin{matrix}\boldsymbol{\bar{\xi}}&\!\boldsymbol{\xi}^\prime\\\end{matrix}\right)
\left(\begin{matrix}\mathbf{R}_{11}&0\\0&-\mathbf{R}_{11}^*\end{matrix}\right)
\left(\begin{matrix}\boldsymbol{\bar{\xi}}\\\boldsymbol{\xi}^\prime\end{matrix}\right)}
\int\text{\textbf{D}}\boldsymbol{\eta}\mathcal{F}\!(\boldsymbol{\xi_i^\prime},\boldsymbol{\bar{\xi}_j},\boldsymbol{\eta})e^{\tfrac{1}{2}\boldsymbol{\eta}^T\boldsymbol{\mathcal{A}}\boldsymbol{\eta}+\boldsymbol{\lambda}^T\!\boldsymbol{\eta} }.
\end{equation}
where $\textbf{D}\boldsymbol{\eta}=\text{d}\bar{\xi}\text{d}\xi$ and
\begin{eqnarray}
\boldsymbol{\eta}^T&=&\left(\begin{matrix}\boldsymbol{\bar{\xi}}&\boldsymbol{\xi}\\\end{matrix}\right),\qquad\boldsymbol{\lambda}^T=\left(\begin{matrix}
-\boldsymbol{\bar{\xi}}\mathbf{R}_{12}&-\boldsymbol{\xi}^\prime\mathbf{R}_{12}^*\\\end{matrix}\right),\\
\boldsymbol{\mathcal{A}}&=&\left(\begin{matrix}
\mathbf{R}_{22}&-\mathbf{I}\\\mathbf{I}&-\mathbf{R}_{22}^*\\
\end{matrix}\right).
\end{eqnarray}
It is much easier to solve the integration in \eqref{eq:App_Trace_2} with this new variables. In the expression of $\mathcal{F}$, the first, second, forth and sixth terms do not depend on the variables of integration. Therefore, for those terms we can write: 
\begin{equation}\label{eq:App_Integ_1}
\rho_{\mathbf{1}}^\beta(1^{\text{st}},2^{\text{nd}},4^{\text{th}},6^{\text{th}},\boldsymbol{\xi},\boldsymbol{\xi}^\prime)=|C^\beta|(\text{det}\big[\boldsymbol{\mathcal{A}}\big])^{\tfrac{1}{2}}\big(1+\beta\boldsymbol{\mathfrak{M}}_1\boldsymbol{\bar{\xi}}\big)\big(1+\beta^*\boldsymbol{\mathfrak{M}}_1^*\boldsymbol{\xi^\prime}\big) e^{\tfrac{1}{2}
	\left(\begin{matrix}
	\bar{\xi}&\!\xi^\prime\\
	\end{matrix}\right)\begin{matrix}
	\boldsymbol{\Omega}
	\end{matrix}\left(\begin{matrix}
	\bar{\xi}\\\xi^\prime
	\end{matrix}\right)},
\end{equation}
where
\begin{equation}
\boldsymbol{\Omega}=\left(\begin{matrix}\mathbf{R}_{11}&0\\0&-{\mathbf{R}_{11}}^*\\\end{matrix}\right)+\left(\begin{matrix}\mathbf{R}_{12}&0\\0&{\mathbf{R}_{12}}^*\\\end{matrix}\right)\boldsymbol{\mathcal{A}}^{-1}\left(\begin{matrix}{\mathbf{R}_{12}}^T&0\\0&{\mathbf{R}_{12}}^\dagger\\\end{matrix}\right).
\end{equation}
For the terms with linear Grassmann variables in the integration we first substitute $\boldsymbol{\eta}\rightarrow\boldsymbol{\eta}+\boldsymbol{\mathcal{A}}^{-1}\boldsymbol{\lambda}$ in \eqref{eq:App_Trace_2}. Using the Berezin integration techniques in presented in \cite{CARACCIOLO2013474}, we are left with
\begin{equation}\label{eq:App_Integ_2}
\rho_{\mathbf{1}}^\beta(3^\text{rd},5^\text{th},7^\text{th},8^\text{th},\boldsymbol{\xi},\boldsymbol{\xi}^\prime)=|C^\beta|(\text{det}\big[\boldsymbol{\mathcal{A}}\big])^{\tfrac{1}{2}}(\text{det}\big[\boldsymbol{\mathcal{W}}\big])^{\tfrac{1}{2}} F(\boldsymbol{\bar{\xi}}_{\mathbf{1}},\boldsymbol{\xi^\prime}_{\mathbf{1}}) e^{\tfrac{1}{2}
	\left(\begin{matrix}
	\bar{\xi}&\!\xi^\prime\\
	\end{matrix}\right)\begin{matrix}
	\boldsymbol{\Omega}
	\end{matrix}\left(\begin{matrix}
	\bar{\xi}\\\xi^\prime
	\end{matrix}\right)},
\end{equation}
with 
\begin{eqnarray}
F(\boldsymbol{\bar{\xi}}_{\mathbf{1}},\boldsymbol{\xi^\prime}_{\mathbf{1}})&=&|\beta|^2\text{Pf}\big[\boldsymbol{\mathcal{W}}\big]+\left(\begin{matrix}-\beta\boldsymbol{\mathfrak{M}}_2&0\\0&\beta^*\boldsymbol{\mathfrak{M}}_2^*\end{matrix}\right)\boldsymbol{\mathcal{A}}^{-1}\boldsymbol{\lambda}+|\beta|^2\boldsymbol{\mathfrak{M}}_1\boldsymbol{\bar{\xi}}\times\left(\begin{matrix}0&0\\0&\boldsymbol{\mathfrak{M}}_2^*\end{matrix}\right)\boldsymbol{\mathcal{A}}^{-1}\boldsymbol{\lambda}\\
&&+|\beta|^2\left(\begin{matrix}-\boldsymbol{\mathfrak{M}}_2&0\\0&0\end{matrix}\right)\boldsymbol{\mathcal{A}}^{-1}\boldsymbol{\lambda}\times\boldsymbol{\mathfrak{M}}_1^*\boldsymbol{\xi^\prime}\nonumber+|\beta|^2\left(\begin{matrix}-\boldsymbol{\mathfrak{M}_2}&0\end{matrix}\right)\boldsymbol{\mathcal{A}}^{-1}\boldsymbol{\lambda}\times\left(\begin{matrix}0&\boldsymbol{\mathfrak{M}_2}^*\end{matrix}\right)\boldsymbol{\mathcal{A}}^{-1}\boldsymbol{\lambda},\\
\boldsymbol{\mathcal{W}}&=&\left(\begin{matrix}
    \boldsymbol{\mathfrak{M}}_2&\mathbf{0}\\ \mathbf{0}&-\boldsymbol{\mathfrak{M}}_2^*\\
    \end{matrix}\right)\boldsymbol{\mathcal{A}}^{-T}\left(\begin{matrix}
    \boldsymbol{\mathfrak{M}}_2^T&\mathbf{0}\\ \mathbf{0}&-\boldsymbol{\mathfrak{M}}_2^\dagger\\
    \end{matrix}\right).
\end{eqnarray}
Putting \eqref{eq:App_Integ_1} and \eqref{eq:App_Integ_2} together, after some algebraic manipulations, we get to \eqref{eq:BetaRDMGrassFin}.

We could have used another trick to solve the Berezin integrations to get RDM. This method ends up in having two exponentials in the final result. The trick is to write the linear Grassmann variable in the integration as exponentials, like:
\begin{equation}\label{eq:GrassProdToExp}
(\boldsymbol{\mathfrak{M}}\bar{\boldsymbol{\xi}})_{k_1}\cdots(\boldsymbol{\mathfrak{M}}\bar{\boldsymbol{\xi}}\big)_{k_N}\!\big(\boldsymbol{\mathfrak{M}}^*\boldsymbol{\xi}^\prime\big)_{k_1}\cdots(\boldsymbol{\mathfrak{M}}^*\boldsymbol{\xi}^\prime\big)_{k_N}=\int\prod_{i}\text{d}\bar{\theta}_i\text{d}\theta_ie^{\sum_{k_j}\theta_{k_j}(\boldsymbol{\mathfrak{M}}\bar{\boldsymbol{\xi}})_{k_j}+\bar{\theta}_{k_j}(\boldsymbol{\mathfrak{M}}^*\boldsymbol{\xi})_{k_j}}.
\end{equation}
In the above relation, $\theta$ and $\bar{\theta}$ are Grassmann variables too.

If we start again from density matrix expression in coherent basis \eqref{eq:DensityMatGrass}, and rewrite it as:
\begin{equation}\label{eq:BetaDensityMat4terms}
\begin{aligned}
    \rho^\beta\!(\boldsymbol{\xi},\boldsymbol{\xi}^\prime)=&|C^\beta|^2\scalebox{1.3}{$e$}^{{}^{\tfrac{1}{2}R_{ij}\bar{\xi}_i\bar{\xi}_j}}(1+\beta\mathfrak{M}_{0k}\bar{\xi}_k+\beta^*\mathfrak{M}_{0l}^*\xi_l^\prime+|\beta|^2\mathfrak{M}_{0k}\mathfrak{M}_{0l}^*\bar{\xi}_k\xi_l^\prime) \scalebox{1.3}{$e$}^{{}^{-\tfrac{1}{2}R^*_{nm}\xi_n^\prime \xi_m^\prime}}\\
    =&\rho^\beta\!(1,\boldsymbol{\xi},\boldsymbol{\xi}^\prime)+\varrho^\beta\!(2,\boldsymbol{\xi},\boldsymbol{\xi}^\prime)+\varrho^\beta\!(3,\boldsymbol{\xi},\boldsymbol{\xi}^\prime)+\rho^\beta\!(4,\boldsymbol{\xi},\boldsymbol{\xi}^\prime),
\end{aligned}
\end{equation}
where the $\rho^\beta\!(1,\boldsymbol{\xi},\boldsymbol{\xi}^\prime)$ and $\rho^\beta\!(4,\boldsymbol{\xi},\boldsymbol{\xi}^\prime)$ are the density matrices corresponding to the vacuum-vacuum state and excited-excited state, and the $\varrho^\beta\!(2,\boldsymbol{\xi},\boldsymbol{\xi}^\prime)$ and $\varrho^\beta\!(3,\boldsymbol{\xi},\boldsymbol{\xi}^\prime)$ are density cross terms corresponding to the vacuum-excited (and exited-vacuum) state terms. We are going to partial trace each term separately and then put the results together afterwards.
For the first term in \eqref{eq:BetaDensityMat4terms}, we have:
\begin{equation}\label{eq:BetaDnstyMatGrass1}
\rho_{\mathbf{1}}^\beta(1,\boldsymbol{\xi},\boldsymbol{\xi}^\prime)=\frac{\mathcal{C}^\beta}{|\beta|^2}e^{\tfrac{1}{2}
	\left(\begin{matrix}
	\bar{\xi}&\!\xi^\prime\\
	\end{matrix}\right)\begin{matrix}
	\boldsymbol{\Omega}^\beta
	\end{matrix}\left(\begin{matrix}
	\bar{\xi}\\\xi^\prime
	\end{matrix}\right)},
\end{equation}
where again
\begin{subequations}%\label{eq:BetaOmega-Ccal}
\begin{equation}%\label{eq:BetaCcalConst}
    \mathcal{C}^\beta=\frac{|\beta|^2\sqrt{\det\big[\mathbf{I}+{\mathbf{R}_{22}}^\dagger\mathbf{R}_{22}\big]}}{(1+|\beta|^2)\sqrt{\det\big[\mathbf{I}+{\mathbf{R}}^\dagger\mathbf{R}\big]}}, \qquad\boldsymbol{\mathcal{A}}=\left(\begin{matrix}
\mathbf{R}_{22}&-\mathbf{I}\\\mathbf{I}&-\mathbf{R}_{22}^*\\
\end{matrix}\right),
\end{equation}
\begin{equation}%\label{eq:BetaOmegaMatSimple}
\boldsymbol{\Omega}=\left(\begin{matrix}\mathbf{R}_{11}&0\\0&-{\mathbf{R}_{11}}^*\\\end{matrix}\right)+\left(\begin{matrix}\mathbf{R}_{12}&0\\0&{\mathbf{R}_{12}}^*\\\end{matrix}\right)\boldsymbol{\mathcal{A}}^{-1}\left(\begin{matrix}{\mathbf{R}_{12}}^T&0\\0&{\mathbf{R}_{12}}^\dagger\\\end{matrix}\right).
\end{equation}
\end{subequations}
For the last term in \eqref{eq:BetaDensityMat4terms}, we can write:
\begin{equation}
\rho^\beta(4,\boldsymbol{\xi},\boldsymbol{\xi}^\prime)=|C^\beta\beta|^2\int\text{d}\bar{\theta}\text{d}\theta e^{\theta(\boldsymbol{\mathfrak{M}}\bar{\boldsymbol{\xi}})+\bar{\theta}(\boldsymbol{\mathfrak{M}}^*\boldsymbol{\xi}^\prime)}e^{\tfrac{1}{2}R_{ij}\bar{\xi_i}\bar{\xi_j}}e^{-\tfrac{1}{2}R^*_{ij}\xi^\prime_i\xi^\prime_j}.
\end{equation}
For the reduced density matrix we get
\begin{equation}
\begin{aligned}
\rho_{\mathbf{1}}^\beta\!(4,\boldsymbol{\xi},\boldsymbol{\xi}^\prime)=&|C^\beta\beta|^2\!\int\text{d}\bar{\theta}\text{d}\theta e^{\tfrac{1}{2}(\mathbf{R}_{11})_{ij}\bar{\xi_i}\bar{\xi}_j-\tfrac{1}{2}(\mathbf{R}_{11}^*)_{ji}\xi_j^\prime\xi_i^\prime}e^{\theta(\boldsymbol{\mathfrak{M}_1}.\bar{\boldsymbol{\xi}})+\bar{\theta}(\boldsymbol{\mathfrak{M}_1^*}.\boldsymbol{\xi}^\prime)}\!\int\prod_{l=k+1}^{L+1}\!\text{d}\bar{\xi_l}\text{d}\xi_l\:e^{-\bar{\xi}_l\xi_l}\\
&\times e^{-\theta(\boldsymbol{\mathfrak{M}_2}\bar{\boldsymbol{\xi}})+\bar{\theta}(\boldsymbol{\mathfrak{M}_2^*}\boldsymbol{\xi})} \big[e^{-(\mathbf{R}_{12})_{in}\bar{\xi}_i\bar{\xi}_n+\tfrac{1}{2}(\mathbf{R}_{22})_{mn}\bar{\xi}_m\bar{\xi}_n}e^{-\tfrac{1}{2}(\mathbf{R}_{22}^*)_{nm}\xi_n\xi_m+(\mathbf{R}_{12}^*)_{jm}\xi^\prime_j\xi_m}\big]
\end{aligned}
\end{equation}
We denote the terms on the left of second integral as $\mathcal{F}(\bar{\boldsymbol{\xi}},\boldsymbol{\xi},\theta,\bar{\theta})$. If we introduce new Grassmann variables  $\boldsymbol{\eta}^T=\left(\begin{matrix}\boldsymbol{\bar{\xi}}&\boldsymbol{\xi}\\\end{matrix}\right)$ and $\boldsymbol{\lambda}^T=\left(\begin{matrix}
-\boldsymbol{\bar{\xi}}\mathbf{R}_{12}-\theta(\boldsymbol{\mathfrak{M}})_0&-\boldsymbol{\xi}^\prime\mathbf{R}_{12}^*+\bar{\theta}(\boldsymbol{\mathfrak{M}^*})_0\\\end{matrix}\right)$, then we can write the integral as
\begin{equation}
\rho_{_\mathbf{1}}^\beta\!(4,\boldsymbol{\xi},\boldsymbol{\xi}^\prime)=\int\text{d}\bar{\theta}\text{d}\theta\mathcal{F}(\bar{\boldsymbol{\xi}},\boldsymbol{\xi}^\prime,\theta,\bar{\theta})\int\text{\textbf{D}}\boldsymbol{\eta}\:e^{\tfrac{1}{2}\boldsymbol{\eta}^T\!\boldsymbol{\mathcal{A}}\boldsymbol{\eta}+\boldsymbol{\lambda}^T\!\boldsymbol{\eta}}
\end{equation}
where $\boldsymbol{\mathcal{A}}$ has been defined previously. Solving the first integral we get:
\begin{equation}\label{eq:BetaRDM4-1}
\rho_{_\mathbf{1}}^\beta\!(4,\boldsymbol{\xi},\boldsymbol{\xi}^\prime)=\mathcal{C}^\beta e^{\tfrac{1}{2}
	\left(\begin{matrix}
	\bar{\xi}&\!\xi^\prime\\
	\end{matrix}\right)\begin{matrix}
	\boldsymbol{\Omega}
	\end{matrix}\left(\begin{matrix}
	\bar{\xi}\\\xi^\prime
	\end{matrix}\right)}\int\text{\textbf{D}}\boldsymbol{\Theta}e^{\tfrac{1}{2}\boldsymbol{\Theta}^T\boldsymbol{\omega}\boldsymbol{\Theta}+\boldsymbol{\eta^\prime}^T\boldsymbol{\mathcal{J}}\boldsymbol{\Theta}}
\end{equation}
where $\boldsymbol{\Theta}^T=\left(\begin{matrix}\bar{\theta}&\!\theta\\ \end{matrix}\right)$, $\textbf{D}\boldsymbol{\Theta}=\text{d}\bar{\theta}\text{d}\theta$, $\boldsymbol{\eta^\prime}^T\!=\!\left(\begin{matrix}\boldsymbol{\bar{\xi}}&\!\boldsymbol{\xi}^\prime\\\end{matrix}\right)$, and $\boldsymbol{\Omega}$ and $\mathcal{C}^\beta$ are given by \eqref{eq:BetaOmegaMatSimple} and \eqref{eq:BetaCcalConst} respectively. Finally, if we take the last integral in the equation \eqref{eq:BetaRDM4-1}, we get 
\begin{equation}\label{eq:BetaRDM4GrassForm}
\rho_{_\mathbf{1}}^\beta\!(4,\boldsymbol{\xi},\boldsymbol{\xi}^\prime)=\mathcal{C}^\beta\text{Pf}[\boldsymbol{\omega}] e^{\tfrac{1}{2}
	\left(\begin{matrix}
	\bar{\xi}&\!\xi^\prime\\
	\end{matrix}\right)\begin{matrix}
	\boldsymbol{\Omega}^\prime
	\end{matrix}\left(\begin{matrix}
	\bar{\xi}\\\xi^\prime
	\end{matrix}\right)}
\end{equation}
where
\begin{subequations}
\begin{equation}
\boldsymbol{\Omega}^\prime=\boldsymbol{\Omega}+\boldsymbol{\mathcal{J}}\boldsymbol{\omega}^{-1}\boldsymbol{\mathcal{J}}^T
\end{equation}
\begin{equation}
\boldsymbol{\omega}=\left(\begin{matrix}0&(\boldsymbol{\mathfrak{M}_2^*})_{0}\\-(\boldsymbol{\mathfrak{M}}_2)_{0}&0\end{matrix}\right)\boldsymbol{\mathcal{A}}^{-1}\left(\begin{matrix}0&-(\boldsymbol{\mathfrak{M}_2})_{0}^T\\(\boldsymbol{\mathfrak{M}}_2)_{0}^\dagger&0\end{matrix}\right),
\end{equation}
\begin{equation}
    \boldsymbol{\mathcal{J}}=\left(\begin{matrix}-\mathbf{R}_{12}&\mathbf{0}\\\mathbf{0}&-{\mathbf{R}_{12}}^*\\\end{matrix}\right)\boldsymbol{\mathcal{A}}^{-1}\left(\begin{matrix}\mathbf{0}&\boldsymbol{\mathfrak{M}}_2^*\\-\boldsymbol{\mathfrak{M}}_2&\mathbf{0}\\\end{matrix}\right)-\left(\begin{matrix}\mathbf{0}&\boldsymbol{\mathfrak{M}}_1^*\\\boldsymbol{\mathfrak{M}}_1&\mathbf{0}\\\end{matrix}\right).
\end{equation}
\end{subequations}

To do the same calculation for cross term $\varrho^\beta(2,\boldsymbol{\xi},\boldsymbol{\xi^\prime})$ in equation \eqref{eq:BetaDensityMat4terms}, we have
\begin{equation}\label{}
	\varrho^\beta(2,\boldsymbol{\xi},\boldsymbol{\xi^\prime})=|C^\beta|^2\beta\:\boldsymbol{\mathfrak{M}}_{_{0l}}\bar{\xi}_{{}_l}\scalebox{1.3}{$e$}^{^{\tfrac{1}{2}R_{ij}\bar{\xi}_i\bar{\xi}_j-\tfrac{1}{2}R_{ij}^*\xi_i^\prime\xi_j^\prime}}
\end{equation}
If we use the equation \eqref{eq:GrassProdToExp} and write the $\xi$ behind the exponential as $(\boldsymbol{\mathfrak{M}}\bar{\boldsymbol{\xi}})_{0}=\int\!\text{d}\theta e^{\theta(\boldsymbol{\mathfrak{M}}\bar{\boldsymbol{\xi}})_{0}}=\int\!\text{d}\theta e^{\theta\boldsymbol{\mathfrak{M}}_{0l}\bar{\xi}_{l}}$, where $\theta$ is a new Grassmann variables. We can write the density matrix as
\begin{equation}
	\begin{aligned}
	\varrho^\beta(2,\boldsymbol{\xi},\boldsymbol{\xi^\prime})=&|C^\beta|^2\beta\int\text{d}\theta e^{\theta(\boldsymbol{\mathfrak{M}}\bar{\boldsymbol{\xi}})_{0}}e^{\tfrac{1}{2}R_{ij}\bar{\xi_i}\bar{\xi_j}}e^{-\tfrac{1}{2}R^*_{ij}\xi^\prime_i\xi^\prime_j}.
	\end{aligned}
\end{equation}
Equivalently, If we introduce new Grassmann variables  $\boldsymbol{\eta}^T\!=\left(\begin{matrix}\boldsymbol{\bar{\xi}}&\boldsymbol{\xi}\\\end{matrix}\right)$ then we can write the integral as
\begin{equation}
	\varrho^\beta(2,\boldsymbol{\xi},\boldsymbol{\xi^\prime})=\int\text{d}\theta\mathcal{F}(\bar{\boldsymbol{\xi}},\boldsymbol{\xi}^\prime,\theta)\int\text{\textbf{D}}\boldsymbol{\eta}\:\scalebox{1.2}{$e$}^{\tfrac{1}{2}\boldsymbol{\eta}^T\!\boldsymbol{\mathcal{A}}\boldsymbol{\eta}+\boldsymbol{\lambda}^T\!\boldsymbol{\eta}},
\end{equation}
where we have denoted the terms on the left of second integral by $\mathcal{F}(\bar{\boldsymbol{\xi}},\boldsymbol{\xi},\theta)$ and
\begin{equation}
	\boldsymbol{\lambda}^T=\left(\begin{matrix}
	-\boldsymbol{\bar{\xi}}\mathbf{R}_{12}-\theta(\boldsymbol{\mathfrak{M}_2})_{0l^\prime}&-\boldsymbol{\xi}^\prime\mathbf{R}_{12}^*\\\end{matrix}\right)
\end{equation}
Taking the integration on $\boldsymbol{\eta}$, we have:
\begin{equation}
	\varrho^\beta(2,\boldsymbol{\xi},\boldsymbol{\xi^\prime})=\frac{\mathcal{C}^\beta}{\beta^*}\scalebox{1.3}{$e$}^{\tfrac{1}{2}
		\left(\begin{matrix}
		\bar{\xi}&\!\xi^\prime\\
		\end{matrix}\right)\begin{matrix}
		\boldsymbol{\Omega}
		\end{matrix}\left(\begin{matrix}
		\bar{\xi}\\\xi^\prime
		\end{matrix}\right)}\int\text{d}\theta\scalebox{1.2}{$e$}^{\theta\boldsymbol{\zeta}}
\end{equation}
where $\boldsymbol{\Omega}$ is the same as \eqref{eq:BetaOmegaMatSimple} and $\mathcal{C}^\beta$ is given by $\eqref{eq:BetaCcalConst}$. Then, we can take the integral over the $\theta$ above and write the final result as 
\begin{equation}\label{eq:BetaRCD1}
	\varrho^\beta(2,\boldsymbol{\xi},\boldsymbol{\xi^\prime})=\frac{\mathcal{C}^\beta}{\beta^*}\big(\boldsymbol{\Upsilon}\boldsymbol{\mathcal{A}}^{-1}\boldsymbol{\mathcal{R}}^T+\boldsymbol{\Upsilon^\prime}\big)\boldsymbol{\eta^\prime}\scalebox{1.3}{$e$}^{^{\tfrac{1}{2}\boldsymbol{\eta^\prime}^T\boldsymbol{\Omega}\boldsymbol{\eta}^\prime}}
\end{equation}
where $\boldsymbol{\eta^\prime}^T=\left(\begin{matrix}\boldsymbol{\bar{\xi}}&\!\boldsymbol{\xi}^\prime\\\end{matrix}\right)$, and
Also we have
\begin{equation}
	\boldsymbol{\Upsilon}=\begin{bmatrix}\boldsymbol{\mathfrak{M}_2}&\mathbf{0}\end{bmatrix}_{{}_{1\times2L^\prime}}
	\!\qquad\boldsymbol{\Upsilon}^{\prime}=\begin{bmatrix}\boldsymbol{\mathfrak{M}_1}&\mathbf{0}\end{bmatrix}_{{}_{1\times2l}}
	\!\qquad\boldsymbol{\mathcal{R}}=\begin{bmatrix}\mathbf{R}_{12}&0\\0&\mathbf{R}_{12}^*\end{bmatrix}_{{}_{2l\times2L^\prime}}
\end{equation}
For the other cross density (third term in equation \eqref{eq:BetaDensityMat4terms}) we have:
\begin{equation}\label{eq:BetaRCD2}
	\varrho^\beta(3,\boldsymbol{\xi},\boldsymbol{\xi^\prime})=\frac{\mathcal{C}^\beta}{\beta}\big(-\boldsymbol{\Upsilon}\boldsymbol{\mathcal{A}}^{-1}\boldsymbol{\mathcal{R}}^T+\boldsymbol{\Upsilon^\prime}\big)\boldsymbol{\eta^\prime}\scalebox{1.3}{$e$}^{^{\tfrac{1}{2}\boldsymbol{\eta^\prime}^T\boldsymbol{\Omega}^\beta\boldsymbol{\eta}^\prime}}
\end{equation}
which
\begin{equation}
	\boldsymbol{\Upsilon}=\begin{bmatrix}\mathbf{0}&\boldsymbol{\mathfrak{M}_2}^*\end{bmatrix}_{{}_{1\times2L^\prime}}
	\!\qquad\boldsymbol{\Upsilon}^{\prime}=\begin{bmatrix}\mathbf{0}&\boldsymbol{\mathfrak{M}_1}^*\end{bmatrix}
	\!\qquad\boldsymbol{\mathcal{R}}=\begin{bmatrix}\mathbf{R}_{12}&0\\0&\mathbf{R}_{12}^*\end{bmatrix}_{{}_{2l\times2L^\prime}}
\end{equation}
Putting equations \eqref{eq:BetaDnstyMatGrass1}, \eqref{eq:BetaRDM4GrassForm}, \eqref{eq:BetaRCD1} and \eqref{eq:BetaRCD2} together and moving and reordering some parts we can write the RDM for the state $\ket{\beta}$ as
\begin{equation}\label{eq:BetaFinalRDMGrass}
\begin{aligned}
    \rho_{_\mathbf{1}}^\beta(\boldsymbol{\xi},\boldsymbol{\xi^\prime})=&\;\mathcal{C}_\beta\Big[\frac{1}{|\beta|^2}+\text{Pf}[\boldsymbol{\Omega_2}]\scalebox{1.3}{$e$}^{\tfrac{1}{2}
	\left(\begin{matrix}
	\bar{\xi}&\!\xi^\prime\\
	\end{matrix}\right)\boldsymbol{\mathcal{J}}\boldsymbol{\omega}^{-1}\boldsymbol{\mathcal{J}}^T\left(\begin{matrix}
	\bar{\xi}\\\xi^\prime
	\end{matrix}\right)}+\boldsymbol{\mathcal{K}}^\beta\left(\begin{matrix}
	\bar{\xi}\\\xi^\prime
	\end{matrix}\right)\Big]\scalebox{1.3}{$e$}^{^{\tfrac{1}{2}(\boldsymbol{\Omega}_{11})_{ij}\bar{\xi}_i\bar{\xi}_j}}\scalebox{1.3}{$e$}^{^{\mathcal{Y}_{ij}\bar{\xi_i}\xi_j^\prime}}\scalebox{1.3}{$e$}^{^{\tfrac{1}{2}(\boldsymbol{\Omega}_{22})_{ij}\xi_i^\prime\xi_j^\prime}},
\end{aligned}
\end{equation}
where $2\boldsymbol{\mathcal{Y}}=\boldsymbol{\Omega}_{12}-{\boldsymbol{\Omega}_{21}}^T$ and
\begin{equation}\label{eq:BetaKcal}
    \boldsymbol{\mathcal{K}}^\beta=\left(\begin{matrix}\tfrac{1}{\beta^*}\boldsymbol{\mathfrak{M}_1}&\tfrac{1}{\beta}\boldsymbol{\mathfrak{M}_1}^*\\\end{matrix}\right)+\left(\begin{matrix}\tfrac{1}{\beta^*}\boldsymbol{\mathfrak{M}_2}&-\tfrac{1}{\beta}\boldsymbol{\mathfrak{M}_2}^*\\\end{matrix}\right)\scalebox{1.2}{${\boldsymbol{\mathcal{A}}}$}^{-1}\left(\begin{matrix}{\mathbf{R}_{12}}^T&0\\0&{\mathbf{R}_{12}}^\dagger\\\end{matrix}\right)
\end{equation}

To write the RDM in operator format, we have to reorder it. Then, the final result is
\begin{equation}\label{eq:BetaRDMFirstVersion}
\begin{aligned}
    \rho_{\mathbf{1}}^\beta(c,c^\dagger)=&\:\mathcal{C}_\beta e^{^{\tfrac{1}{2}(\boldsymbol{\Omega}_{11})_{ij} c_i^\dagger c_j^\dagger}}\Big[\frac{1}{|\beta|^2}e^{^{(\ln\frac{\boldsymbol{\Omega}_{12}-{\boldsymbol{\Omega}_{21}}^T}{2})_{ij} c_i^\dagger c_j}}+(\boldsymbol{\mathcal{K}}^\beta_1)_{k}c^\dagger_k e^{^{(\ln\frac{\boldsymbol{\Omega}_{12}-{\boldsymbol{\Omega}_{21}}^T}{2})_{ij} c_i^\dagger c_j}}\\
    &+e^{^{(\ln\frac{\boldsymbol{\Omega}_{12}-{\boldsymbol{\Omega}_{21}}^T}{2})_{ij} c_i^\dagger c_j}}(\boldsymbol{\mathcal{K}}^\beta_2)_{l}c_l\Big]e^{^{\tfrac{1}{2}(\boldsymbol{\Omega}_{22})_{ij} c_i c_j}}\\
	&+\mathcal{C}^\beta\text{Pf}[\boldsymbol{\Omega_2}]e^{^{\tfrac{1}{2}(\boldsymbol{\Omega}_{11}^\prime)_{ij} c_i^\dagger c_j^\dagger}} e^{^{(\ln\frac{\boldsymbol{\Omega}_{12}^\prime-{\boldsymbol{\Omega}_{21}^\prime}^T}{2})_{ij} c_i^\dagger c_j}}e^{^{\tfrac{1}{2}(\boldsymbol{\Omega}_{22}^\prime)_{ij} c_i c_j}}.
\end{aligned}
\end{equation}
The $\boldsymbol{\mathcal{K}}^\beta_1$ stands for the first block part of vector $\boldsymbol{\mathcal{K}}^\beta$ (and the same argument for $\boldsymbol{\mathcal{K}}^\beta_2$). To find a shorter notation for the $\rho_{\mathbf{1}}^\beta(c,c^\dagger)$ we use the relation \cite{Balian1969}
\begin{equation}
e^{{}^{\frac{1}{2}
	\left(\begin{matrix} 
	\mathbf{c}^\dagger&\!\mathbf{c}\\
	\end{matrix}\right)\begin{matrix}
	\boldsymbol{\mathcal{M}_1}
	\end{matrix}\left(\begin{matrix}
	\!\mathbf{c}\\\mathbf{c}^\dagger
	\end{matrix}\right)}}\:e^{{}^{\frac{1}{2}
	\left(\begin{matrix}
	\mathbf{c}^\dagger&\!\mathbf{c}\\
	\end{matrix}\right)\begin{matrix}
	\boldsymbol{\mathcal{M}_2}
	\end{matrix}\left(\begin{matrix}
	\!\mathbf{c}\\\mathbf{c}^\dagger
	\end{matrix}\right)}}\:e^{{}^{\frac{1}{2}
	\left(\begin{matrix}
	\mathbf{c}^\dagger&\!\mathbf{c}\\
	\end{matrix}\right)\begin{matrix}
	\boldsymbol{\mathcal{M}_3}
	\end{matrix}\left(\begin{matrix}
	\!\mathbf{c}\\\mathbf{c}^\dagger
	\end{matrix}\right)}}=e^{{}^{\frac{1}{2}
	\left(\begin{matrix}
	\mathbf{c}^\dagger&\!\mathbf{c}\\
	\end{matrix}\right)\begin{matrix}
	\boldsymbol{\mathcal{M}}
	\end{matrix}\left(\begin{matrix}
	\!\mathbf{c}\\\mathbf{c}^\dagger
	\end{matrix}\right)}}
\end{equation}
where $e^{\boldsymbol{\mathcal{M}}}=e^{\boldsymbol{\mathcal{M}}_1}\:e^{\boldsymbol{\mathcal{M}_2}}\:e^{\boldsymbol{\mathcal{M}_3}}$. Therefore, we can write \eqref{eq:BetaRDMFirstVersion} as
\begin{equation}\label{eq:BetaRDMSecondVersion}
\begin{aligned}
\rho_{\mathbf{1}}^\beta(c,c^\dagger)=&\:\mathcal{C}_\beta\:\Big[\frac{1}{|\beta|^2} e^{\frac{1}{2}
	\left(\begin{matrix}
	\mathbf{c}^\dagger&\!\mathbf{c}\\
	\end{matrix}\right)\begin{matrix}
	\boldsymbol{\mathcal{M}}
	\end{matrix}\left(\begin{matrix}
	\!\mathbf{c}\\\mathbf{c}^\dagger
	\end{matrix}\right)} e^{^{\tfrac{1}{2}\text{tr}\ln\boldsymbol{\mathcal{Y}}}}+\text{Pf}[\boldsymbol{\Omega_2}] e^{\frac{1}{2}
	\left(\begin{matrix}
	\mathbf{c}^\dagger&\!\mathbf{c}\\
	\end{matrix}\right)\begin{matrix}
	\boldsymbol{\mathcal{M}^\prime}
	\end{matrix}\left(\begin{matrix}
	\!\mathbf{c}\\\mathbf{c}^\dagger
	\end{matrix}\right)} e^{^{\tfrac{1}{2}\text{tr}\ln(\boldsymbol{\mathcal{Y}}+\boldsymbol{\mathfrak{Y}})}}\\
	&+(\boldsymbol{\mathcal{K}}^\beta_1)_{k}c^\dagger_k e^{\frac{1}{2}
	\left(\begin{matrix}
	\mathbf{c}^\dagger&\!\mathbf{c}\\
	\end{matrix}\right)\begin{matrix}
	\boldsymbol{\mathcal{M}}
	\end{matrix}\left(\begin{matrix}
	\!\mathbf{c}\\\mathbf{c}^\dagger
	\end{matrix}\right)} e^{^{\tfrac{1}{2}\text{tr}\ln\boldsymbol{\mathcal{Y}}}}+e^{\frac{1}{2}
	\left(\begin{matrix}
	\mathbf{c}^\dagger&\!\mathbf{c}\\
	\end{matrix}\right)\begin{matrix}
	\boldsymbol{\mathcal{M}}
	\end{matrix}\left(\begin{matrix}
	\!\mathbf{c}\\\mathbf{c}^\dagger
	\end{matrix}\right)} e^{^{\tfrac{1}{2}\text{tr}\ln\boldsymbol{\mathcal{Y}}}}(\boldsymbol{\mathcal{K}}^\beta_2)_{l}c_l\Big].
\end{aligned}
\end{equation}
To move the exponentials to one side (left or right) of the \eqref{eq:BetaRDMSecondVersion}, we use the relation \eqref{eq:BeckHausd_Relation}. Therefore, we can write the \eqref{eq:BetaRDMSecondVersion} as
\begin{equation}\label{eq:BetaRDMFinalVersion}
\begin{aligned}
    \rho_{\mathbf{1}}^\beta(c,c^\dagger)=&\:\mathcal{C}_\beta\:
	e^{\frac{1}{2} 
	\left(\begin{matrix}
	\mathbf{c}^\dagger&\!\mathbf{c}\\
	\end{matrix}\right)\begin{matrix}
	\boldsymbol{\mathcal{M}}
	\end{matrix}\left(\begin{matrix}
	\!\mathbf{c}\\\mathbf{c}^\dagger
	\end{matrix}\right)}e^{^{\tfrac{1}{2}\text{tr}\ln(\tfrac{1}{2}\boldsymbol{\Omega}_{12}-\tfrac{1}{2}\boldsymbol{\Omega}_{21}^T)}}\Big[\frac{1}{|\beta|^2}+\left(
	\begin{matrix} \boldsymbol{\mathcal{L}}^\beta_1 & \boldsymbol{\mathcal{L}}^\beta_2\end{matrix}
	\right)\left(
	\begin{matrix}\mathbf{c}^\dagger\\\mathbf{c} \end{matrix}
	\right)\Big]\\
	&\:+\mathcal{C}_\beta\:\text{Pf}[\boldsymbol{\Omega_2}] e^{\frac{1}{2}
	\left(\begin{matrix}
	\mathbf{c}^\dagger&\!\mathbf{c}
	\end{matrix}\right)\begin{matrix}
	\boldsymbol{\mathcal{M}^\prime}
	\end{matrix}\left(\begin{matrix}
	\!\mathbf{c}\\\mathbf{c}^\dagger
	\end{matrix}\right)} e^{^{\tfrac{1}{2}\text{tr}\ln(\tfrac{1}{2}\boldsymbol{\Omega}_{12}^\prime-\tfrac{1}{2}{\boldsymbol{\Omega}_{21}^\prime}^T)}}
\end{aligned}
\end{equation}
where we have:
\begin{subequations}
\begin{equation}
    \boldsymbol{\mathcal{M}}=\ln{\left(\begin{matrix} 
    \tfrac{1}{2}\boldsymbol{\Omega}_{12}-\tfrac{1}{2}{\boldsymbol{\Omega}_{21}}^T+2\boldsymbol{\Omega}_{11}(\boldsymbol{\Omega}_{12}^T-{\boldsymbol{\Omega}_{21}})^{-1}\boldsymbol{\Omega}_{22}&\;2\boldsymbol{\Omega}_{11}(\boldsymbol{\Omega}_{12}^T-{\boldsymbol{\Omega}_{21}})^{-1}\\2(\boldsymbol{\Omega}_{12}^T-\boldsymbol{\Omega}_{21})^{-1}\boldsymbol{\Omega}_{22}&2(\boldsymbol{\Omega}_{12}^T-{\boldsymbol{\Omega}_{21}})^{-1}
    \end{matrix}\right)},
\end{equation}    
\begin{equation}
    \boldsymbol{\mathcal{M}}^\prime=\ln\left(\begin{matrix} 
    \tfrac{1}{2}\boldsymbol{\Omega}_{12}^\prime-\tfrac{1}{2}{\boldsymbol{\Omega}_{21}^\prime}^T+2\boldsymbol{\Omega}_{11}^\prime({\boldsymbol{\Omega}_{12}^\prime}^T-{\boldsymbol{\Omega}_{21}}^\prime)^{-1}\boldsymbol{\Omega}_{22}^\prime&\;2\boldsymbol{\Omega}_{11}^\prime({\boldsymbol{\Omega}_{12}^\prime}^T-{\boldsymbol{\Omega}_{21}^\prime})^{-1}\\2({\boldsymbol{\Omega}_{12}^\prime}^T-\boldsymbol{\Omega}_{21}^\prime)^{-1}\boldsymbol{\Omega}_{22}^\prime&2({\boldsymbol{\Omega}_{12}^\prime}^T-{\boldsymbol{\Omega}_{21}^\prime})^{-1}
    \end{matrix}\right)
\end{equation}
\begin{equation}
\boldsymbol{\Omega}=\left(\begin{matrix}\mathbf{R}_{11}&0\\0&-{\mathbf{R}_{11}}^*\\\end{matrix}\right)+\left(\begin{matrix}\mathbf{R}_{12}&0\\0&{\mathbf{R}_{12}}^*\\\end{matrix}\right){\boldsymbol{\mathcal{A}}}^{-1}\left(\begin{matrix}{\mathbf{R}_{12}}^T&0\\0&{\mathbf{R}_{12}}^\dagger\\\end{matrix}\right);\qquad
\boldsymbol{\mathcal{A}}=\left(\begin{matrix}
\mathbf{R}_{22}&-\mathbf{I}\\\mathbf{I}&-\mathbf{R}_{22}^*\\
\end{matrix}\right)
\end{equation}
\begin{equation}
\boldsymbol{\Omega}^\prime=\boldsymbol{\Omega}+\boldsymbol{\mathcal{J}}\boldsymbol{\omega}^{-1}\boldsymbol{\mathcal{J}}^T,
\qquad
\boldsymbol{\omega}=\left(\begin{matrix}\boldsymbol{\mathfrak{M}_2^*}(\boldsymbol{\mathcal{A}}^{{-1}})_{22}\boldsymbol{\mathfrak{M}}_2^\dagger&-\boldsymbol{\mathfrak{M}_2^*}(\boldsymbol{\mathcal{A}}^{{-1}})_{21}\boldsymbol{\mathfrak{M}}_2^T\\-\boldsymbol{\mathfrak{M}}_2(\boldsymbol{\mathcal{A}}^{{-1}})_{12}\boldsymbol{\mathfrak{M}}_2^\dagger&\boldsymbol{\mathfrak{M}_2^*}(\boldsymbol{\mathcal{A}}^{{-1}})_{11}\boldsymbol{\mathfrak{M}}_2^T\end{matrix}\right)
\end{equation}
\begin{equation}
\boldsymbol{\mathcal{J}}=\left(\begin{matrix}-\mathbf{R}_{12}&\mathbf{0}\\\mathbf{0}&-{\mathbf{R}_{12}}^*\\\end{matrix}\right){\boldsymbol{\mathcal{A}}}^{-1}\left(\begin{matrix}\mathbf{0}&\boldsymbol{\mathfrak{M}}_2^*\\-\boldsymbol{\mathfrak{M}}_2&\mathbf{0}\\\end{matrix}\right)-\left(\begin{matrix}\mathbf{0}&\boldsymbol{\mathfrak{M}}_1^*\\\boldsymbol{\mathfrak{M}}_1&\mathbf{0}\\\end{matrix}\right),
\end{equation}
\begin{equation}
    \boldsymbol{\mathcal{K}}^\beta=\left(\begin{matrix}\tfrac{1}{\beta^*}\boldsymbol{\mathfrak{M}_1}&\tfrac{1}{\beta}\boldsymbol{\mathfrak{M}_1}^*\\\end{matrix}\right)+\left(\begin{matrix}\tfrac{1}{\beta^*}\boldsymbol{\mathfrak{M}_2}&-\tfrac{1}{\beta}\boldsymbol{\mathfrak{M}_2}^*\\\end{matrix}\right){\boldsymbol{\mathcal{A}}}^{-1}\left(\begin{matrix}{\mathbf{R}_{12}}^T&0\\0&{\mathbf{R}_{12}}^\dagger\\\end{matrix}\right)
\end{equation}
\begin{equation}
    \boldsymbol{\mathcal{L}}^\beta_1=2\boldsymbol{\mathcal{K}}^\beta_1(\boldsymbol{\Omega}_{12}^T-{\boldsymbol{\Omega}_{21}})^{-1},\qquad \boldsymbol{\mathcal{L}}_2^\beta=2\boldsymbol{\mathcal{K}}^\beta_1(\boldsymbol{\Omega}_{12}^T-{\boldsymbol{\Omega}_{21}})^{-1}\boldsymbol{\Omega}_{22}+\boldsymbol{\mathcal{K}}^\beta_2.
\end{equation}
\end{subequations}
$\boldsymbol{\mathcal{K}}^\beta_1$ stands for the first block part of vector $\boldsymbol{\mathcal{K}}^\beta$. Since, we can think of $\boldsymbol{\mathcal{K}}^\beta$ as a vector made from two block vectors (and the same argument for $\boldsymbol{\mathcal{K}}^\beta_2$).

\section{Entanglement calculations of general \texorpdfstring{$\beta$}{} parity broken state}\label{sec:App_Ent_alpha_Beta_calculation}
\setcounter{equation}{0}

To calculate the $n=2$ R\'enyi EE, we start by:
\begin{equation}
    \text{tr}({\rho_{A}^\beta}^2)=\text{tr}(\rho_{A}^\beta\mathbf{I}\rho_{A}^\beta).
\end{equation}
We can use the Identity resolution and tracing formula of Grassmann variables to write:
\begin{subequations}
\begin{equation}\label{eq:GrassIdentity}
    \mathbf{I}=\int\prod_{l}\text{d}\bar{\xi}_l\text{d}\xi_l e^{-\boldsymbol{\bar{\xi}.\xi}}\ket{\boldsymbol{\xi}}\!\bra{\boldsymbol{\xi}},
\end{equation}
\begin{equation}\label{eq:GrassTrace}
    \text{tr} \mathcal{O} =\int\prod_{i}\text{d}\bar{\xi_i}\text{d}\xi_ie^{-\boldsymbol{\bar{\xi}.\xi}}\bra{-\boldsymbol{\xi}}\mathcal{O}\ket{\boldsymbol{\xi}}.
\end{equation}
\end{subequations}
Therefore, it is possible to calculate the trace of $\rho^2$ in terms of Berezin integrals of Grassmann variables:
\begin{equation}\label{eq:Ent_Beta_Integral1}
    \text{tr}({\rho_{A}^\beta}^2)=\int\begingroup\footnotesize
    \prod_{i}\text{d}\bar{\xi_i}\text{d}\xi_i\prod_{l}\text{d}\bar{\eta}_l\text{d}\eta_l 
    \endgroup
    e^{\begingroup\small-\boldsymbol{\bar{\xi}.\xi}\endgroup} e^{-\boldsymbol{\bar{\eta}.\eta}}\bra{-\boldsymbol{\xi}}\rho_{A}^\beta\ket{\boldsymbol{\eta}}\!\bra{\boldsymbol{\eta}}\rho_{A}^\beta\ket{\boldsymbol{\xi}}.
\end{equation}
We can use the result of \eqref{eq:BetaRDMGrassFin} to write the integrand in the expression above as: 
\begin{equation}\label{eq:Ent_Beta_Integral1.5}
\begin{aligned}
    \text{tr}({\rho_{A}^\beta}^2)=&{\mathcal{C}^\beta}^2\!\int\begingroup\footnotesize\prod_{i}\!\text{d}\bar{\xi_i}\text{d}\xi_i\prod_{l}\!\text{d}\bar{\eta}_l\text{d}\eta_l\endgroup 
    e^{-\boldsymbol{\bar{\xi}.\xi}} e^{-\boldsymbol{\bar{\eta}.\eta}}
    {\begingroup\footnotesize \big[(\frac{1}{\beta}-\boldsymbol{\mathcal{L}}_1.\bar{\boldsymbol{\xi}}+\boldsymbol{\mathcal{L}}_2.\boldsymbol{\eta})(\frac{1}{\beta^*}-\boldsymbol{\mathcal{L}}_3.\bar{\boldsymbol{\xi}}+\boldsymbol{\mathcal{L}}_4.\boldsymbol{\eta})-\text{Pf}[\boldsymbol{\mathcal{W}}]\big]\endgroup}\\
    &e^{\begingroup\tiny\tfrac{1}{2}\left(\begin{matrix}
	\bar{\xi}&\!\eta \\
	\end{matrix}\right)\begin{matrix}
	\boldsymbol{\Omega}^\prime
	\end{matrix}\left(\begin{matrix}
	\bar{\xi}\\\eta
	\end{matrix}\right)\endgroup}
	{\begingroup\footnotesize \big[(\frac{1}{\beta}+\boldsymbol{\mathcal{L}}_1.\bar{\boldsymbol{\eta}}+\boldsymbol{\mathcal{L}}_2.\boldsymbol{\xi})(\frac{1}{\beta^*}+\boldsymbol{\mathcal{L}}_3.\bar{\boldsymbol{\eta}}+\boldsymbol{\mathcal{L}}_4.\boldsymbol{\xi})-\text{Pf}[\boldsymbol{\mathcal{W}}]\big]\endgroup}
	e^{\begingroup\tiny\tfrac{1}{2}
	\left(\begin{matrix}
	\bar{\eta}&\!\xi\\
	\end{matrix}\right)\begin{matrix}
	\boldsymbol{\Omega}
	\end{matrix}\left(\begin{matrix}
	\bar{\eta}\\\xi
	\end{matrix}\right)\endgroup},
\end{aligned}
\end{equation}
By defining a new Grassmann variable such as:
\begin{equation}
    \boldsymbol{\theta} = \left(\!\begin{matrix}
    \bar{\boldsymbol{\xi}}\\ \bar{\boldsymbol{\eta}}\\ \boldsymbol{\eta}\\ \boldsymbol{\xi}
    \end{matrix}\!\right),
\end{equation}
the integral \eqref{eq:Ent_Beta_Integral1} can be written in a simpler form as:
\begin{equation}\label{eq:Ent_Beta_Integral2}
    \text{tr}({\rho_{A}^\beta}^2)={\mathcal{C}^\beta}^2\int\boldsymbol{\mathrm{D}\theta}
    \mathrm{f}(\boldsymbol{\theta})e^{\begingroup\small\tfrac{1}{2}\boldsymbol{\theta}^T\!.\boldsymbol{\mathcal{B}}.\boldsymbol{\theta} \endgroup},
\end{equation}
where in the above $\boldsymbol{\mathrm{D}\theta}=\prod_{i}\!\text{d}\bar{\theta_i}\text{d}\theta_i$, and also
\begin{equation*}
    \mathrm{f}(\boldsymbol{\theta})=\big[(\frac{1}{\beta}+\mathbb{C}.\boldsymbol{\theta}|_{1})(\frac{1}{\beta^*}+\mathbb{C}.\boldsymbol{\theta}|_{2})+\text{Pf}[\boldsymbol{\mathcal{W}}]\big]\big[(\frac{1}{\beta}+\mathbb{C}.\boldsymbol{\theta}|_{3})(\frac{1}{\beta^*}+\mathbb{C}.\boldsymbol{\theta}|_{4})+\text{Pf}[\boldsymbol{\mathcal{W}}]\big].
\end{equation*}
The new $\boldsymbol{\mathcal{B}}$ and $\boldsymbol{\mathbb{C}}$ matrices are defined below
\begin{equation}
    \boldsymbol{\mathcal{B}}=\left(\begin{matrix}
    \boldsymbol{\Omega}_{_{11}}&\boldsymbol{0}&-\boldsymbol{\Omega}_{_{12}}&\mathbf{I}\\
    \boldsymbol{0}&\boldsymbol{\Omega}_{_{11}}&\mathbf{I}&\boldsymbol{\Omega}_{_{12}}\\
    -\boldsymbol{\Omega}_{_{21}}&-\mathbf{I}&\boldsymbol{\Omega}_{_{22}}&\boldsymbol{0}\\
    -\mathbf{I}&\boldsymbol{\Omega}_{_{21}}&\boldsymbol{0}&\boldsymbol{\Omega}_{_{21}}\\
    \end{matrix} \right),\qquad
    \mathbb{C}=\left(\begin{matrix}
    -\boldsymbol{\mathcal{L}}_1&\boldsymbol{0}&\boldsymbol{\mathcal{L}}_2&\boldsymbol{0}\\
    -\boldsymbol{\mathcal{L}}_3&\boldsymbol{0}&\boldsymbol{\mathcal{L}}_4&\boldsymbol{0}\\
    \boldsymbol{0}&\boldsymbol{\mathcal{L}}_1&\boldsymbol{0}&\boldsymbol{\mathcal{L}}_2\\
    \boldsymbol{0}&\boldsymbol{\mathcal{L}}_3&\boldsymbol{0}&\boldsymbol{\mathcal{L}}_4\\
    \end{matrix} \right).
\end{equation}
The rest of matrices are given by \eqref{subeq:Beta_Consts_Matrixs}. Before proceeding to solve the integral above, we have to clarify couple of points here. First, the notation $\mathbb{C}.\boldsymbol{\theta}|_{r}$ stand for the $r^{\text{th}}$ row of the matrix product $\mathbb{C}.\boldsymbol{\theta}$. Second, the $\mathrm{f}(\boldsymbol{\theta})$ produce terms with product of odd number of Grassmann variables such as $\mathbb{C}.\boldsymbol{\theta}|_{r}$ or $\mathbb{C}.\boldsymbol{\theta}|_{r_3}\mathbb{C}.\boldsymbol{\theta}|_{r_2}\mathbb{C}.\boldsymbol{\theta}|_{r_1}$, the integration on these terms will be zero automatically. Therefore, The above integral is straightforward to solve, and the final result of tracing is given by \eqref{eq:Trace_Beta_Final}.

In some cases $\boldsymbol{\mathcal{B}}$ does not have an inverse for odd $n$. However, it is still possible to take the Grassmann integrations.

\section{Exact diagonalization of A=B=0}\label{sec:App_A_B_Zero}
\setcounter{equation}{0}

In the case where $\mathbf{A}$ and $\mathbf{B}$ are zero then the Hamiltonian \eqref{eq:FermionHamil} becomes
\begin{equation}\label{eq:HamABZero}
\begin{aligned}
H=&\sum_{j=1}^{L}\big[\alpha_j^0(c_{0}c_{j}-c_{0}^\dagger c_{j})+\alpha_j^{L+1}(c_{j}c_{L+1}^{\dagger}+c_{j}c_{L+1})+{\alpha_{j}^{0}}^*(c_{j}^{\dagger}c_{0}^\dagger-c_{j}^{\dagger}c_{0})+{\alpha_{j}^{L+1}}^*(c_{L+1}c_{j}^{\dagger}+c_{L+1}^{\dagger}c_{j}^{\dagger})\big].
\end{aligned}
\end{equation}
This model can be related to a linear fermionic model such as the model discussed in \cite{Colpa_1979}
\begin{equation}\label{eq:App_Ham_Colpa}
    H=\:\sum_{j=1}^{L}\alpha_{j}c_j+\alpha_{j}^*c_j^\dagger.
\end{equation}
In fact, with a specific projection, one can find the eigenstates of \eqref{eq:App_Ham_Colpa} in the Hilbert space of  \eqref{eq:HamABZero}. We can write the new form of the $\mathbf{M}$ matrix \eqref{eq:M matrix}\tiny
\begin{equation}\label{eq:M matform}
\mathbf{M}=\left(\begin{array}{ccccccc|ccccccc}
0 & -{\alpha_1^{0}} & -{\alpha_2^{0}} & \cdots & -{\alpha_{L-1}^{0}} & -{\alpha_L^{0}} & 0 &0&-{\alpha_{1}^{0}}^*&-{\alpha_{2}^{0}}^*&\cdots&-{\alpha_{L-1}^{0}}^*&-{\alpha_{L}^{0}}^*&0\\
-{\alpha_1^{0}}^* & {0} & {0} & \cdots & {0} & {0} & -{\alpha_{1}^{L+1}}^*&{\alpha_{1}^{0}}^*& 0 & 0 & \cdots & 0 & 0 & -{\alpha_{1}^{L+1}}^*\\
-{\alpha_2^{0}}^* &0 & {} & {} & {} &0& -{\alpha_{2}^{L+1}}^*&{\alpha_{2}^{0}}^*&0& {} & {} & {} &0& -{\alpha_{2}^{L+1}}^*\\
\vdots &\vdots & {} &{}& {} & {} & \vdots & \vdots &\vdots& {} & & {} &\vdots& \vdots\\
-{\alpha_{L-1}^{0}}^* &0&&&&0& -{\alpha_{L-1}^{L+1}}^* & {\alpha_{L-1}^{0}}^* &0&& &&0& -{\alpha_{L-1}^{L+1}}^*\\
-{\alpha_L^{0}}^* &0&0&\cdots&0&0& -{\alpha_{L}^{L+1}}^*&{\alpha_L^{0}}^* &0&0&\cdots&0&0& -{\alpha_{L}^{L+1}}^*\\
0 & -{\alpha_{1}^{L+1}} & -{\alpha_{2}^{L+1}} & \cdots & -{\alpha_{L-1}^{L+1}} & -{\alpha_{L}^{L+1}} & 0&0 & {\alpha_{1}^{L+1}}^* & {\alpha_{2}^{L+1}}^* & \cdots & {\alpha_{L-1}^{L+1}}^* & {\alpha_{L}^{L+1}}^* & 0\\\hline
0 & {\alpha_1^{0}} & {\alpha_2^{0}} & \cdots & {\alpha_{L-1}^{0}} & {\alpha_L^{0}} & 0&0 & {\alpha_1^{0}}^* & {\alpha_2^{0}}^* & \cdots & {\alpha_{L-1}^{0}}^* & {\alpha_L^{0}}^* & 0\\
-{\alpha_1^{0}} & 0 & 0 & \cdots & 0 & 0 & {\alpha_{1}^{L+1}}&{\alpha_1^{0}} & 0 & 0 & \cdots & 0 & 0 & {\alpha_{1}^{L+1}}\\
-{\alpha_2^{0}} &0& {} & {} & {} &0& {\alpha_{2}^{L+1}}&{\alpha_2^{0}} &0& {} & {} & {} &0& {\alpha_{2}^{L+1}}\\
\vdots &\vdots& {} & {} & {} &\vdots& \vdots&\vdots & \vdots& {} & & {} &\vdots&\vdots\\
-{\alpha_{L-1}^{0}} &0& {} & {} & {} &0& {\alpha_{L-1}^{L+1}}&{\alpha_{L-1}^{0}} &0& {} & {} & {} &0& {\alpha_{L-1}^{L+1}}\\
-{\alpha_L^{0}} & 0 & 0 & \cdots & 0 & 0 & {\alpha_{L}^{L+1}}&{\alpha_L^{0}} & 0 & 0 & \cdots & 0 & 0 & {\alpha_{L}^{L+1}}\\
0 & -{\alpha_{1}^{L+1}} & -{\alpha_{2}^{L+1}} & \cdots & -{\alpha_{L-1}^{L+1}} & -{\alpha_{L}^{L+1}} & 0 & 0 & {\alpha_{1}^{L+1}}^* & {\alpha_{2}^{L+1}}^* & \cdots & {\alpha_{L-1}^{L+1}}^* & {\alpha_{L}^{L+1}}^* & 0
\end{array}\right).
\end{equation}\normalsize
It can be shown that $\mathbf{\Lambda}$ has only two nonzero diagonal elements, $\lambda_1$ and $\lambda_2$ and the rest of eigenvalues are zero. For nonzero eigenvalues, we have
\begin{equation}\label{eq:nonzeroEgVal}
\lambda_1,\lambda_2=\sqrt{2}\sqrt{\sum_{j=1}^{L}(|\alpha_{j}^{0}|^2+|\alpha_{j}^{L+1}|^2)\pm\sqrt{\Big|\sum_{j=1}^{L}(\alpha_{j}^{0}+\alpha_{j}^{L+1})(\alpha_{j}^{0*}-\alpha_{j}^{L+1*})\Big|^2}}.
\end{equation}
Then, $\mathbf{M}$ would have only four nonzero eigenvalues and the rest are zero independent of size. The next step is to find the eigenvectors to construct the $\mathbf{U}$ matrix. In \eqref{eq:zeromode evec 1}, we have already introduced two eigenvectors of zero mode. Using the orthogonality condition, and the condition that if $\ket{u}$ is an eigenvector with eigenvalue $\lambda$ then $\mathbf{J}\ket{u}$ is also an eigenvector with eigenvalue $-\lambda$, therefore, we can find and present only half of the eigenvectors. For simplicity, we assume $\alpha_j^0=\alpha_j^{L+1}=\alpha_j$, then eigenvectors of positive modes. we have:
\begin{equation}
\ket{u^+_1}=\left(\begin{array}{c}
u\\v
\end{array}\right);\qquad\quad\begin{array}{l}
u_{L+1}=u_0=-v_0=v_{L+1}=1\\
u_i=-\frac{2\alpha_{i}^*}{\sqrt{\sum_{j=1}^{L}|\alpha_j|^2}}\\
\text{else}=0
\end{array}
\end{equation}
\begin{equation}
\ket{u^+_2}=\left(\begin{array}{c}
u\\v
\end{array}\right);\qquad\quad\begin{array}{l}
u_{L+1}=-u_0=v_0=v_{L+1}=\frac{\alpha_{L}}{|\alpha_{L}|}\\
v_i=\frac{2\alpha_{i}}{\sqrt{\sum_{j=1}^{L}|\alpha_j|^2}}\frac{\alpha_{L}}{|\alpha_{L}|}\\
\text{else}=0
\end{array}
\end{equation}
For eigenvectors of zero modes, we have:
\begin{equation}
\ket{u^0_1}=\left(\begin{array}{c}
u\\v
\end{array}\right);\qquad\quad\begin{array}{l}
u_{L+1}=u_0=v_0=-v_{L+1}=1\\
\text{else}=0
\end{array}
\end{equation}
\begin{equation}
\qquad\;\;\ket{u^0_k}=\left(\begin{array}{c}
u\\v
\end{array}\right);\qquad\quad\begin{array}{l}
u_i=-\frac{\alpha_{k}\alpha_i^*}{\sum_{j=1}^{k-1}|\alpha_j|^2};\quad i<k\neq0\hspace{1.0cm}\\
u_k=1\\
\text{else}=0,
\end{array}
\end{equation}
where $k=2,3,\cdots,L$. With this expressions for eigenvectors, we can construct the unitary matrices $\mathbf{U}$. Having the exact $\mathbf{U}$, we can calculate correlation matrices (see section \ref{sec:Correlation}). putting these eigenvectors together, one can construct the $\mathbf{U}$.

Having the $\mathbf{U}$ matrix, we can directly calculate the $\delta_{+}$ introduced in \eqref{eq:deltaplus}. The calculation of $\delta_{+}$ shows that
\begin{equation}\label{eq:DeltaPA,B=0}
\delta_{+}=\begin{cases}
+1& \text{if}\;\:L\;\:\text{even}\\
-1& \text{if}\;\:L\;\:\text{odd}\\
\end{cases}
\end{equation}

\subsection{Vacuum state in configuration basis}%\label{subsec:etta_vacuum}

Another method to calculate the entanglement of subsystem in a particular state is to use a more direct way. We could use the equation \eqref{eq:VacEtaVacC} to calculate the entanglement. Using the results of subsection \ref{subsec:Configur_Basis} for $\mathbf{A},\mathbf{B}=0$ we can find the form of $\mathbf{R}$ matrix as:. 
{\arraycolsep=1.5pt\def\arraystretch{1.5}
\begin{equation}
	\mathbf{R}=\left(\begin{array}{c|c|c}
	0&\frac{\alpha_j}{\sqrt{|\alpha_{1}|^2+\cdots+|\alpha_{L}|^2}}&1\\\hline
	\frac{-\alpha_i}{\sqrt{|\alpha_{1}|^2+\cdots+|\alpha_{L}|^2}}&0&\frac{\alpha_i}{\sqrt{|\alpha_{1}|^2+\cdots+|\alpha_{L}|^2}}\\\hline
	-1&\frac{-\alpha_j}{\sqrt{|\alpha_{1}|^2+\cdots+|\alpha_{L}|^2}}&0 \\
	\end{array}\right)
\end{equation}}
where, 
\begin{equation*}
\ket{0}_\eta=\frac{1}{\sqrt{\mathrm{det}[\mathbf{I}+\mathbf{R}^\dagger\mathbf{R}]}}\scalebox{1.5}{$e$}^{\tfrac{1}{2}\sum_{i,j}R_{ij}c^\dagger_i c_j^\dagger}\ket{0}_c.
\end{equation*}
Numerical investigations suggest that for even sizes we get exactly the same expression, and for odd sizes we get the minus of the vacuum ($\alpha_{i}\in\mathbb{R}$).

\subsection{Correlations}\label{subsec:CorrAB=0}
%\subsubsection{Vacuum}
For the special case of $\mathbf{A},\mathbf{B}=0$, the correlation matrices look pretty much simple in terms of the parameters $\alpha_i$'s. We are going to demonstrate the matrix form of correlations for general $\alpha_i\in\mathbb{C}$. For start, in the vacuum state, correlations looks like below. (In the following expressions indexes $n$ and $m$ are the number of rows and columns respectively.)
{\arraycolsep=2.0pt\def\arraystretch{1.8}
\begin{equation*}
\mathbf{C}^{0}=\left(\begin{array}{c|c|c}
\frac{1}{2}&\frac{\alpha_{m}}{4\sqrt{\sum_{k}|\alpha_{k}|^2}}&-\frac{1}{4}\\\hline
\frac{\alpha_{n}^*}{4\sqrt{\sum_{k}|\alpha_{k}|^2}}&\frac{\alpha_{n}^*\alpha_{m}}{2(\sum_{k}|\alpha_{k}|^2)}&\frac{\alpha_{n}^*}{4\sqrt{\sum_{k}|\alpha_{k}|^2}} \\\hline
-\frac{1}{4}&\frac{\alpha_{m}}{4\sqrt{\sum_{k}|\alpha_{k}|^2}}&\frac{1}{2} \\
\end{array}\right),
\quad
\mathbf{F}^{0}=\left(\begin{array}{c|c|c}
0&\frac{\alpha_{m}^*}{4\sqrt{\sum_{k}|\alpha_{k}|^2}}&\frac{1}{4}\\\hline
\frac{-\alpha_{n}^*}{4\sqrt{\sum_{k}|\alpha_{k}|^2}}&0&\frac{\alpha_{n}^*}{4\sqrt{\sum_{k}|\alpha_{k}|^2}} \\\hline
-\frac{1}{4}&\frac{-\alpha_{m}^*}{4\sqrt{\sum_{k}|\alpha_{k}|^2}}&0\\
\end{array}\right)
\end{equation*}}
{\arraycolsep=2.0pt\def\arraystretch{1.8}
\begin{equation*}
\mathbf{K}^{0}=\left(\begin{array}{c|c|c}
1&0&\qquad0\qquad\\\hline
0&\;\delta_{nm}-\frac{m-n}{|m-n|}\frac{i\Im[\alpha_{n}^*\alpha_{m}]}{\sum_{k}|\alpha_{k}|^2}\;&\frac{-i\Im[\alpha_{n}]}{\sqrt{\sum_{k}|\alpha_{k}|^2}}\\\hline
0&\;\frac{i\Im[\alpha_{m}]}{\sqrt{\sum_{k}|\alpha_{k}|^2}}\;&1 \\
\end{array}\right),\quad
\mathbf{\bar{K}}^{0}=\left(\begin{array}{c|c|c}
	1&\frac{i\Im[\alpha_{m}]}{\sqrt{\sum_{k}|\alpha_{k}|^2}}&0\\\hline
\;\frac{-i\Im[\alpha_{n}]}{\sqrt{\sum_{k}|\alpha_{k}|^2}}\;&\;\delta_{nm}-\frac{m-n}{|m-n|}\frac{i\Im[\alpha_{n}^*\alpha_{m}]}{\sum_{k}|\alpha_{k}|^2}\;&0\\\hline
	0&0&1 \\
\end{array}\right)
\end{equation*}
\begin{equation*}
\mathbf{G}^{0}=\left(\begin{array}{c|c|c}
0&\frac{\Re[\alpha_{m}]}{\sqrt{\sum_{k}|\alpha_{k}|^2}}&\qquad0\qquad\\\hline
0&\;-\delta_{nm}+\frac{\Re[\alpha_{n}^*\alpha_{m}]}{\sum_{k}|\alpha_{k}|^2}\;&\;\frac{\Re[\alpha_{n}]}{\sqrt{\sum_{k}|\alpha_{k}|^2}}\;\\\hline
\qquad-1\qquad&0&0 \\
\end{array}\right)
\end{equation*}}

%\subsubsection{ZME state}
For the case where the state is ZME state ($\ket{\emptyset}$) then the correlation matrices look like below.
{\arraycolsep=2.0pt\def\arraystretch{1.8}
\begin{equation*}
	\mathbf{C}^{\emptyset}=\left(\begin{array}{c|c|c}
	\frac{1}{2}&\frac{\alpha_{m}}{4\sqrt{\sum_{k}|\alpha_{k}|^2}}&\frac{1}{4}\\\hline
	\frac{\alpha_{n}^*}{4\sqrt{\sum_{k}|\alpha_{k}|^2}}&\frac{\alpha_{n}^*\alpha_{m}}{2(\sum_{k}|\alpha_{k}|^2)}&\frac{\alpha_{n}^*}{4\sqrt{\sum_{k}|\alpha_{k}|^2}} \\\hline
	\frac{1}{4}&\frac{\alpha_{m}}{4\sqrt{\sum_{k}|\alpha_{k}|^2}}&\frac{1}{2} \\
	\end{array}\right),
\quad
	\mathbf{F}^{\emptyset}=\left(\begin{array}{c|c|c}
	0&\frac{\alpha_{m}^*}{4\sqrt{\sum_{k}|\alpha_{k}|^2}}&-\frac{1}{4}\\\hline
	\frac{-\alpha_{n}^*}{4\sqrt{\sum_{k}|\alpha_{k}|^2}}&0&\frac{\alpha_{n}^*}{4\sqrt{\sum_{k}|\alpha_{k}|^2}} \\\hline
	\frac{1}{4}&\frac{-\alpha_{m}^*}{4\sqrt{\sum_{k}|\alpha_{k}|^2}}&0\\
	\end{array}\right)
\end{equation*}}
{\arraycolsep=2.0pt\def\arraystretch{1.8}
\begin{equation*}
	\mathbf{G}^{\emptyset}=\left(\begin{array}{c|c|c}
	0&\frac{\Re[\alpha_{m}]}{\sqrt{\sum_{k}|\alpha_{k}|^2}}&\qquad0\qquad\\\hline
	0&\;-\delta_{nm}+\frac{\Re[\alpha_{n}^*\alpha_{m}]}{\sum_{k}|\alpha_{k}|^2}\;&\;\frac{\Re[\alpha_{n}]}{\sqrt{\sum_{k}|\alpha_{k}|^2}}\;\\\hline
	\;1\;&0&0 \\
	\end{array}\right),\qquad\mathbf{K}^{\emptyset}=\mathbf{K}^{0},\qquad\bar{\mathbf{K}}^{\emptyset}=\bar{\mathbf{K}}^{0}.
\end{equation*}}
%\begin{equation*}
%	\mathbf{K}^{\emptyset}=\left(\begin{array}{c|c|c}
%	\;1\;&0&\;0\\\hline
%	0&\;\delta_{nm}-\frac{m-n}{|m-n|}\frac{i\Im[\alpha_{n}^*\alpha_{m}]}{\sum_{k}|\alpha_{k}|^2}\;&\frac{-i\Im[\alpha_{n}]}{\sqrt{\sum_{k}|\alpha_{k}|^2}}\\\hline
%	0&\;\frac{i\Im[\alpha_{m}]}{\sqrt{\sum_{k}|\alpha_{k}|^2}}\;&1 \\
%	\end{array}\right),\quad
%	\mathbf{\bar{K}}^{\emptyset}=\left(\begin{array}{c|c|c}
%	1&\frac{i\Im[\alpha_{m}]}{\sqrt{\sum_{k}|\alpha_{k}|^2}}&0\;\\\hline
%	\frac{-i\Im[\alpha_{n}]}{\sqrt{\sum_{k}|\alpha_{k}|^2}}\;&\;\delta_{nm}-\frac{m-n}{|m-n|}\frac{i\Im[\alpha_{n}^*\alpha_{m}]}{\sum_{k}|\alpha_{k}|^2}\;&0\;\\\hline
%	0&0&1 \\
%	\end{array}\right)
%\end{equation*}

%\subsubsection{Zero parity state}
Finally, for this case when the state is $\ket{G_{\pm}}$ then the correlations look like below.
{\arraycolsep=1.9pt\def\arraystretch{1.8}
\begin{equation*}
	\mathbf{C}^{\pm}=\left(\begin{array}{c|c|c}
	\frac{1}{2}&\frac{\alpha_{m}}{4\sqrt{\sum_{k}|\alpha_{k}|^2}}&0\\\hline
	\frac{\alpha_{n}^*}{4\sqrt{\sum_{k}|\alpha_{k}|^2}}&\frac{\alpha_{n}^*\alpha_{m}}{2(\sum_{k}|\alpha_{k}|^2)}&\frac{\alpha_{n}^*}{4\sqrt{\sum_{k}|\alpha_{k}|^2}} \\\hline
	0&\frac{\alpha_{m}}{4\sqrt{\sum_{k}|\alpha_{k}|^2}}&\frac{1}{2} \\
	\end{array}\right),
\quad
	\mathbf{F}^{\pm}=\left(\begin{array}{c|c|c}
	0&\frac{\alpha_{m}^*}{4\sqrt{\sum_{k}|\alpha_{k}|^2}}&0\\\hline
	\frac{-\alpha_{n}^*}{4\sqrt{\sum_{k}|\alpha_{k}|^2}}&0&\frac{\alpha_{n}^*}{4\sqrt{\sum_{k}|\alpha_{k}|^2}} \\\hline
	0&\frac{-\alpha_{m}^*}{4\sqrt{\sum_{k}|\alpha_{k}|^2}}&0\\
	\end{array}\right),
\end{equation*}}
{\arraycolsep=2.0pt\def\arraystretch{1.8}
\begin{equation*}
	\mathbf{G}^{\pm}=\left(\begin{array}{c|c|c}
	0&\frac{\Re[\alpha_{m}]}{\sqrt{\sum_{k}|\alpha_{k}|^2}}&\qquad0\qquad\\\hline
	0&\;\:-\delta_{nm}+\frac{\Re[\alpha_{n}^*\alpha_{m}]}{\sum_{k}|\alpha_{k}|^2}&\;\frac{\Re[\alpha_{n}]}{\sqrt{\sum_{k}|\alpha_{k}|^2}}\;\\\hline
	0\;&0&0 \\
	\end{array}\right)
	,\qquad\mathbf{K}^{\pm}=\mathbf{K}^{0},\qquad\bar{\mathbf{K}}^{\pm}=\bar{\mathbf{K}}^{0}.
\end{equation*}}
%\begin{equation*}
%	\mathbf{K}^{\pm}=\left(\begin{array}{c|c|c}
%	\;1\;&0&\qquad0\qquad\\\hline
%	0&\delta_{nm}-\frac{m-n}{|m-n|}\frac{i\Im[\alpha_{n}^*\alpha_{m}]}{\sum_{k}|\alpha_{k}|^2}&\frac{-\iota\Im[\alpha_{n}]}{\sqrt{\sum_{k}|\alpha_{k}|^2}}\\\hline
%	0&\;\frac{i\Im[\alpha_{m}]}{\sqrt{\sum_{k}|\alpha_{k}|^2}}\;&1 \\
%	\end{array}\right),\quad
%	\mathbf{\bar{K}}^{\pm}=\left(\begin{array}{c|c|c}
%	1&\frac{i\Im[\alpha_{m}]}{\sqrt{\sum_{k}|\alpha_{k}|^2}}&\;0\\\hline
%	\frac{-i\Im[\alpha_{n}]}{\sqrt{\sum_{k}|\alpha_{k}|^2}}\;&\;\delta_{nm}-\frac{m-n}{|m-n|}\frac{i\Im[\alpha_{n}^*\alpha_{m}]}{\sum_{k}|\alpha_{k}|^2}&0\\\hline
%	0&0&1 \\
%	\end{array}\right)
%\end{equation*}

\section{\textbf{M} matrix for modified \texorpdfstring{$XY$}{} chain}\label{sec:App_M_XY}
\setcounter{equation}{0}
In this appendix, the exact forms of the matrices $\textbf{A}$ and $\mathbf{B}$ for modified $XY$ spin chain is demonstrated. For $|\Vec{b}_1|=1$ and $|\Vec{b}_L|=1$, they can be written as:\small
\begin{equation}%\label{eq:M matrix}
\mathbf{A}=\left(
\begin{array}{cccccccc}
0 & -\frac{\sin\theta_{1}e^{i\varphi_{1}}}{2}& 0 & ...\\
-\frac{\sin\theta_{1}e^{-i\varphi_{1}}}{2}& -h+\cos\theta_{1} & -\frac{J}{2} & 0 & ...\\
0 & -\frac{J}{2} & -h & -\frac{J}{2} & 0 & ...\\
0 & 0 & -\frac{J}{2} & -h & -\frac{J}{2} & 0 & ...\\
. & . & .\\
. & . & .\\
&&&& . & . & -h+\cos\theta_{L}&-\frac{\sin\theta_{L}e^{-i\varphi_{L}}}{2}\\
&&&& \cdots & 0 & -\frac{\sin\theta_{L}e^{i\varphi_{L}}}{2}&0
\end{array}\right),
\end{equation}
\begin{equation}
\mathbf{B}=\begin{pmatrix}
0&-\frac{\sin\theta_{1}e^{-i\varphi_{1}}}{2}&0&\cdots\\
\frac{\sin\theta_{1}e^{-i\varphi_{1}}}{2}& 0 & -\frac{J\gamma}{2}&0&\cdots\\
0 & \frac{J\gamma}{2} & 0 & -\frac{\gamma J}{2} & 0 & ...\\
\vdots& 0 & \frac{J\gamma}{2} & 0 & -\frac{\gamma J}{2} & 0 & ...\\
& . &.\\
& . &.\\
&.&.&.&.&.&\frac{J\gamma}{2}&0&-\frac{\sin\theta_{L}e^{-i\varphi_{L}}}{2}\\
&.& . & . & . & . & 0 & \frac{\sin\theta_{L}e^{-i\varphi_{L}}}{2} & 0
\end{pmatrix}.
\end{equation}\normalsize
With the help of \eqref{eq:Ham matrix form} and \eqref{eq:M matrix}, one can construct the actual $\mathbf{M}$ matrix.
%\end{landscape}

\end{appendices}

%%\section*{References}
%\bibstyle{SciPost_bibstyle}
%\bibliography{Bibliography.bib}

\begin{thebibliography}{10}
\providecommand{\url}[1]{\texttt{#1}}
\providecommand{\urlprefix}{URL }
\expandafter\ifx\csname urlstyle\endcsname\relax
  \providecommand{\doi}[1]{doi:\discretionary{}{}{}#1}\else
  \providecommand{\doi}{doi:\discretionary{}{}{}\begingroup
  \urlstyle{rm}\Url}\fi
\providecommand{\eprint}[2][]{\url{#2}}

\bibitem{LIEB1961407}
E.~Lieb, T.~Schultz and D.~Mattis,
\newblock \emph{Two soluble models of an antiferromagnetic chain},
\newblock Annals of Physics \textbf{16}(3), 407 (1961),
\newblock \doi{https://doi.org/10.1016/0003-4916(61)90115-4}.

\bibitem{Peschel_2001}
M.-C. Chung and I.~Peschel,
\newblock \emph{Density-matrix spectra of solvable fermionic systems},
\newblock Phys. Rev. B \textbf{64}, 064412 (2001),
\newblock \doi{10.1103/PhysRevB.64.064412}.

\bibitem{Peschel_2004}
I.~Peschel,
\newblock \emph{On the reduced density matrix for a chain of free electrons},
\newblock Journal of Statistical Mechanics: Theory and Experiment
  \textbf{2004}(06), P06004 (2004),
\newblock \doi{10.1088/1742-5468/2004/06/p06004}.

\bibitem{Vidal2001}
G.~Vidal, J.~I. Latorre, E.~Rico and A.~Kitaev,
\newblock \emph{Entanglement in quantum critical phenomena},
\newblock Phys. Rev. Lett. \textbf{90}, 227902 (2003),
\newblock \doi{10.1103/PhysRevLett.90.227902}.

\bibitem{jin2004quantum}
B.-Q. Jin and V.~E. Korepin,
\newblock \emph{Quantum spin chain, toeplitz determinants and the
  fisher—hartwig conjecture},
\newblock Journal of statistical physics \textbf{116}(1), 79 (2004).

\bibitem{Keating2004}
J.~P. Keating and F.~Mezzadri,
\newblock \emph{Random matrix theory and entanglement in quantum spin chains},
\newblock Communications in Mathematical Physics \textbf{252}(1), 543 (2004),
\newblock \doi{10.1007/s00220-004-1188-2}.

\bibitem{Peschel_Eisler_2009}
I.~Peschel and V.~Eisler,
\newblock \emph{Reduced density matrices and entanglement entropy in free
  lattice models},
\newblock Journal of Physics A: Mathematical and Theoretical \textbf{42}(50),
  504003 (2009),
\newblock \doi{10.1088/1751-8113/42/50/504003}.

\bibitem{Franchini_2005}
F.~Franchini and A.~G. Abanov,
\newblock \emph{Asymptotics of toeplitz determinants and the emptiness
  formation probability for the {XY} spin chain},
\newblock Journal of Physics A: Mathematical and General \textbf{38}(23), 5069
  (2005),
\newblock \doi{10.1088/0305-4470/38/23/002}.

\bibitem{NajafiRajab2015}
K.~Najafi and M.~A. Rajabpour,
\newblock \emph{Formation probabilities and shannon information and their time
  evolution after quantum quench in the transverse-field xy chain},
\newblock Phys. Rev. B \textbf{93}, 125139 (2016),
\newblock \doi{10.1103/PhysRevB.93.125139}.

\bibitem{Najaf_Rajab_2020}
M.~N. Najafi and M.~A. Rajabpour,
\newblock \emph{Formation probabilities and statistics of observables as defect
  problems in free fermions and quantum spin chains},
\newblock Phys. Rev. B \textbf{101}, 165415 (2020),
\newblock \doi{10.1103/PhysRevB.101.165415}.

\bibitem{Alba_2009}
V.~Alba, M.~Fagotti and P.~Calabrese,
\newblock \emph{Entanglement entropy of excited states},
\newblock Journal of Statistical Mechanics: Theory and Experiment
  \textbf{2009}(10), P10020 (2009),
\newblock \doi{10.1088/1742-5468/2009/10/p10020}.

\bibitem{Alcaraz_Sierra_2011}
F.~C. Alcaraz, M.~I.~n. Berganza and G.~Sierra,
\newblock \emph{Entanglement of low-energy excitations in conformal field
  theory},
\newblock Phys. Rev. Lett. \textbf{106}, 201601 (2011),
\newblock \doi{10.1103/PhysRevLett.106.201601}.

\bibitem{Berganza_2012}
M.~I. Berganza, F.~C. Alcaraz and G.~Sierra,
\newblock \emph{Entanglement of excited states in critical spin chains},
\newblock Journal of Statistical Mechanics: Theory and Experiment
  \textbf{2012}(01), P01016 (2012),
\newblock \doi{10.1088/1742-5468/2012/01/p01016}.

\bibitem{Ares_2014}
F.~Ares, J.~G. Esteve, F.~Falceto and E.~S{\'{a}}nchez-Burillo,
\newblock \emph{Excited state entanglement in homogeneous fermionic chains},
\newblock Journal of Physics A: Mathematical and Theoretical \textbf{47}(24),
  245301 (2014),
\newblock \doi{10.1088/1751-8113/47/24/245301}.

\bibitem{Rigol_2017}
L.~Vidmar, L.~Hackl, E.~Bianchi and M.~Rigol,
\newblock \emph{Entanglement entropy of eigenstates of quadratic fermionic
  hamiltonians},
\newblock Phys. Rev. Lett. \textbf{119}, 020601 (2017),
\newblock \doi{10.1103/PhysRevLett.119.020601}.

\bibitem{PhysRevB.99.075123}
L.~Hackl, L.~Vidmar, M.~Rigol and E.~Bianchi,
\newblock \emph{Average eigenstate entanglement entropy of the xy chain in a
  transverse field and its universality for translationally invariant quadratic
  fermionic models},
\newblock Phys. Rev. B \textbf{99}, 075123 (2019),
\newblock \doi{10.1103/PhysRevB.99.075123}.

\bibitem{Rajabpour_2019}
A.~Jafarizadeh and M.~A. Rajabpour,
\newblock \emph{Bipartite entanglement entropy of the excited states of free
  fermions and harmonic oscillators},
\newblock Phys. Rev. B \textbf{100}, 165135 (2019),
\newblock \doi{10.1103/PhysRevB.100.165135}.

\bibitem{2020arXiv201013973Z}
J.~{Zhang} and M.~A. {Rajabpour},
\newblock \emph{{Universal R{\'e}nyi Entropy of Quasiparticle Excitations}},
\newblock arXiv e-prints arXiv:2010.13973 (2020),
\newblock \eprint{2010.13973}.

\bibitem{2020arXiv201016348Z}
J.~{Zhang} and M.~A. {Rajabpour},
\newblock \emph{{Universal R{\'e}nyi entropy in quasiparticle excited states of
  quantum chains}},
\newblock arXiv e-prints arXiv:2010.16348 (2020),
\newblock \eprint{2010.16348}.

\bibitem{PhysRevLett.96.100603}
N.~Laflorencie, E.~S. S\o{}rensen, M.-S. Chang and I.~Affleck,
\newblock \emph{Boundary effects in the critical scaling of entanglement
  entropy in 1d systems},
\newblock Phys. Rev. Lett. \textbf{96}, 100603 (2006),
\newblock \doi{10.1103/PhysRevLett.96.100603}.

\bibitem{PhysRevA.74.050305}
H.-Q. Zhou, T.~Barthel, J.~O. Fj\ae{}restad and U.~Schollw\"ock,
\newblock \emph{Entanglement and boundary critical phenomena},
\newblock Phys. Rev. A \textbf{74}, 050305 (2006),
\newblock \doi{10.1103/PhysRevA.74.050305}.

\bibitem{PhysRevLett.99.087203}
O.~Legeza, J.~S\'olyom, L.~Tincani and R.~M. Noack,
\newblock \emph{Entropic analysis of quantum phase transitions from uniform to
  spatially inhomogeneous phases},
\newblock Phys. Rev. Lett. \textbf{99}, 087203 (2007),
\newblock \doi{10.1103/PhysRevLett.99.087203}.

\bibitem{PhysRevB.77.045106}
E.~Szirmai, O.~Legeza and J.~S\'olyom,
\newblock \emph{Spatially nonuniform phases in the one-dimensional
  $\mathrm{SU}(n)$ hubbard model for commensurate fillings},
\newblock Phys. Rev. B \textbf{77}, 045106 (2008),
\newblock \doi{10.1103/PhysRevB.77.045106}.

\bibitem{Castro_Doyon_2009}
O.~A. Castro-Alvaredo and B.~Doyon,
\newblock \emph{Bi-partite entanglement entropy in massive qft with a boundary:
  the ising model},
\newblock Journal of Statistical Physics \textbf{134} (2009),
\newblock \doi{10.1007/s10955-008-9664-2}.

\bibitem{Affleck_2009}
I.~Affleck, N.~Laflorencie and E.~S. S{\o}rensen,
\newblock \emph{Entanglement entropy in quantum impurity systems and systems
  with boundaries},
\newblock Journal of Physics A: Mathematical and Theoretical \textbf{42}(50),
  504009 (2009),
\newblock \doi{10.1088/1751-8113/42/50/504009}.

\bibitem{PhysRevB.88.075112}
L.~Taddia, J.~C. Xavier, F.~C. Alcaraz and G.~Sierra,
\newblock \emph{Entanglement entropies in conformal systems with boundaries},
\newblock Phys. Rev. B \textbf{88}, 075112 (2013),
\newblock \doi{10.1103/PhysRevB.88.075112}.

\bibitem{Fagotti_2011}
M.~Fagotti and P.~Calabrese,
\newblock \emph{Universal parity effects in the entanglement entropy of xx
  chains with open boundary conditions},
\newblock Journal of Statistical Mechanics: Theory and Experiment
  \textbf{2011}(01), P01017 (2011),
\newblock \doi{10.1088/1742-5468/2011/01/p01017}.

\bibitem{Igloi2010}
F.~Igl{\'{o}}i and I.~Peschel,
\newblock \emph{On reduced density matrices for disjoint subsystems},
\newblock {EPL} (Europhysics Letters) \textbf{89}(4), 40001 (2010),
\newblock \doi{10.1209/0295-5075/89/40001}.

\bibitem{BRAVYI_2002}
S.~B. Bravyi and A.~Y. Kitaev,
\newblock \emph{Fermionic quantum computation},
\newblock Annals of Physics \textbf{298}(1), 210 (2002),
\newblock \doi{https://doi.org/10.1006/aphy.2002.6254}.

\bibitem{Banuls_Wolf_2007}
M.-C. Ba\~nuls, J.~I. Cirac and M.~M. Wolf,
\newblock \emph{Entanglement in fermionic systems},
\newblock Phys. Rev. A \textbf{76}, 022311 (2007),
\newblock \doi{10.1103/PhysRevA.76.022311}.

\bibitem{moriya_2006}
H.~Moriya,
\newblock \emph{On separable states for composite systems of distinguishable
  fermions},
\newblock Journal of Physics A: Mathematical and General \textbf{39}(14), 3753
  (2006).

\bibitem{Vega_1993}
H.~J. de~Vega and A.~G. Ruiz,
\newblock \emph{Boundary k-matrices for the six vertex and the
  n(2n-1)an-1vertex models},
\newblock Journal of Physics A: Mathematical and General \textbf{26}(12), L519
  (1993),
\newblock \doi{10.1088/0305-4470/26/12/007}.

\bibitem{NEPOMECHIE2002}
R.~I. Nepomechie,
\newblock \emph{Solving the open xxz spin chain with nondiagonal boundary terms
  at roots of unity},
\newblock Nuclear Physics B \textbf{622}(3), 615 (2002),
\newblock \doi{https://doi.org/10.1016/S0550-3213(01)00585-5}.

\bibitem{Nepomechie_2003}
R.~I. Nepomechie,
\newblock \emph{Bethe ansatz solution of the open {XXZ} chain with nondiagonal
  boundary terms},
\newblock Journal of Physics A: Mathematical and General \textbf{37}(2), 433
  (2003),
\newblock \doi{10.1088/0305-4470/37/2/012}.

\bibitem{CAO_2003}
J.~Cao, H.-Q. Lin, K.-J. Shi and Y.~Wang,
\newblock \emph{Exact solution of xxz spin chain with unparallel boundary
  fields},
\newblock Nuclear Physics B \textbf{663}(3), 487 (2003),
\newblock \doi{https://doi.org/10.1016/S0550-3213(03)00372-9}.

\bibitem{CAO_2013}
J.~Cao, W.-L. Yang, K.~Shi and Y.~Wang,
\newblock \emph{Off-diagonal bethe ansatz solution of the xxx spin chain with
  arbitrary boundary conditions},
\newblock Nuclear Physics B \textbf{875}(1), 152 (2013),
\newblock \doi{https://doi.org/10.1016/j.nuclphysb.2013.06.022}.

\bibitem{CAO_2013_2}
J.~Cao, W.-L. Yang, K.~Shi and Y.~Wang,
\newblock \emph{Off-diagonal bethe ansatz solutions of the anisotropic spin-12
  chains with arbitrary boundary fields},
\newblock Nuclear Physics B \textbf{877}(1), 152 (2013),
\newblock \doi{https://doi.org/10.1016/j.nuclphysb.2013.10.001}.

\bibitem{NICCOLI_2013}
G.~Niccoli,
\newblock \emph{Antiperiodic spin-1/2 xxz quantum chains by separation of
  variables: Complete spectrum and form factors},
\newblock Nuclear Physics B \textbf{870}(2), 397 (2013),
\newblock \doi{https://doi.org/10.1016/j.nuclphysb.2013.01.017}.

\bibitem{belliard_2013}
S.~Belliard, N.~Cramp{\'e} \emph{et~al.},
\newblock \emph{Heisenberg xxx model with general boundaries: eigenvectors from
  algebraic bethe ansatz},
\newblock SIGMA. Symmetry, Integrability and Geometry: Methods and Applications
  \textbf{9}, 072 (2013).

\bibitem{Nepomechie_2013_3}
R.~I. Nepomechie and C.~Wang,
\newblock \emph{Boundary energy of the open {XXX} chain with a non-diagonal
  boundary term},
\newblock Journal of Physics A: Mathematical and Theoretical \textbf{47}(3),
  032001 (2013),
\newblock \doi{10.1088/1751-8113/47/3/032001}.

\bibitem{Faldella_2014}
S.~Faldella, N.~Kitanine and G.~Niccoli,
\newblock \emph{The complete spectrum and scalar products for the open spin-1/2
  {XXZ} quantum chains with non-diagonal boundary terms},
\newblock Journal of Statistical Mechanics: Theory and Experiment
  \textbf{2014}(1), P01011 (2014),
\newblock \doi{10.1088/1742-5468/2014/01/p01011}.

\bibitem{Kitanine_2014}
N.~Kitanine, J.~M. Maillet and G.~Niccoli,
\newblock \emph{Open spin chains with generic integrable boundaries: Baxter
  equation and bethe ansatz completeness from separation of variables},
\newblock Journal of Statistical Mechanics: Theory and Experiment
  \textbf{2014}(5), P05015 (2014),
\newblock \doi{10.1088/1742-5468/2014/05/p05015}.

\bibitem{LI_2014}
Y.-Y. Li, J.~Cao, W.-L. Yang, K.~Shi and Y.~Wang,
\newblock \emph{Thermodynamic limit and surface energy of the xxz spin chain
  with arbitrary boundary fields},
\newblock Nuclear Physics B \textbf{884}, 17 (2014),
\newblock \doi{https://doi.org/10.1016/j.nuclphysb.2014.04.010}.

\bibitem{Pozsgay_2018}
B.~Pozsgay and O.~R{\'{a}}kos,
\newblock \emph{Exact boundary free energy of the open {XXZ} chain with
  arbitrary boundary conditions},
\newblock Journal of Statistical Mechanics: Theory and Experiment
  \textbf{2018}(11), 113102 (2018),
\newblock \doi{10.1088/1742-5468/aae5a5}.

\bibitem{Colpa_1979}
J.~H.~P. Colpa,
\newblock \emph{Diagonalisation of the quadratic fermion hamiltonian with a
  linear part},
\newblock Journal of Physics A: Mathematical and General \textbf{12}(4), 469
  (1979),
\newblock \doi{10.1088/0305-4470/12/4/008}.

\bibitem{BARIEV_1991}
R.~Bariev and I.~Peschel,
\newblock \emph{Non-universal critical behaviour in a two-dimensional ising
  model with a field},
\newblock Physics Letters A \textbf{153}(4), 166 (1991),
\newblock \doi{https://doi.org/10.1016/0375-9601(91)90786-8}.

\bibitem{Bilstein_1999}
U.~Bilstein and B.~Wehefritz,
\newblock \emph{The -model with boundaries: Part i. diagonalization of the
  finite chain},
\newblock Journal of Physics A: Mathematical and General \textbf{32}(2), 191
  (1999),
\newblock \doi{10.1088/0305-4470/32/2/001}.

\bibitem{Campostrini_2015}
M.~Campostrini, A.~Pelissetto and E.~Vicari,
\newblock \emph{Quantum ising chains with boundary fields},
\newblock Journal of Statistical Mechanics: Theory and Experiment
  \textbf{2015}(11), P11015 (2015),
\newblock \doi{10.1088/1742-5468/2015/11/p11015}.

\bibitem{XavierRajabpour2020}
J.~C. Xavier and M.~A. Rajabpour,
\newblock \emph{Entanglement and boundary entropy in quantum spin chains with
  arbitrary direction of the boundary magnetic fields},
\newblock Phys. Rev. B \textbf{101}, 235127 (2020),
\newblock \doi{10.1103/PhysRevB.101.235127}.

\bibitem{Wick_1950}
G.~C. Wick,
\newblock \emph{The evaluation of the collision matrix},
\newblock Phys. Rev. \textbf{80}, 268 (1950),
\newblock \doi{10.1103/PhysRev.80.268}.

\bibitem{Vidal_Kitaev_2003}
G.~Vidal, J.~I. Latorre, E.~Rico and A.~Kitaev,
\newblock \emph{Entanglement in quantum critical phenomena},
\newblock Phys. Rev. Lett. \textbf{90}, 227902 (2003),
\newblock \doi{10.1103/PhysRevLett.90.227902}.

\bibitem{latorre_2003}
J.~I. Latorre, E.~Rico and G.~Vidal,
\newblock \emph{Ground state entanglement in quantum spin chains},
\newblock arXiv preprint quant-ph/0304098  (2003).

\bibitem{Peschel_2003}
I.~Peschel,
\newblock \emph{Calculation of reduced density matrices from correlation
  functions},
\newblock Journal of Physics A: Mathematical and General \textbf{36}(14), L205
  (2003),
\newblock \doi{10.1088/0305-4470/36/14/101}.

\bibitem{vanHemmen1980}
J.~L. van Hemmen,
\newblock \emph{A note on the diagonalization of quadratic boson and fermion
  hamiltonians},
\newblock Zeitschrift f{\"u}r Physik B Condensed Matter \textbf{38}(3), 271
  (1980),
\newblock \doi{10.1007/BF01315667}.

\bibitem{Zanardi_2007}
M.~Cozzini, P.~Giorda and P.~Zanardi,
\newblock \emph{Quantum phase transitions and quantum fidelity in free fermion
  graphs},
\newblock Phys. Rev. B \textbf{75}, 014439 (2007),
\newblock \doi{10.1103/PhysRevB.75.014439}.

\bibitem{CARACCIOLO2013474}
S.~Caracciolo, A.~D. Sokal and A.~Sportiello,
\newblock \emph{Algebraic/combinatorial proofs of cayley-type identities for
  derivatives of determinants and pfaffians},
\newblock Advances in Applied Mathematics \textbf{50}(4), 474 (2013),
\newblock \doi{https://doi.org/10.1016/j.aam.2012.12.001}.

\bibitem{Balian1969}
R.~Balian and E.~Brezin,
\newblock \emph{Nonunitary bogoliubov transformations and extension of wick's
  theorem},
\newblock Il Nuovo Cimento B (1965-1970) \textbf{64}(1), 37 (1969),
\newblock \doi{10.1007/BF02710281}.

\end{thebibliography}

%%%%%%%%%%%%%%%%%%%%%%%%%%%%\end{document}

\end{document}